\documentclass[preprint]{elsarticle}
\usepackage[colorlinks=true,linkcolor=blue,urlcolor=darkblue,plainpages=false]{hyperref}\usepackage{color}
\usepackage{comment}
\usepackage{gensymb}
\usepackage{graphicx}
\usepackage{amssymb}

\usepackage{lineno}
\newcommand{\snn}{{\sqrt{s_{\rm NN}}}}
\newcommand{\pp}{{p+p}}
\newcommand{\auau}{{\rm Au+Au}}
\newcommand{\pau}{{p+{\rm Au}}}

\begin{document}

\begin{frontmatter}

\title{The sPHENIX Micromegas Outer Tracker}
\date{\today}
% LANL: 
% Hugo Pereira Da Costa
% Eric Renner
% Bade Sayki
% Walter Sondheim

% Saclay:
% Stephan Aune
% Arnaud Bonenfant
% Audrey Francisco
% Cyril Goblin
% Aude Grabas
% Irakli Mandjavidze
% Maxence Vandenbroucke
% Anna Wils

% MIT:
% Jim Kelsey
% Chris Vidal

% BNL:
% Bob Azmoun
% Stephen Boose
% Dan Cacace
% Russ Feder
% John Haggerty
% Jin Huang
% Ivan Kotov
% John Kuczewski
% Jim Mills
% Rob Pisani
% Rich Ruggiero 
% Takao Sakaguchi
% Joel Vasquez
% Joe Mead 
% Chris Pinkenburg

% Lehigh university
% Tristan Protzman

% University Sao Paulo 
% W. A. M. Van Noije (1) 
% M. Bregant (1)
% B. C. Sanches (1)
% H. D. H. Herrera (1)+
% R. A. Hernandez (1)
% R. W. da Silva (1)
% T. A. Martins (1)

% LUND:
% David Silvermyr
% Anders Oskarsson

%% aknowledgments
% Sal Pollizzo
% Mike Lenz, 
% Carter Biggs, 
% Jim Labounty, 
% Frank Toldo

\author[CEA]{S.~Aune}
\author[BNL]{B.~Azmoun}
\author[CEA]{A.~Bonenfant}
\author[BNL]{S.~Boose}
\author[USP]{M.~Bregant}
\author[BNL]{D.~Cacace}
\author[USP]{R.~W.~da~Silva}
\author[BNL]{R.~Feder}
\author[CEA]{A.~Francisco}
\author[CEA]{C.~Goblin}
\author[CEA]{A.~Grabas}
\author[BNL]{J.~S.~Haggerty}
\author[USP]{R.~A.~Hernandez}
\author[USP]{H.~D.~H.~Herrera\corref{cor2}}
\author[BNL]{J.~Huang}
\author[MIT]{J.~Kelsey}
\author[BNL]{I.~Kotov}
\author[BNL]{J.~Kuczewski}
\author[CEA]{I.~Mandjavidze}
\author[USP]{T.~A.~Martins}
\author[BNL]{J.~Mead}
\author[BNL]{J.~Mills}
\author[LUND]{A.~Oskarsson}
\author[LANL]{H.~Pereira~Da~Costa\corref{cor1}}
\ead{hugo.pereira-da-costa@lanl.gov}
\author[BNL]{C.~Pinkenburg}
\author[BNL]{R.~Pisani}
\author[LEHIGH]{T.~Protzman}
\author[BNL]{M.~L.~Purschke}
\author[LANL]{E.~Renner}
\author[BNL]{R.~Ruggiero}
\author[BNL]{T.~Sakaguchi}
\author[USP]{B.~C.~S.~Sanches}
\author[LANL]{B.~Sayki}
\author[LUND]{D.~Silvermyr}
\author[LANL]{W.~Sondheim}
\author[CEA]{M.~Vandenbroucke}
\author[USP]{W.~A.~M.~Van Noije}
\author[BNL]{J.~Vasquez}
\author[MIT]{C.~Vidal}
\author[CEA]{A.~Wils}

\cortext[cor2]{\,now SLAC - Stanford University}
\cortext[cor1]{\,Corresponding author.}

\affiliation[CEA]{
  organization={CEA, Université Paris-Saclay},
  city={Gif-sur-Yvette},
  country={France}
}

\affiliation[BNL]{
  organization={Brookhaven National Laboratory},
  city={Upton},
  state={New York}
}

\affiliation[LANL]{
  organization={Los Alamos National Laboratory},
  city={Los Alamos},
  state={New Mexico}
}

\affiliation[LEHIGH]{
  organization={Lehigh University},
  city={Bethlehem},
  state={Pennsylvania}
}

\affiliation[MIT]{
  organization={Massachusetts Institute of Technology},
  city={Cambridge},
  state={Massachusetts}
}

\affiliation[USP]{
  organization={Universidade de São Paulo},
  city={São Paulo},
  country={Brazil}
}

\affiliation[LUND]{
  organization={Lund University},
  city={Lund},
  country={Sweden}
}

\begin{abstract}
The sPHENIX Time Projection Chamber Outer Tracker (TPOT) is a Micromegas based detector. It is a part of the sPHENIX experiment that aims to facilitate the calibration of the Time Projection Chamber, in particular the correction of the time-averaged and beam-induced distortions of the electron drift. This paper describes the detector mission, setup, construction, installation, commissioning and performance during the first year of sPHENIX data taking.

\end{abstract}


\end{frontmatter}

\section{\label{sec:introduction}Introduction}
The sPHENIX detector~\cite{PHENIX:2015siv, Belmont:2023fau} is an experiment located at the Relativistic Heavy Ion Collider (RHIC)~\cite{HARRISON2003235} in Brookhaven National Laboratory (BNL).
%The sPHENIX experiment~\cite{PHENIX:2015siv, Belmont:2023fau} is being conducted at the Relativistic Heavy Ion Collider (RHIC)~\cite{HARRISON2003235} at the Brookhaven National Laboratory (BNL). 
It focuses on measuring jets as well as open and hidden heavy flavor production in heavy ion collisions and to study the properties of the Quark Gluon Plasma created in gold on gold ($\auau$) collisions at a center of mass energy per nucleon-nucleon collision $\snn = 200$\,GeV. It will also collect data using proton-proton ($\pp$) and possibly proton-gold ($\pau$) collisions to serve as a reference to the $\auau$ data and to study the cold nuclear matter that constitutes the initial state of such collisions.

\begin{figure}[htb]
 \centering
 \includegraphics[width=0.95\textwidth]{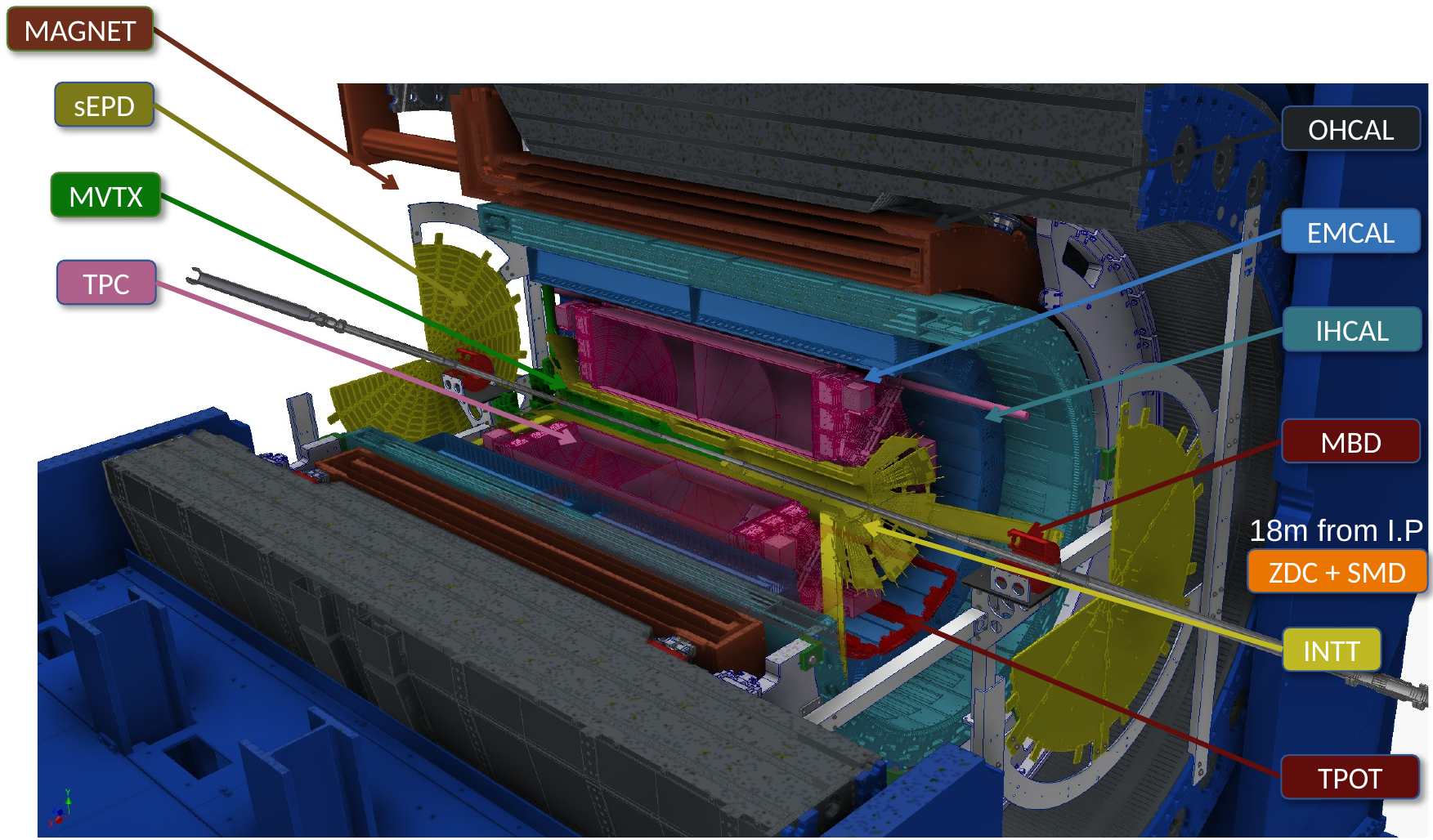}
 \caption{Schematic view of the sPHENIX detector, including all subsystems. The TPOT detector appears in red. It is located between the EMCAL and the TPC, at the bottom of the TPC.} 
 \label{fig:sphenix_layout}
\end{figure}

Figure~\ref{fig:sphenix_layout} presents a schematic view of the sPHENIX detector. The two beams provided by RHIC collide at the center of the experiment. The products of the collisions are measured in the various subsystems that constitute the detector. Going outwards starting from the beam line, sPHENIX is comprised of the following subsystems: the MAPS-Based Vertex Detector (MVTX); the INTermediate Tracker (INTT); the Time Projection Chamber (TPC); the Time Projection Chamber Outer Tracker (TPOT), which is the topic of this paper; the Electromagnetic Calorimeter (EMCAL); the Inner Hadronic Calorimeter (IHCAL); the solenoid superconducting magnet that delivers a longitudinal magnetic field of intensity 1.4\,T and the Outer Hadronic Calorimeter (OHCAL). 
The tracking subsystems are the MVTX, INTT, TPC and TPOT. The calorimetry subsystems are the EMCAL, IHCAL and OHCAL. sPHENIX also contains a number of forward detectors, namely the Minimum Bias Detectors (MBD), the sPHENIX Event Plane detectors (sEPD), the Zero Degree Calorimeters (ZDC) and the Shower Maximum Detector (SMD), which is part of the ZDC. 

\begin{figure}[htb]
 \centering
 \includegraphics[width=0.4\textwidth]{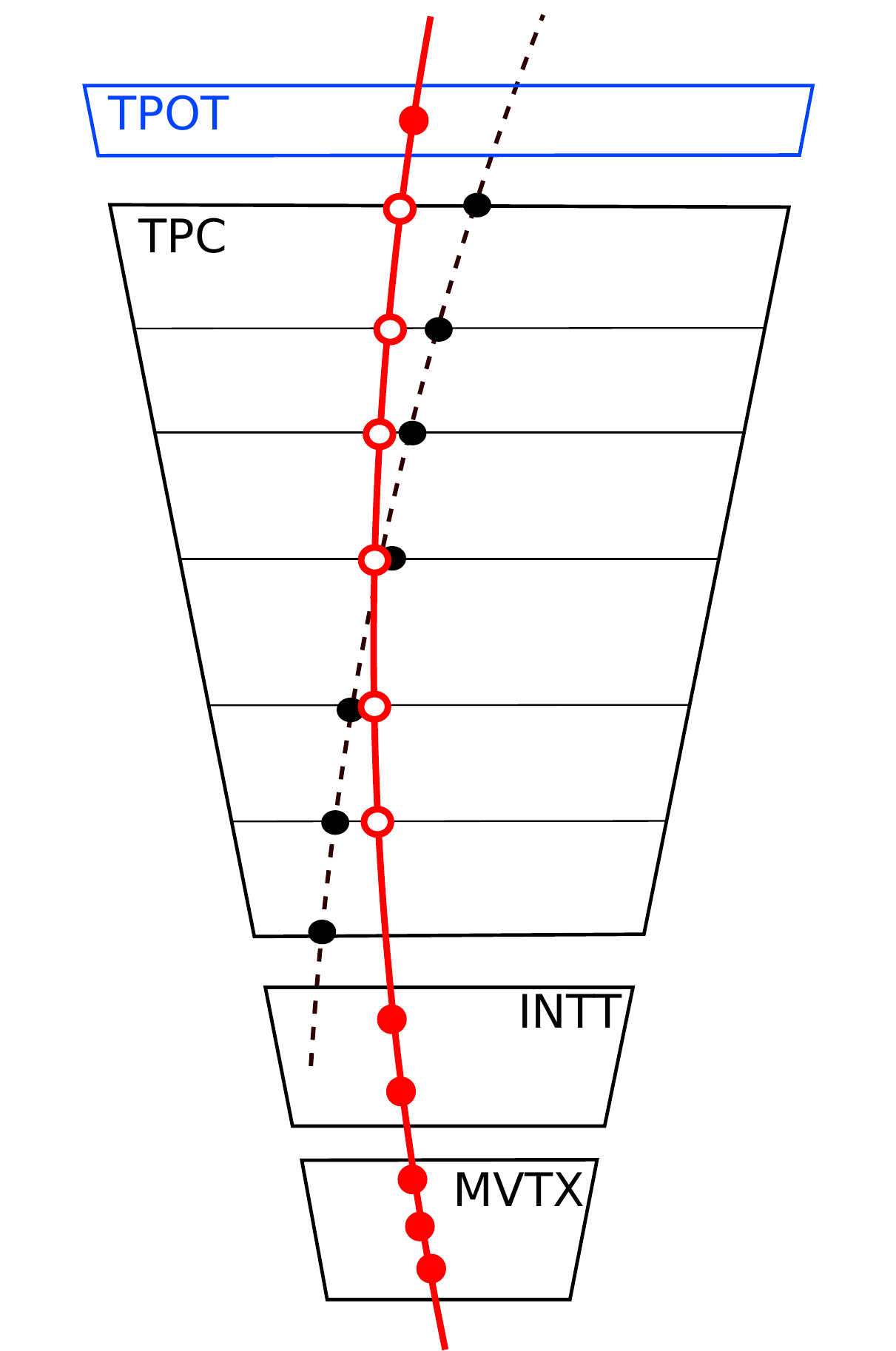}
 \caption{Schematic view of the TPC calibration procedure using all available outside detectors (INTT, MVTX and TPOT).} 
 \label{fig:sc_reconstruction}
\end{figure}

TPOT is a late addition to the sPHENIX apparatus. It is installed on the outside of the TPC and greatly facilitates the calibration of the TPC.
In particular, it allows to make the maximum use of tracks for measuring in near real time the distortions of the electron drift in the TPC during data taking. It consists of 16 Micromesh Gaseous detectors (Micromegas)~\cite{GIOMATARIS199629} grouped two-by-two to provide an additional space point on the outside of the TPC in a fraction of the TPC acceptance. Figure~\ref{fig:sc_reconstruction} presents a schematic view of how the TPC can be calibrated using all of the other tracking detectors. A reference track (in red) is built using the clusters measured in the MVTX, INTT and TPOT (filled red point). It is interpolated inside the TPC volume and the expected cluster positions are inferred (open red circles). For each TPC layer, the expected position is compared to the measured clusters in the TPC and the difference between the two is used to calibrate the TPC. This calibration is then extrapolated to the full acceptance of the TPC. 

This paper describes the design, construction, testing, installation, and commissioning of the TPOT detector. Section~\ref{sec:system_description} is dedicated to the description of the TPOT system (detector, electronics, and services); section~\ref{sec:installation} to the detector installation inside sPHENIX and survey; and section~\ref{sec:detector_performance} to the detector characterization and performance.

\section{\label{sec:system_description}System description}
\subsection{Requirements}
The requirements on TPOT for it to be installed and operated as a part of sPHENIX are listed below. 

\textit{Available space:} the purpose of TPOT is to provide an additional position measurement to the reconstructed particles' trajectories on the outside of the TPC. It must therefore be located as close as possible to the outer radius of the TPC. In the initial sPHENIX layout, there is a radial gap of 11\,cm on average between the TPC and the covers of the EMCAL sectors. This is the maximum space available for TPOT. It puts strong constrains on the maximum thickness of each detector layer, the number of layers that can be installed and on the mechanical support structure for the detector. 

\textit{Performance:} because TPOT is not used as a measurement device critical for the sPHENIX physics program but rather as a calibration device for the TPC, the requirements on the detector in terms of detection efficiency and spatial resolution are somewhat loose. A minimum of 90\% detection efficiency is required, together with a spatial resolution of a few 100\,$\mu$m in both azimuthal and longitudinal directions. The requirement on the detector spatial resolution is driven by the amount of time it takes to achieve an accuracy of order 10\,$\mu$m for reconstructing the distortions of the electron drift in the TPC with sufficient granularity (a few square centimeters). With 500\,$\mu$m resolution, a trigger rate of 15\,kHz and an expected mean number of particles per Micromegas detector and $\auau$ collision around eight, this integration time is estimated to be less than a minute, much smaller than the typical time over which the distortions of the electron drift in the TPC is expected to vary significantly, beyond event by event fluctuations. The positioning of the detector itself is not critical, and an accuracy of 1\,mm or less is required. This is provided that the real position (alignment) of the detector can be measured {\em a posteriori} with a precision of 100\,$\mu$m or less, significantly smaller than the expected spatial resolution of the detector.

\textit{Cost and schedule:} the decision to construct and install TPOT as a part of the sPHENIX apparatus was taken late in the sPHENIX design process, about eighteen months before the beginning of sPHENIX operations, and based on design studies started approximately six months before that. This compressed schedule, together with the total budget allocated to the project put strong constraints on the amount of prototyping that could be performed before starting the construction of the detector, as well as the number of modules that could be built. In particular, it was decided early on that TPOT would only cover a fraction of the TPC acceptance (Section~\ref{subsec:tpot_description}). Extrapolation methods would then be developed offline to extrapolate the calibrations provided by TPOT to the rest of the TPC acceptance.

\textit{Magnetic field:} TPOT is installed inside sPHENIX superconducting magnet, which delivers an approximately uniform longitudinal magnetic field of intensity 1.4\,T. The magnetic field directly impacts TPOT operation, in particular the choice of gas that is used to operate the Micromegas chambers and the electric field in the detector's drift gap (Section~\ref{subsubsec:principle}). In addition, its presence prevents the use of any magnetic component for both the detector and the mechanical support structure.

\textit{Particle rates and discharges:} 
because of the requirements above in terms of performance, cost, schedule and available space, it was realized early on that the Micromegas technology~\cite{GIOMATARIS199629} is a good candidate for building TPOT (section~\ref{subsec:micromegas_detectors}). However such detectors are known to be susceptible to electrostatic discharges when the number of electrons created in the avalanche process responsible for the signal amplification exceeds the Raether limit~\cite{raether1964electron} of $2 \times 10^{7}$ electrons. Based on simulations of the expected particle rates in TPOT, the mean deposited energy of these particles, and for a typical amplification gain of $6\times10^{3}$, the rate of such electrostatic discharge was estimated to be about 100\,Hz. To prevent these discharges from generating a prohibitively high dead-time in the detector or even damaging the readout electronics, they must be quenched, in the TPOT case by adding a resistive layer on top of the readout plane.

\subsection{\label{subsec:tpot_description}Detector segmentation}

TPOT is made up of eight modules each consisting of two Micromegas chambers stacked radially, one to measure the coordinate along the longitudinal direction ($z$) that follows the beam axis and the second along the azimuthal direction ($\phi$), as defined in polar coordinates around $z$. They are referred to as the $z$ and $\phi$ views, respectively. The dimension of each chamber's active area is 256\,mm along $\phi$ and 512\,mm along $z$. The signal on the $\phi$ views is collected on 256 strips oriented along $z$ and with a pitch of 1\,mm. The signal on the $z$ views is collected on 256 strips oriented along $\phi$ and with a pitch of 2\,mm. Four of the eight TPOT modules cover the full $z$ extent of the bottom-most sector of the TPC (out of twelve sectors total), and approximately one half of its azimuthal extent. The other four modules are located along the neighbor two sectors (two modules each). With this configuration, the eight TPOT modules cover approximately 8\% of the TPC acceptance.

\subsection{\label{subsec:micromegas_detectors}The Micromegas chambers}
\subsubsection{\label{subsubsec:principle}Principle}

\begin{figure}[htb]
 \centering
 \includegraphics[width=0.95\textwidth]{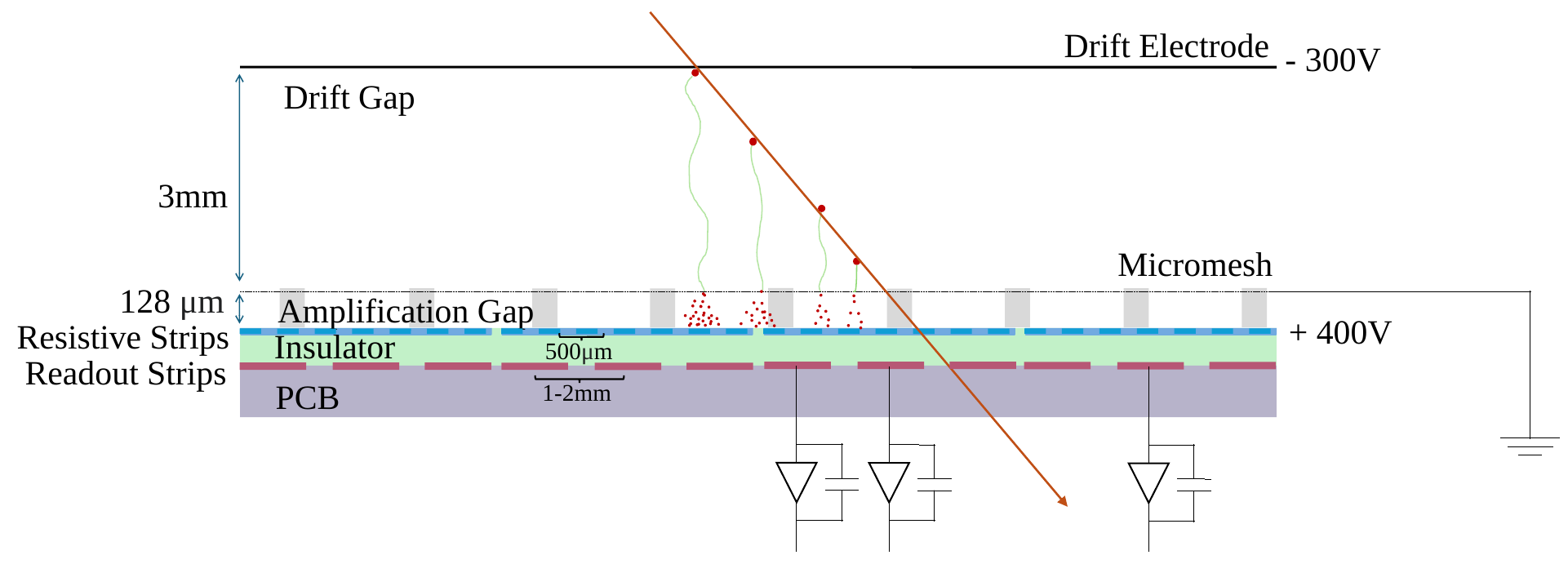}
 \caption{Schematic view of one bulk, resistive Micromegas chamber.} 
 \label{fig:micromegas_diagram}
\end{figure}

\begin{figure}[htb]
 \centering
 \includegraphics[width=0.7\textwidth]{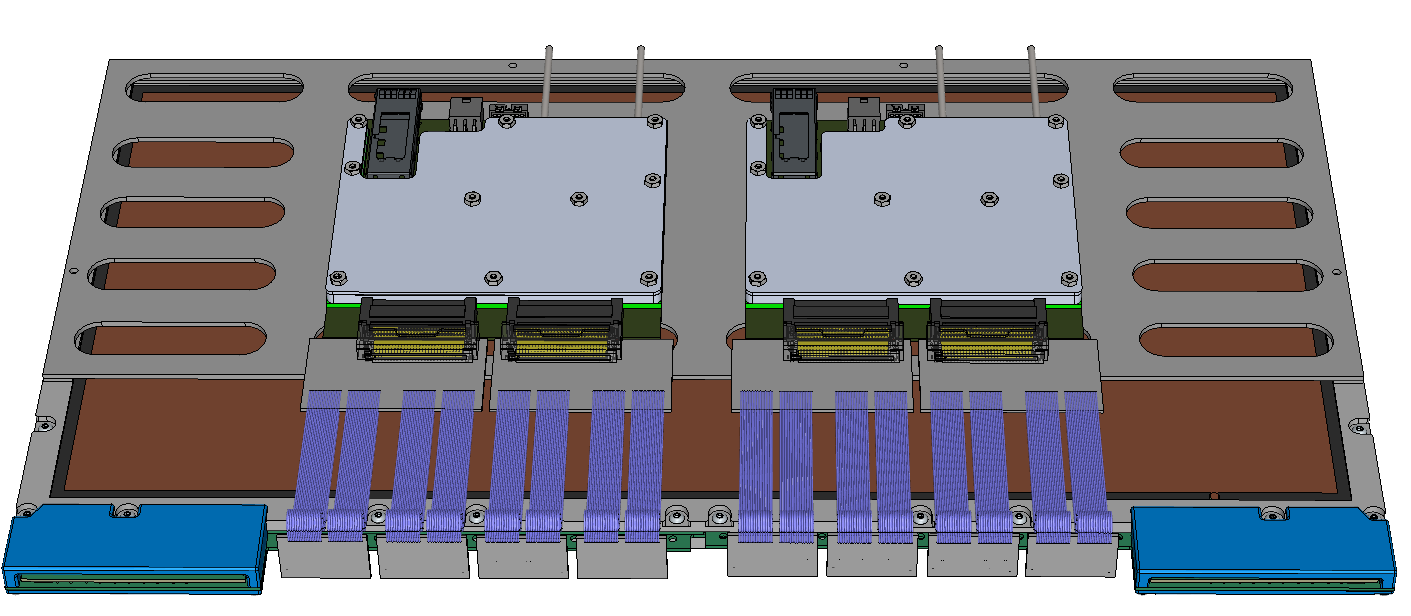}
 \caption{Schematics of a fully assembled TPOT module.} 
 \label{fig:tpot_module}
\end{figure}

Figure~\ref{fig:micromegas_diagram} shows a schematic view of a bulk resistive Micromegas chamber, and Figure~\ref{fig:tpot_module} a schematic view of a fully assembled TPOT module (out of eight modules total), with electronic boards and cooling plates mounted.

Each chamber consists of a drift electrode, a micromesh, a resistive layer and a readout layer arranged in parallel top to bottom. In order to guide electrons through the drift (or conversion) gap, a negative electric field is applied through the layers. The drift electrode is operated at negative voltage. The micromesh is located 3\,mm away from the drift electrode and connected to the electrical ground. The resistive layer is located 128\,$\mu$m away from the micromesh and is operated at a positive voltage. It is segmented into four groups to be connected to the High Voltage (HV) independently. This segmentation allows for mitigation of a HV short in one part of the chamber while allowing the other 3/4\textsuperscript{th} of the chamber to remain active. The readout layer is located immediately below the resistive layer and consists of 256 parallel strips per chamber. The strips read the signal collected on the resistive layer through capacitive coupling and are connected to the Front-End Electronics (FEE) boards.

To optimize the fabrication process, the chambers have no components soldered on the readout board. The thickness of the readout Printed Circuit Board (PCB) is 1.6\,mm to match the SAMTEC MEC8 edge connector standard. One side of the readout PCB is designed to form 4 MEC8 edge connectors to connect the 256 strips to the front-end electronics (Figure~\ref{fig:micromegas_pcb}). 

\begin{figure}[htb]
 \centering
 \includegraphics[width=0.5\textwidth]{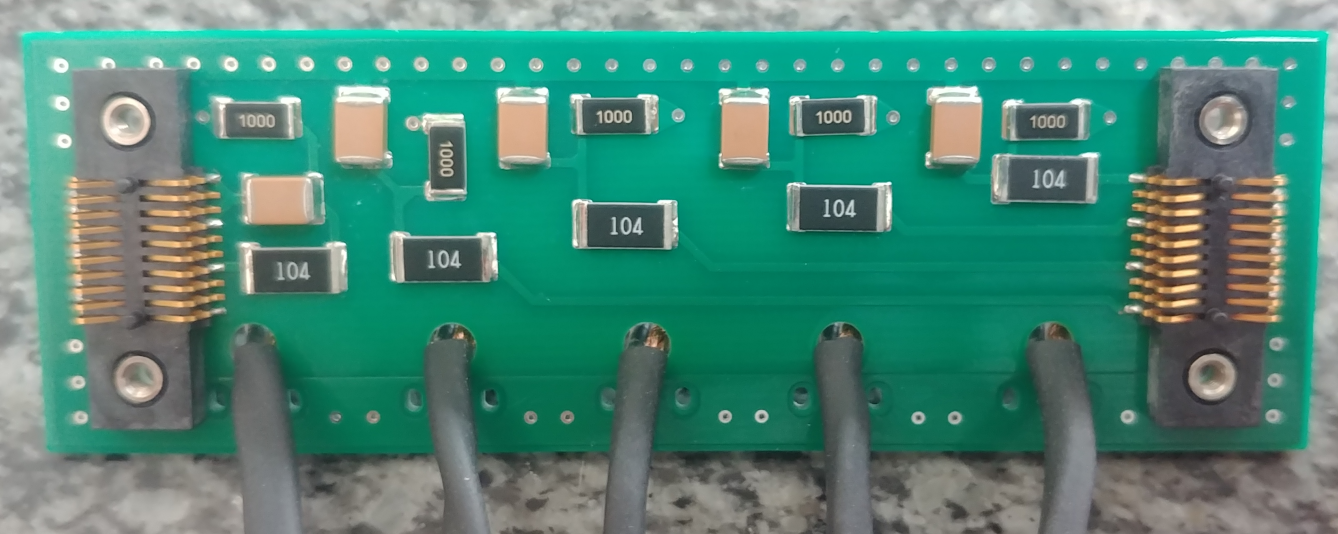}\\
 \includegraphics[width=0.5\textwidth]{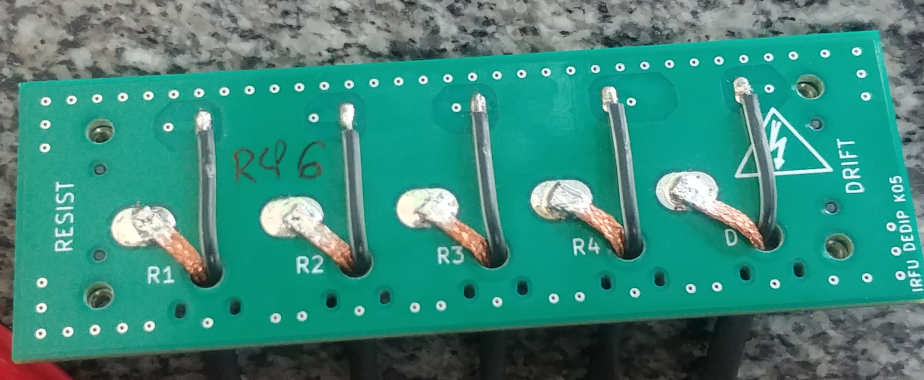}\\
 \includegraphics[width=0.5\textwidth]{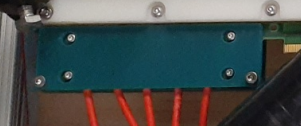}
 \caption{Pictures of the High Voltage boards. Top: Top view; Center: bottom view; bottom: HV board mounted on a Micromegas chamber.} 
 \label{fig:hv_board}
\end{figure}

Connection to HV is established using a dedicated board, a FSI one-piece compression connector from SAMTEC and its corresponding footprint on the readout PCB. Such a system can stably support up to 2\,kV voltage. The HV board is shown in Figure~\ref{fig:hv_board}. It houses the low-pass filter for five HV channels, four of which are connected to the resistive layer and one to the drift electrode, and the corresponding cables connected on the other end to the TPOT patch panel.

\begin{figure}[htb]
 \centering
 \includegraphics[width=0.49\textwidth]{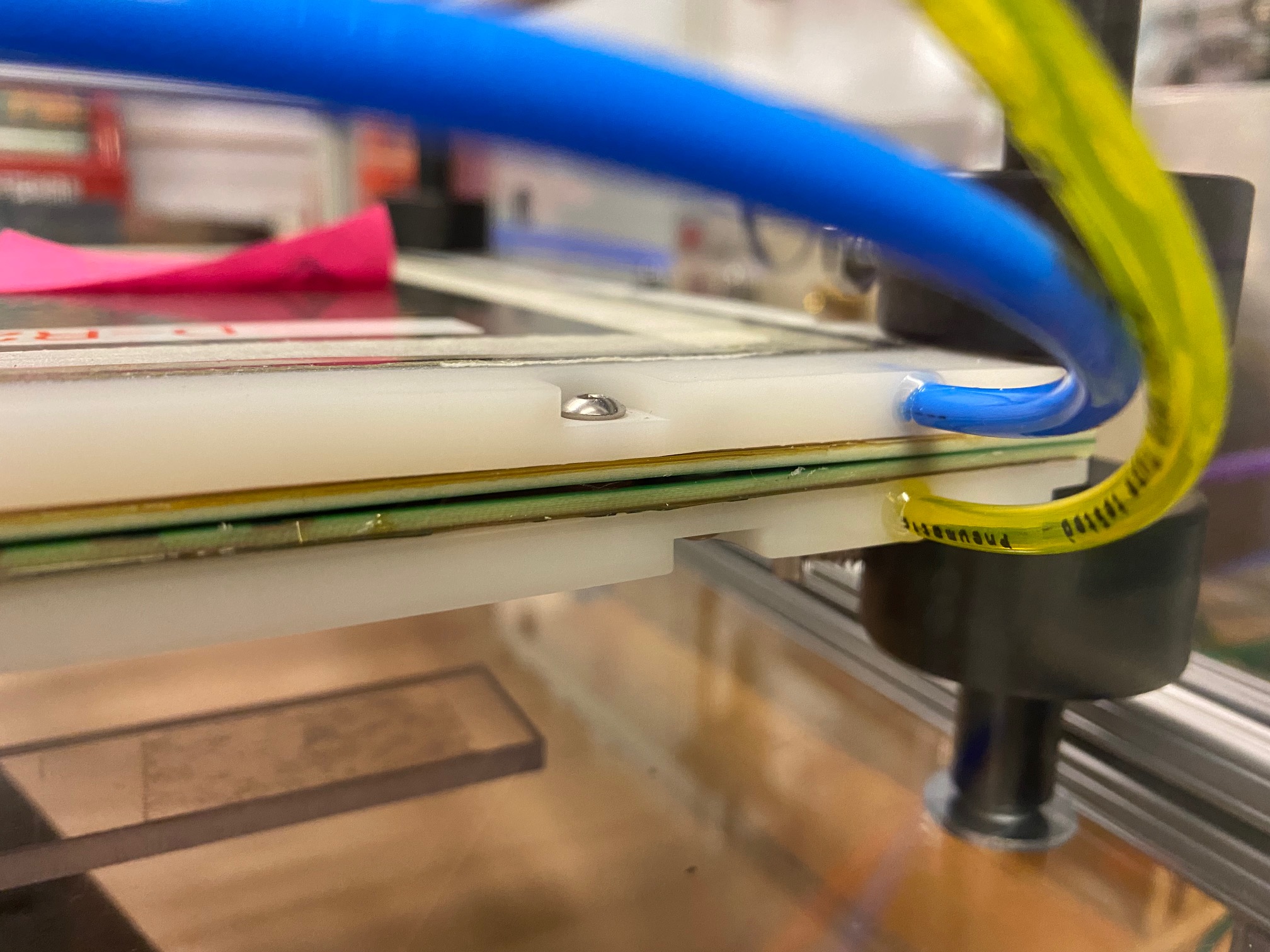}
 \includegraphics[width=0.49\textwidth]{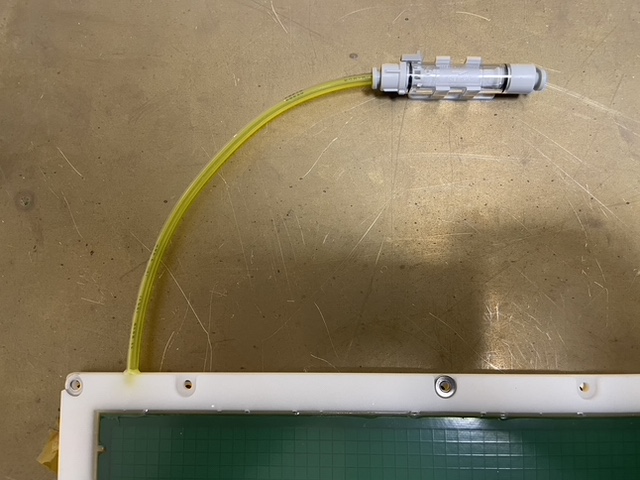}
 \caption{Left: picture of the gas inlet (yellow) and outlet (blue) on the two chambers of a TPOT module. The chambers are connected in series, so that the outlet of the first chamber is connected to the inlet of the second. Right: gas inlet tube with its filter.} 
 \label{fig:gas_inlet}
\end{figure}

The chambers are circulated with a 95/5\% mixture of Ar/iC\textsubscript{4}H\textsubscript{10} which provides good gain and stability for TPOT operation. In addition, this mixture has a small Lorentz angle, estimated to be less than 15$\degree$ in sPHENIX conditions, and defined as the angle between the drift direction of the electrons and the direction of the electric field in presence of a longitudinal magnetic field. It also allows to operate at a large drift electric field (about 1\,kV/cm for TPOT).
%This is important for TPOT considering the wide range of energy deposit that comes from heavy ion collisions. 
The distribution of the gas to the inside of the chamber is carried out by the 3D-printed frame that also holds the drift electrode and an O-ring seal for gas tightness. This 5\,mm thick frame cannot accommodate standard L-shape gas connectors. 3\,mm diameter polyurethane tubes glued directly inside the frame are used instead. The tubes are capped with filters to ensure cleanliness inside the chambers at all times (Figure~\ref{fig:gas_inlet}). Each chamber is operated at a gas flow rate ranging from $50$ to 100\,cc/min (cubic centimeter per minute) and slightly above atmospheric pressure. The material used for the 3D-printed frame is a high quality plastic, Accura ClearVue Free, SL 7870. 

\subsubsection{\label{subsubsec:readout}The readout electrode}
\begin{figure}[htb]
 \centering
 \includegraphics[width=0.49\textwidth]{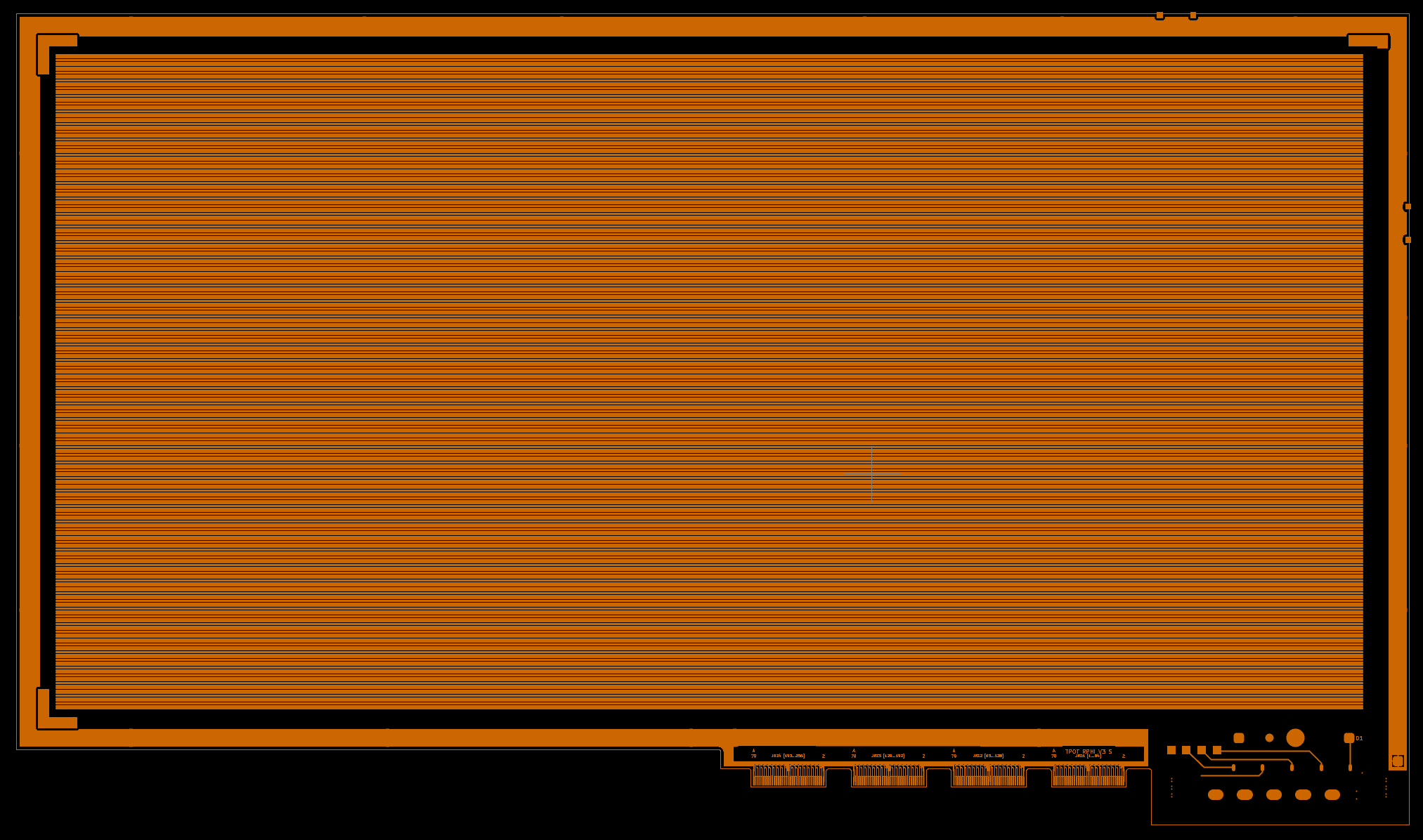}
 \includegraphics[width=0.49\textwidth]{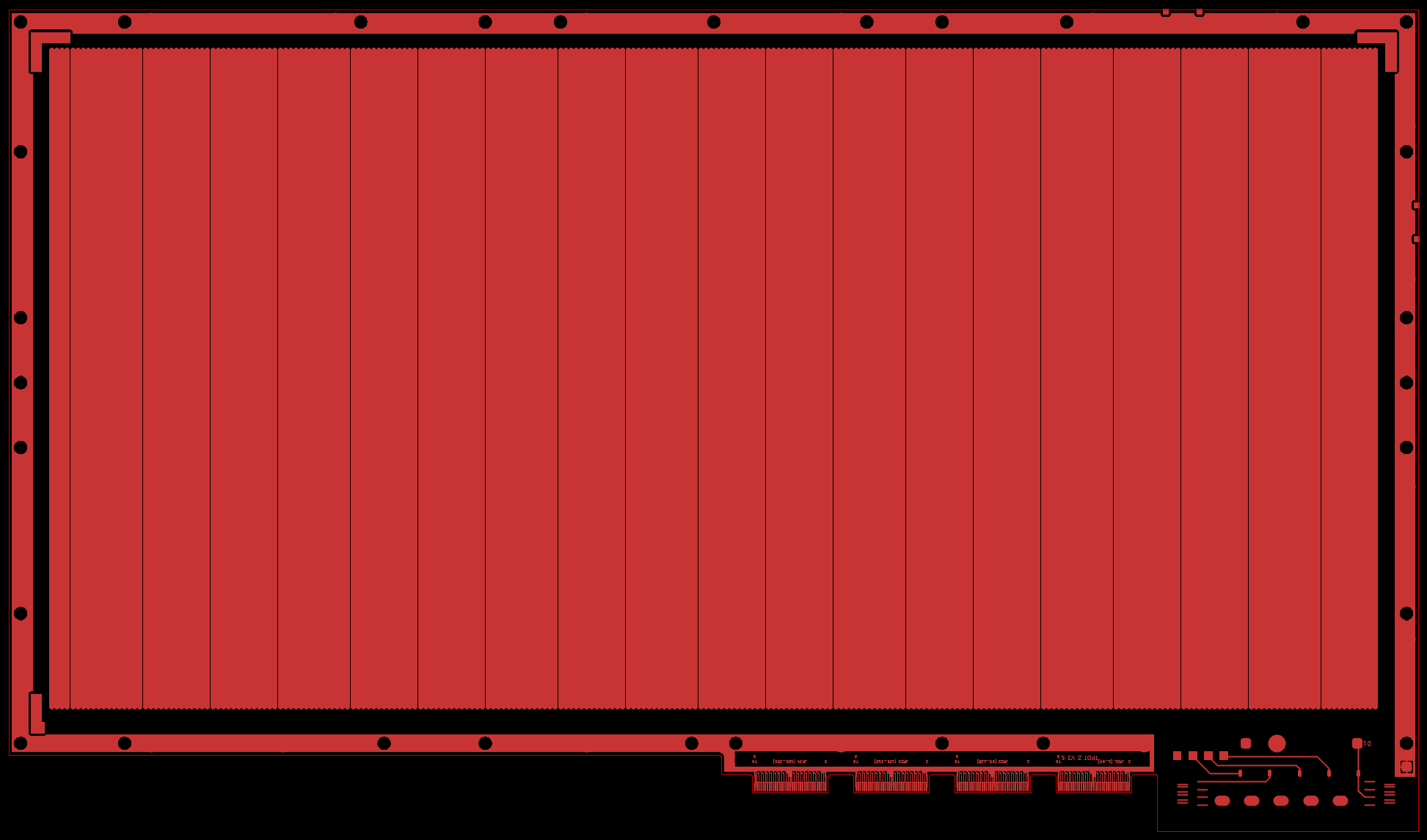}
 \caption{Layout of the readout PCB for the $\phi$ (left) and the $z$ (right) view. Shown are the strips that collect the signal, the four MEC8 connectors located at the bottom, as well as the location where the HV board from Fig.~\ref{fig:hv_board} connects to the detector. The actual pitch of the detector strips (1\,mm for the $\phi$ view and 2\,mm for the $z$) is not properly represented on the figures, due to their finite resolution.} 
 \label{fig:micromegas_pcb}
\end{figure}

The readout electrode is a multi-layer PCB of thickness 1.6\,mm and size 316$\times$542\,mm$^2$, with an active area of 256$\times$512\,mm$^2$.
The top layer corresponds to the chamber readout strips. 
For $\phi$ views, there are 256 straight strips of length $512$\,mm along the long dimension of the PCB ($z$) and with a pitch of 1\,mm and an interstrip distance of 100\,$\mu$m.
For the $z$ views, there are 256 strips of length $256$\,mm along the short dimension of the PCB ($\phi$) and with a pitch of 2\,mm and an interstrip distance of 500\,$\mu$m.
%along $\phi$.
A zigzag pattern is chosen for the $z$ view strips in order to increase the electric charge sharing between neighbor strips and thus the cluster size in the chamber, despite the large pitch. 
The other layers of the PCB are used for routing the readout strips to 4 MEC8 connectors of 64 (active) + 6 (ground) pads each, located on the side of the PCB (Figure~\ref{fig:micromegas_pcb}).
A ground layer is also added to the PCB layout to provide appropriate electromagnetic shielding to the chamber.

\subsubsection{\label{subsubsec:resist}The resistive layer}
\begin{figure}[htb]
 \centering
 \includegraphics[width=0.23\textwidth]{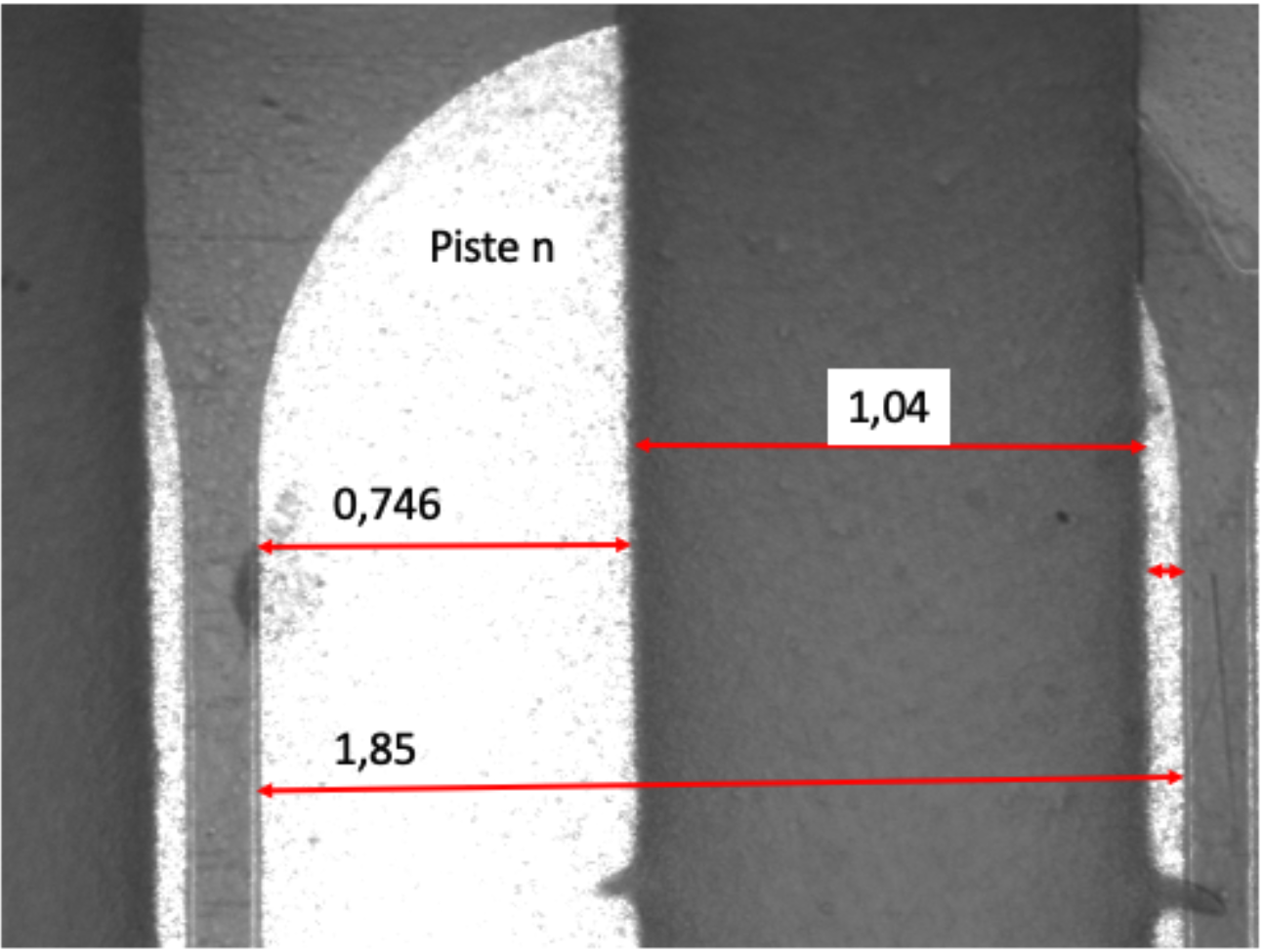}
 \includegraphics[width=0.23\textwidth]{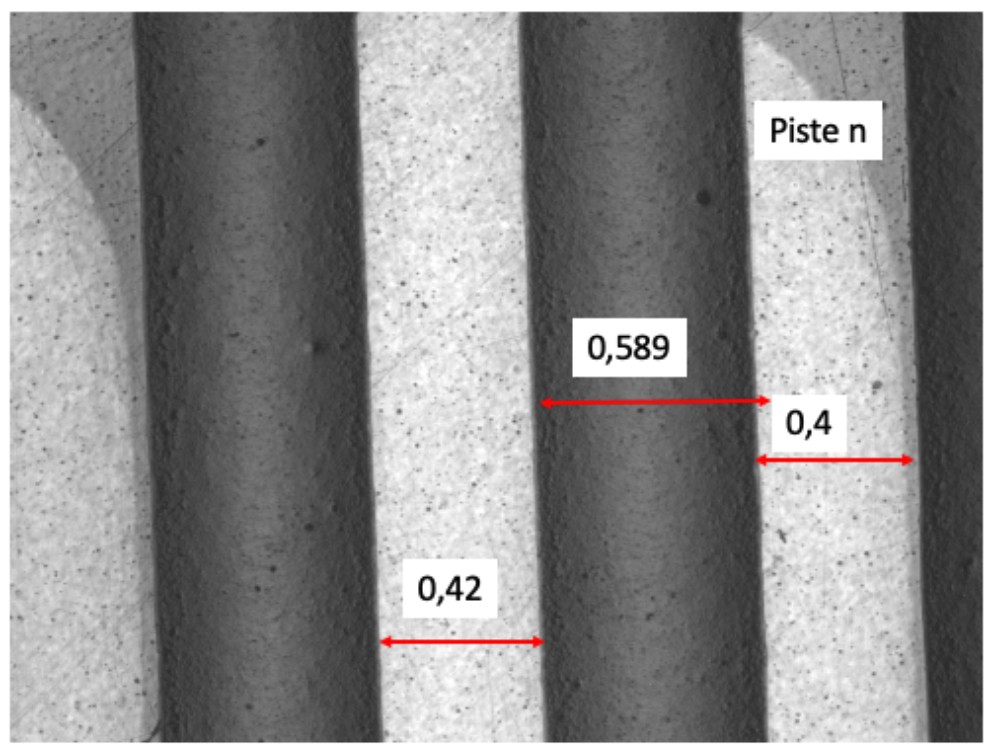}
 \includegraphics[width=0.23\textwidth]{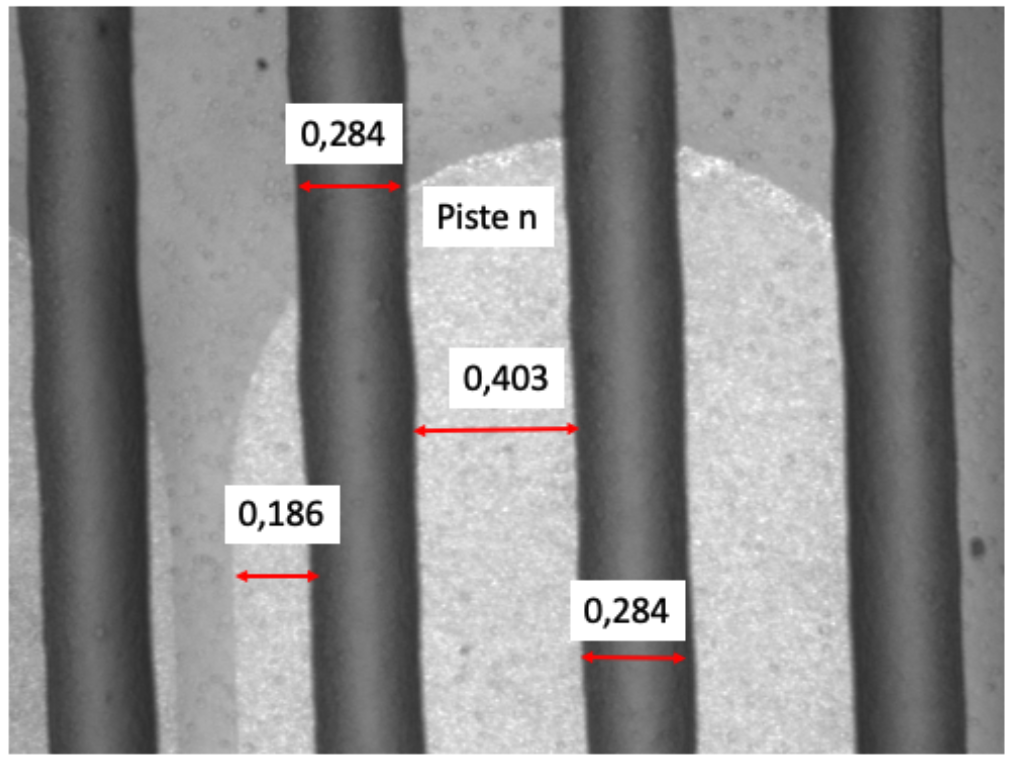}
 \includegraphics[width=0.26\textwidth]{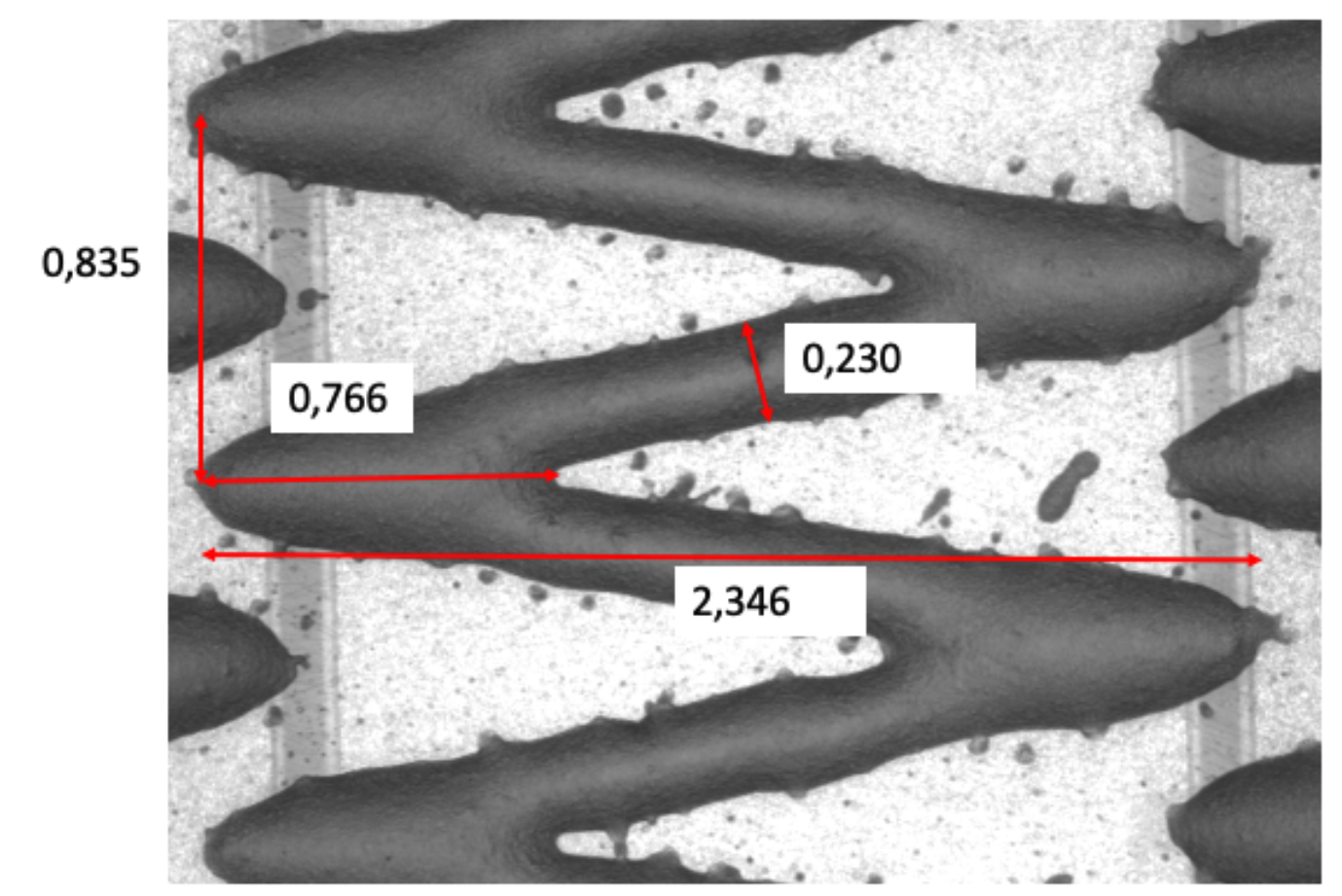}
 \caption{Examples of resistive layer patterns tested on a 2\,mm straight strip readout pattern. From left to right : (i) 2\,mm pitch and 1\,mm straight strip; (ii) 1\,mm pitch and 600\,$\mu$m straight strip; (iii) 700\,$\mu$m pitch and 290\,$\mu$m straight strip and (iv) 2\,mm zigzag strips.} 
 \label{fig:resist_layers}
\end{figure}

The resistive layer has several purposes: protection against electromagnetic discharges, protection of the readout electronics and spread of the charge.
Here the main concern is caused by the many highly ionizing particles created in heavy ion collisions. 
Early estimates of the charge deposited in the drift volume of the chamber and assuming a $\auau$ collision rate of 50\,kHz gave a 100\,Hz rate for crossing the Raether limit~\cite{raether1964electron} and likely causing discharges. Hence, the resistive layer was deemed necessary to protect the SAMPA electronics.
%which are not designed to support non ESD-like discharges. 
At 50\,kHz collision rate, the expected particle rate in a given Micromegas module is about 400\,kHz, corresponding to an average of eight particles passing through each module and for each $\auau$ collision. 
The resistive layer must evacuate the corresponding electric charges quickly.
A mix of 100\,k$\ohm/\square$ and 1\,M$\ohm/\square$ polymer resistors (RS12116 and RS12115 by FERRO) is used for the resistive layer material, to achieve a final resistivity of 270\,k$\ohm/\square$ after curing. 
%which forbids the use of Diamond-Like Carbon (DLC)~\cite{ROBERTSON1992185} as a resistive layer. 
%With low resistivity comes large cluster sizes. To prevent this, 
The resistive layer is segmented in the same direction as the readout strips. 
A number of pitch values and patterns have been tested for the resistive layer segmentation, illustrated in Figure~\ref{fig:resist_layers}. 
The one with the smallest pitch (700\,$\mu$m) was found to lead to the best (smallest) spatial resolution and was adopted for both views. 
%The material used for the resistive strips is a mix of 100\,k$\ohm/\square$ and 1\,M$\ohm/\square$ polymer resistors (RS12116 and RS12115 by FERRO), to achieve a final resistivity after curing of 270\,k$\ohm/\square$. 
The glue used to assemble the resistive layer to the readout electrode is EM-370 from Elite Material Co., Ltd.

\subsubsection{Fabrication}
The fabrication process of a typical bulk Micromegas chamber is described in~\cite{Giomataris_2006}. For TPOT additional steps are needed to include the resistive layer (Section~\ref{subsubsec:resist}). The final stack consists of the PCB, a kapton with the printed resisitive ink, two layers of 64\,$\mu$m photoresistive film (Pyralux), the mesh, a third layer of photoresistive film of thickness 64\,$\mu$m and the drift plate. The resistive layer itself is fabricated using a thin layer of isolating material (Kapton) on which the strip pattern is printed using ink screening technique. The readout PCB (Section~\ref{subsubsec:readout}) and the resistive layer are then glued (EM-370 pre-preg from Elite Material Co., Ltd.) and pressed together. The gluing and pressing is performed at CERN's Micro-Pattern Gaseous Detector (MPGD) laboratory. 
The resulting assembly is laminated twice at high temperature.
%The resulting assembly, the three layers of photoresistive film and the woven wire micromesh are then laminated together at high temperature. 
The photoresistive material is subsequently etched using photolitography to produce the pillars that keep the micromesh at the proper distance (128\,$\mu$m) from the resistive layer (Figure~\ref{fig:micromegas_diagram}). The pillars are cylindrical, with a diameter of 400\,$\mu$m, and located every 4\,mm (3\,mm) along (perpendicular to) $z$. 
The woven wire micromesh is then laminated on top of the stack and the third photoresistive layer is etched so to only keep the frame of the mesh and hold it into position. These steps are performed in the MPGD laboratory at the CEA research center of Saclay. A 45-18 stainless steel mesh from BOPP SD is used, with 18\,$\mu$m diameter wires and 45\,$\mu$m aperture, corresponding to 400 lines per inch. In parallel, the carbon fiber supported drift electrode is mounted on the 3D-printed frame which defines the active volume of the chamber and provides gas circulation. The drift electrode is made of a polyimide film laminated with a copper foil, glued onto a carbon plate for sturdiness. A second polyimide film laminated with a copper foil is glued to the top of the carbon plate, for grounding. The drift electrode mounted on the 3D-printed frame is then assembled to the Micromegas bulk, using a set of screws and an O-ring seal between the bulk and the printed frame to ensure gas tightness. 

\subsubsection{Conditioning}
The conditioning of the Micromegas chambers consists of incrementally raising the voltage of the resistive layer and the drift electrode, each time monitoring the current drawn on the corresponding channel and waiting for it to stabilize to a small, near-zero value before raising to the next step. This procedure is performed at multiple stages of the chamber fabrication and before installation.

At Saclay, the Micromegas bulk-assembly (readout plane, resistive layer and micromesh, Figure~\ref{fig:micromegas_diagram}) is first conditioned in air and without a drift electrode, to a voltage up to 800\,V. HV boards (Section~\ref{subsec:micromegas_detectors}, Figure~\ref{fig:hv_board}) are tested up to 1000\,V for approximately twelve hours. The drift electrodes are  conditioned independently up to 1500\,V. This conditioning is repeated once the chamber is assembled and the gas tightness is verified, using the nominal 95/5\% Ar/iC\textsubscript{4}H\textsubscript{10} gas mixture, up to a value of 450\,V for the resistive layer, and 500\,V for the drift electrode. This corresponds to a detector amplification gain of approximately $5\times10^{4}$, a value significantly larger than needed for routine data taking in sPHENIX (Sec.~\ref{subsubsec:operation_point}).
Gas tightness is checked at 2\,l/h gas flow and a leak rate tolerance of 0.05\,l/h.

Upon reception at BNL, the same conditioning of the assembled modules is repeated, to ensure that no damage occurred during transportation.

\subsection{\label{subsec:mechanical_structure}Mechanical structure}

\begin{figure}[htb]
 \centering
 \includegraphics[width=0.55\textwidth]{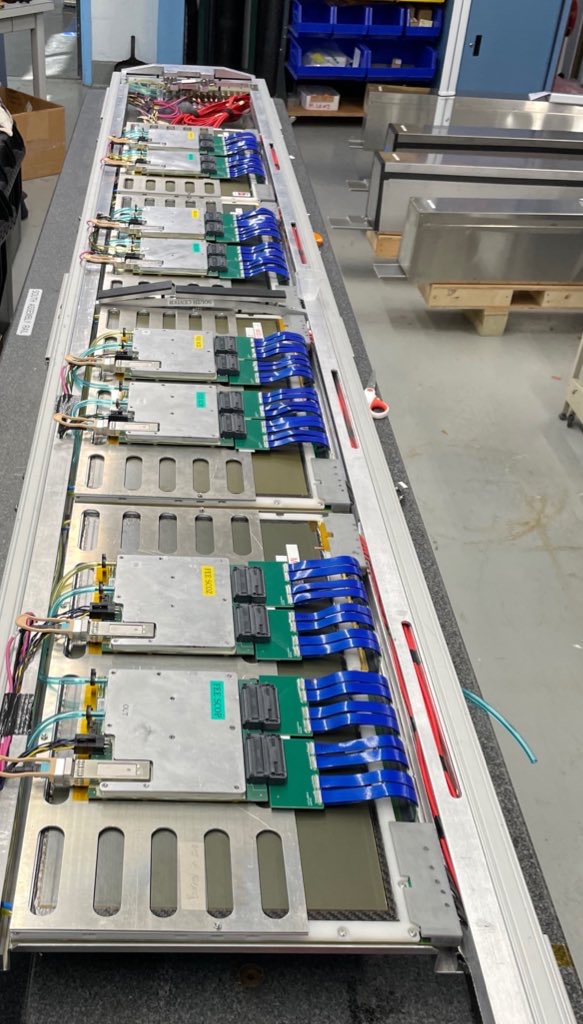}
 \caption{Picture of the central TPOT sector, fully assembled, with four Micromegas modules, readout electronics, cooling plates and services.} 
 \label{fig:tpot_sector_assembled}
\end{figure}

Figure~\ref{fig:tpot_sector_assembled} shows a picture of one fully assembled TPOT sector, out of three sectors total. The modules are supported on 80/20 10 series aluminum beams that form the frame for each of the three sectors. These are attached to cable trays that house the HV cables, low voltage cables, optical fibers as well as cooling and gas tubing, relevant for TPOT services (Section~\ref{subsec:services}). 
Aluminum plates are mounted at each end of the sector to support the extra cable, tube and fiber lengths between the end of the cable tray and the patch panel to which they are connected.
The three sectors are held together at a $150\degree$ angle by a total of fourteen trapezoidal connector brackets. 

Each sector has mounting systems at both ends that are connected to aluminum frames attached to the IHCAL. These enable precision movement in three directions during and after installation, for precise positioning of the detector. They consist of steel turnbuckles with ball joint ends as well as swivel tip set screws. 

To minimize deflection, adjustable support brackets are added to the midpoint of each sector, which rest on the EMCAL.
They consist of soft pads, an aluminum frame, as well as steel swivel tip set screws for precision leveling.

\subsection{\label{subsec:electronics}Electronics}
\subsubsection{Readout system architecture}

\begin{figure}[htb]
 \centering
 \includegraphics[width=0.95\textwidth]{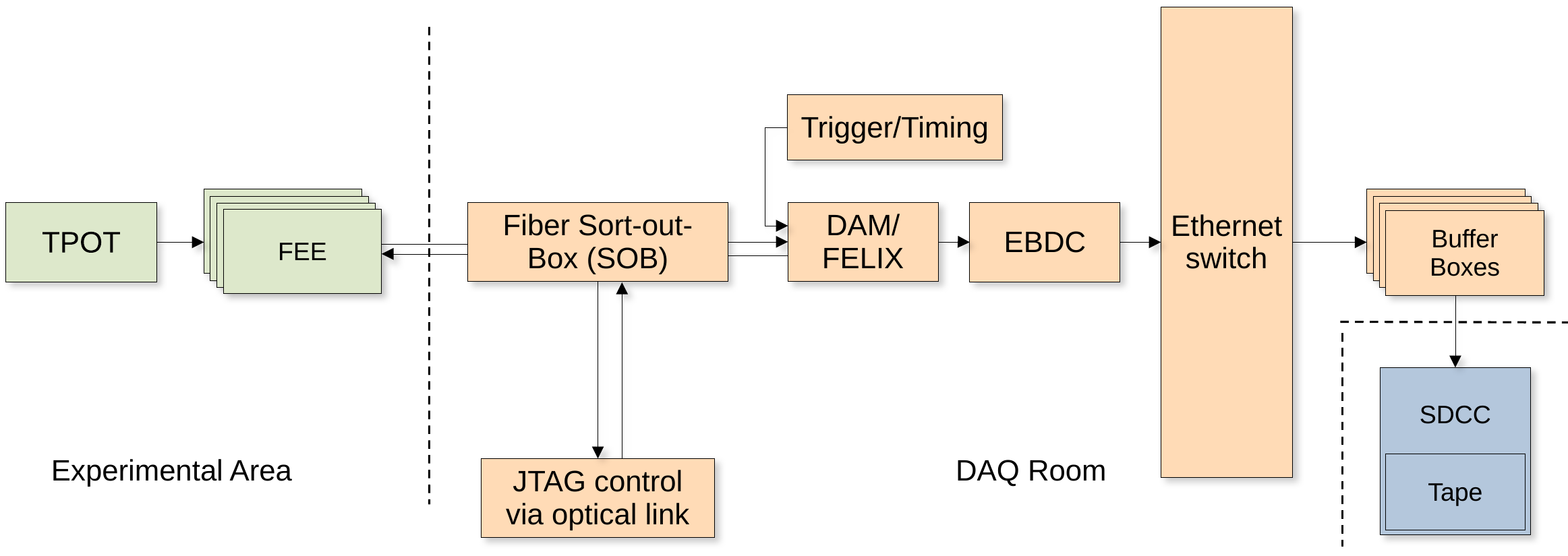}
 \caption{Block diagram of the sPHENIX readout system for TPOT.} 
 \label{fig:acquisition}
\end{figure}

Figure~\ref{fig:acquisition} shows the block diagram of the readout system for TPOT. The signal from TPOT is fed into the FEE which amplifies, shapes and digitizes it into data. Digital data is sent to the back-end electronics which consists of a Data Aggregation Module (DAM) installed inside the Event Buffering and Data Compressor (EBDC). The DAM is a specialized version of the FELIX board developed for the ATLAS experiment at the Large Hadron Collider (LHC)~\cite{Ryu_2017}. The EBDC is a high-end rack-mounted commodity server. Compressed data from the EBDC is sent to Buffer Boxes, which are high capacity, high input-output disks, and serve as a buffer before it is transferred to the Scientific Data and Computing Center (SDCC) at BNL and recorded on permanent storage (tapes).

The FEE has an optical transceiver with four Tx/Rx channels (Quad Small Form factor Plugable, QSFP). Two channels are dedicated to data transfer and clock/trigger distribution and one channel is dedicated to the JTAG (Joint Test Action Group) interface implemented over optical communication. The optical fibers from the FEE are sorted out in the Sort-Out-Box (SOB) in which the JTAG controller board called Mighty-JACK is installed. The TPOT readout system uses 16 FEE, one SOB/Mighty-Jack, one DAM, and one EBDC.

\subsubsection{\label{subsubsec:sampa}The 32-channel SAMPA ASIC}
The ASIC used for the FEE is called SAMPA v5~\cite{Barboza_2016,8772086}. Primarily designed for the ALICE TPC and Muon Chamber upgrades~\cite{Adolfsson_2021,9507849}, the ASIC was modified for the sPHENIX TPC requirement, the main modification consisting of including a shaping time of 80\,ns in place of 320\,ns. A description and block diagram of the ASIC is found in~\cite{Barboza_2016}.

For TPOT, the SAMPA operation parameters are: (i) 160\,ns shaping time, which is the longest available value, chosen to increase the collection of the ion tail in the Micromegas; (ii) 20\,mV/fC gain, which is the smallest available gain, chosen to minimize the fraction of signal that would saturate the ADC range and (iii) 20\,MHz sampling rate.

The dynamic range of the SAMPA ADC is 1024 units, with a conversion factor of 2.15\,mV/ADU (ADC unit). Thus the full range of the ADC corresponds to 2.2\,V, or 110\,fC at a nominal gain of 20\,mV/fC.

The quoted gain of 20\,mV/fC is measured assuming a small input capacitance from the detector of order 10-20\,pF. The effective gain becomes smaller when the input capacitance increases. The input capacitance from the Micromegas chambers, measured at Saclay, is $\sim 150$\,pF ($\sim 300$\,pF) for the $\phi$ ($z$) view, respectively. The gain of the SAMPA ASIC for such capacitance drops to 16\,mV/fC ($13.5$~mV/fC), corresponding to 82\% (68\%) of the nominal gain.

\subsubsection{\label{subsubsec:fee_board}The 256-channel Front-End Electronics (FEE) Board}

\begin{figure}[htb]
 \centering
 \includegraphics[width=0.9\textwidth]{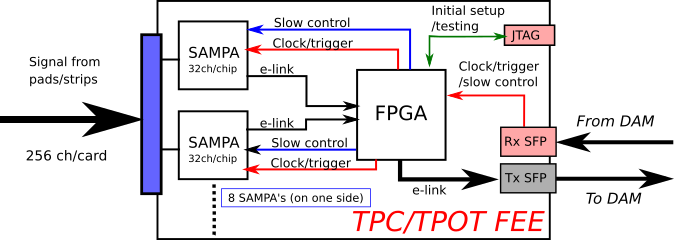}
 \caption{Circuit diagram of the TPOT FEE board.} 
 \label{fig:fee_diagram}
\end{figure}

\begin{figure}[htb]
 \centering
 \includegraphics[width=0.9\textwidth]{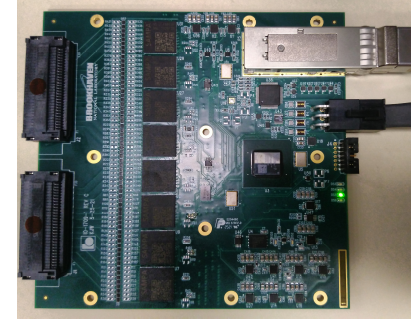}
 \caption{Picture of one TPOT FEE board.} 
 \label{fig:fee_board}
\end{figure}

Each FEE board has eight SAMPA v5 chips that accept 256 inputs from the detector. The block diagram of the FEE board is shown in Figure~\ref{fig:fee_diagram}. A picture of the board is shown in Figure~\ref{fig:fee_board}. The two SAMTEC SEAF connectors on the left collect the signal. They connect to the Micromegas chamber strips using a dedicated transition cable, shown in blue in Figure~\ref{fig:tpot_module}. Connectors on the right side of the board in Figure~\ref{fig:fee_board}, top-to-bottom, correspond to the optical transceiver, low voltage input, and a JTAG interface for board configuration redundant with the interface implemented over the optical link. A Xilinx Artix-7 200T FPGA (XC7A200T) mounted on the FEE has multiple functions: it collects data from the SAMPA chips, formats them into a data packet, and sends the packet off to the back-end electronics (DAM/EBDC) via the optical transceiver.

\subsubsection{Magnetic field and radiation tolerance}
The FEE boards are mounted directly on the TPOT modules (Figure~\ref{fig:tpot_module}) and sit inside sPHENIX longitudinal $1.4$\,T magnetic field. 
This field might affect the performance of the optical transceiver, due to the presence of inductor coils.
During preliminary tests performed at the Massachusetts Institute of Technology (MIT) and BNL Alternating Gradient Synchrotron (AGS) magnetic field facilities, the FEE was placed in magnetic fields oriented along three orthogonal directions and the transmission capability of the optical transceiver was quantified using the optical connection eye-diagram. A variation of $\sim25$\% in transmission capability was measured, acceptable for optical communication.

Radiation tolerance for the FEE is another key factor for validating its design and selection of individual electronic parts. The same FEE is used for TPOT and the TPC. For the TPC, some boards are installed close to the beam pipe, where the radiation dose is maximum. The Total Ionization Dose (TID) at a 16\,cm distance from the beam pipe and for 5 years of sPHENIX data taking is estimated to be 25\,kRad, using data from the PHENIX experiment~\cite{ADCOX2003469}. It is significantly smaller at the TPOT location. 
TID tests performed on the optical transceiver and the full FEE board using a $^{60}$Co $\gamma$ source show that all components of the FEE can sustain a 100\,kRad dose except the EEPROM (42\,kRad) and the Phase-Lock Loop (PLL, $\sim50$\,kRad). These limits are well above the dose expected during sPHENIX running.

\subsubsection{Mighty-Jack and Sort-out-Box}
\begin{figure}[htb]
 \centering
 \includegraphics[width=0.8\textwidth]{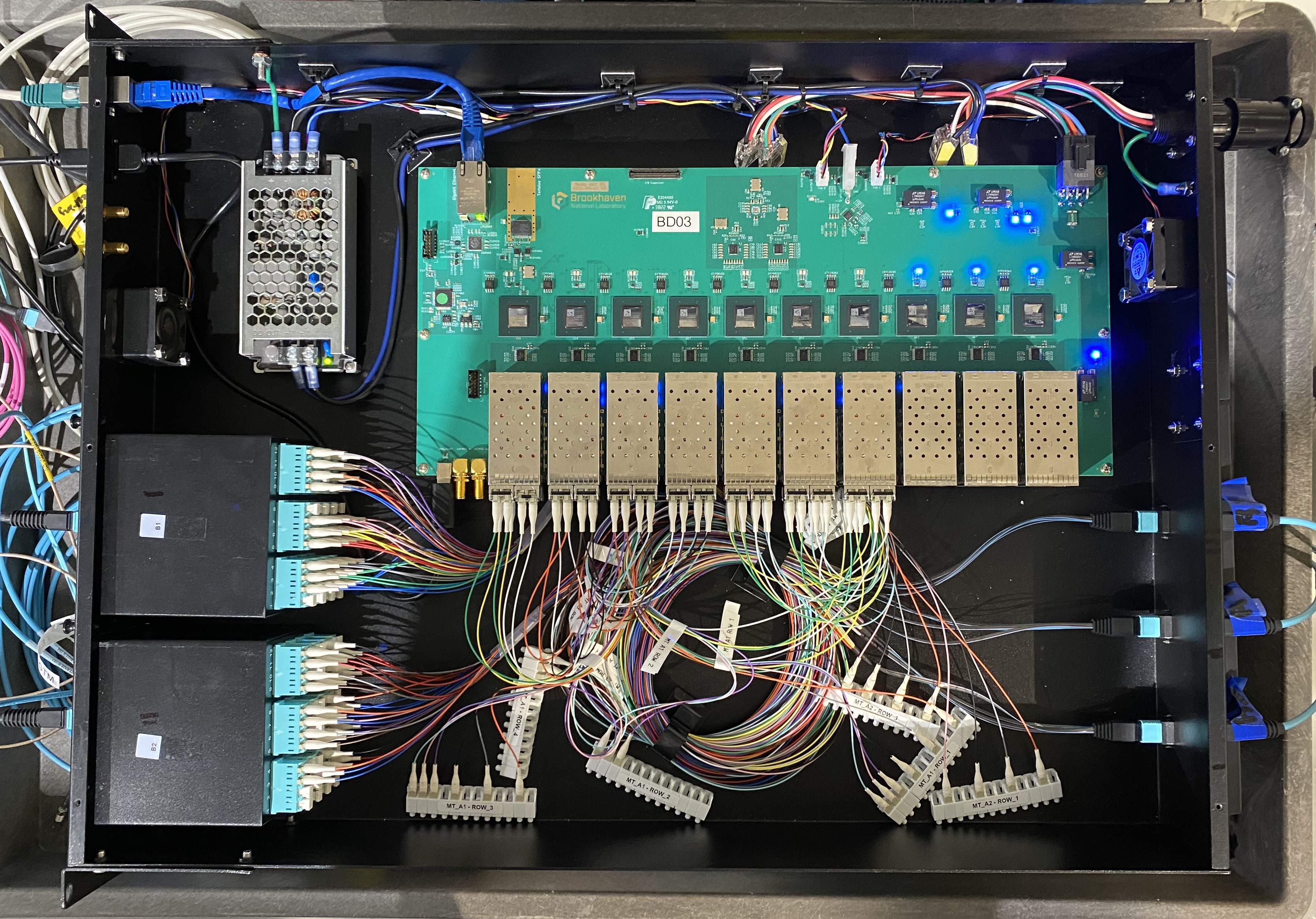}
 \caption{Picture of the Mighty-JACK board integrated into the Sort-Out-Box (SOB).} 
 \label{fig:Mighty-JACK}
\end{figure}

\begin{figure}[htb]
 \centering
 \includegraphics[width=0.8\textwidth]{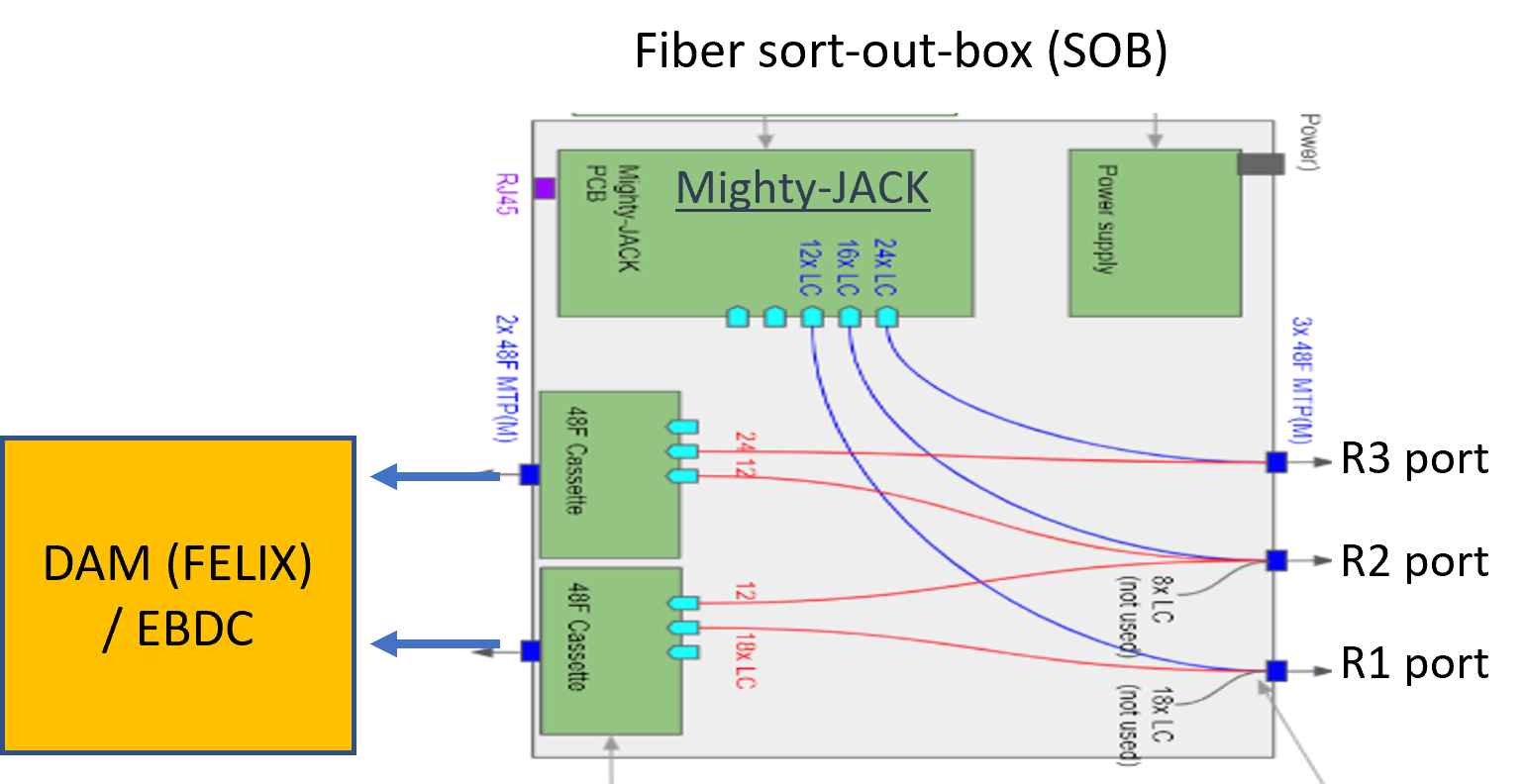}
 \caption{Fiber configuration in the Sort-out-box.} 
 \label{fig:SOB_diagram}
\end{figure}

The FEE board implements the JTAG control functionality over optical communication so that (i) the FPGA EEPROM can be reprogrammed as needed and (ii) the FPGA configuration memory can be monitored continuously and repaired in case of damage (e.g. single event upsets). The Mighty-JACK board and the fiber Sort-Out-Box (SOB) were developed to utilize this functionality. Pictures of those are shown in Figure~\ref{fig:Mighty-JACK}. The fiber configuration is shown in Figure~\ref{fig:SOB_diagram}.

\subsubsection{The back-end electronics (DAM and EBDC)}
\begin{figure}[htb]
 \centering
 \includegraphics[width=0.9\textwidth]{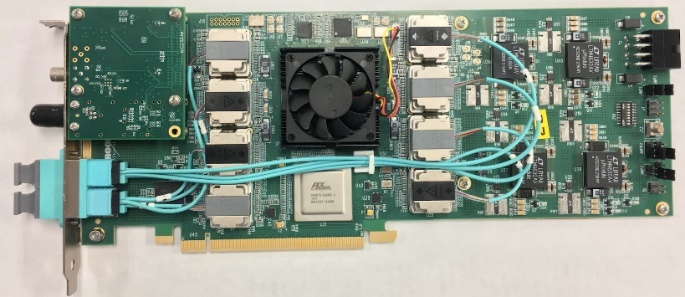}
 \caption{Picture of the DAM board.} 
 \label{fig:felix_board}
\end{figure}

\begin{figure}[htb]
\centering
\includegraphics[width = 0.8\textwidth]{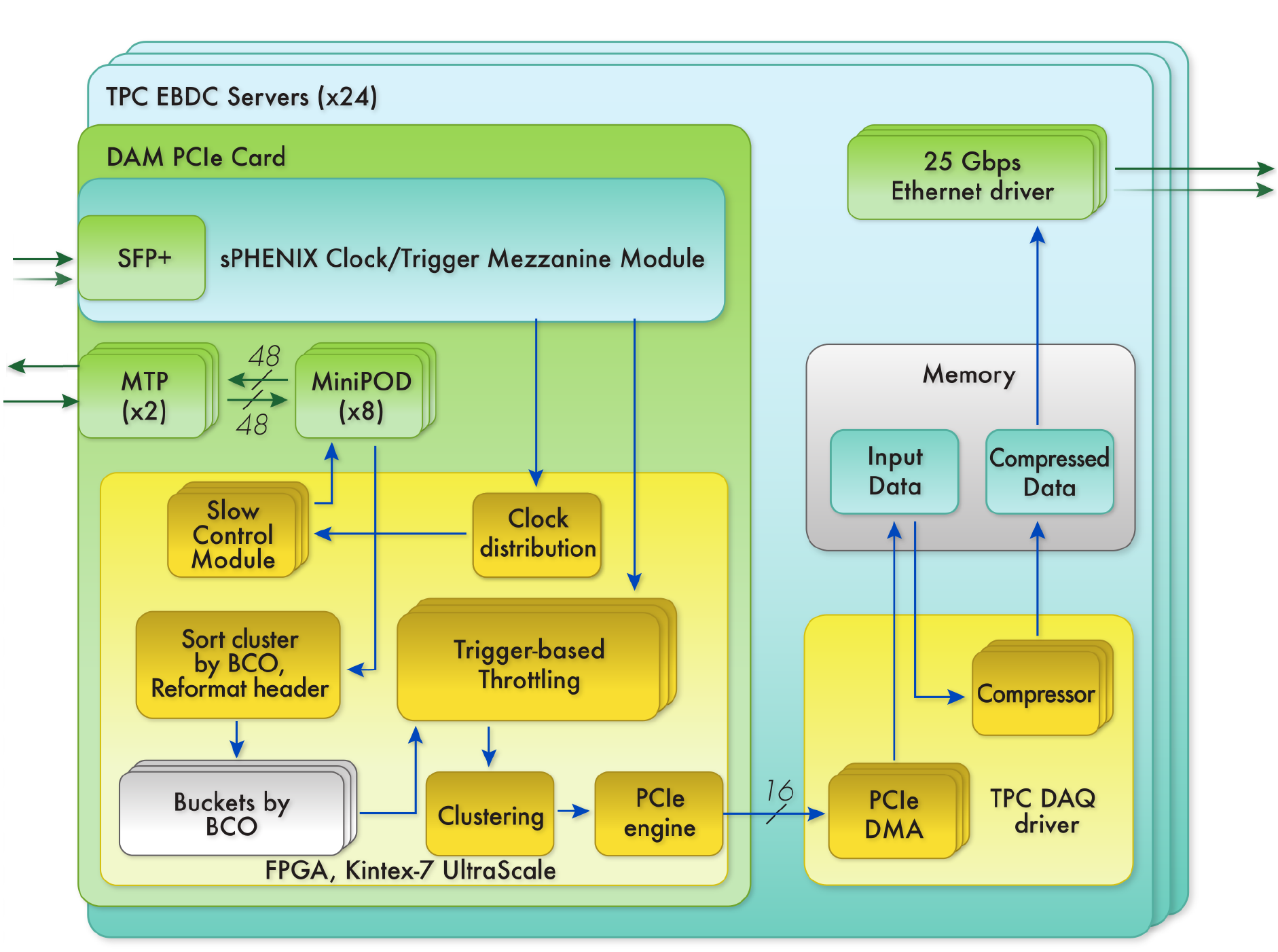}
\caption{Block diagram for DAM and EBDC.}
\label{fig:18-ElectronicsDetails}
\end{figure}

The FELIX board was developed primarily for the ATLAS experiment at CERN~\cite{Ryu_2017}, with the help of the BNL Instrumentation Division and the ATLAS group at BNL. It satisfies all the requirements for interfacing with the SAMPA FEE. A picture of the DAM board is shown in Figure~\ref{fig:felix_board}, and Figure~\ref{fig:18-ElectronicsDetails} shows the corresponding block diagram, both using ATLAS FELIX v2.1 card.

A Xilinx Kintex Ultrascale XCKU115-2FLVF1924E FPGA is on board of each DAM. To the FEE side, the FPGA is linked with four pairs of Mini Parallel Optical Device (MiniPOD) with 12 channels each. Each of the 48 pairs of fiber link supports up to 12.8\,Gb/s bi-directional data rate. The DAM FPGA is also linked to the EBDC server using a 16-lane PCI Express Gen3 connection which supports more than 100\,Gb/s. For sPHENIX, one of the 48 fiber links is redirected to a clock/trigger mezzanine module that provides an optical connection with sPHENIX Global Timing Module (GTM) using an SFP+ (Small Form factor Plugable) transceiver. 

The raw data is buffered in the DAM for up to 20\,$\mu$s. Only the data that falls within an adjustable time window matching a given trigger is sent for further processing via a throttling algorithm in the DAM FPGA. For the TPC, this reduces the data volume by about a factor four. A significantly larger reduction is expected for TPOT. It is possible to implement some clustering of the data on the DAM FPGA for additional data reduction. 
After transmitting the data to the EBDC, lossless compression is performed on the CPU before sending it out to sPHENIX DAQ for further processing and storage.

\subsection{\label{subsec:services}Services}
\subsubsection{High voltage}
The High Voltage (HV) system of TPOT consists of one negative drift electrode and four positive resistive electrodes for each chamber, for a total of eighty HV channels. HV is supplied by CAEN HV units A7030SN and A7030SP through appropriate Safe High Voltage (SHV) cables. Each channel can deliver voltage of up to 3\,kV and currents up to 5\,mA, well above the typical operation range necessary for Micromegas (voltage less than 1\,kV and currents of at most 10\,$\mu$A). A total of six 24-channel HV units are used (two negative and four positive), controlled by CAEN SY4527 universal multichannel power supply system.

To protect the chambers from discharge-induced damage, TPOT is equipped with a spark protection system through the HV units. If a given channel draws more current than an adjustable limit (typically a few $\mu$A, channel-dependent), for more than a given adjustable amount of time (typically a few seconds), the voltage is ramped down safely, and the channel is turned off. 

TPOT HV is operated with three modes: OFF, SAFE and ON. When in the OFF mode, the channels all hold 0\,V. When in SAFE mode, the channels hold low enough voltage that there is no amplification in the chambers. SAFE mode is typically used when there is no stable beam in the accelerator. When in ON mode, the chambers are all brought up to operating voltage to collect data from collisions with optimal gain and stability.

\subsubsection{Low voltage}
The same power system provides the Low Voltage (LV) for the TPC and TPOT FEE. 
Three LV channels are needed to power the FEE: $0.5$\,A at $4$\,V (digital), $2.4$\,A at $2$\,V (digital) and $2.4$\,A at $2$\,V (analog), corresponding to a total power consumption of about 12\,W per FEE board and 192\,W total, for TPOT. 

\begin{figure}[htb]
\centering
\includegraphics[width = 0.95\textwidth]{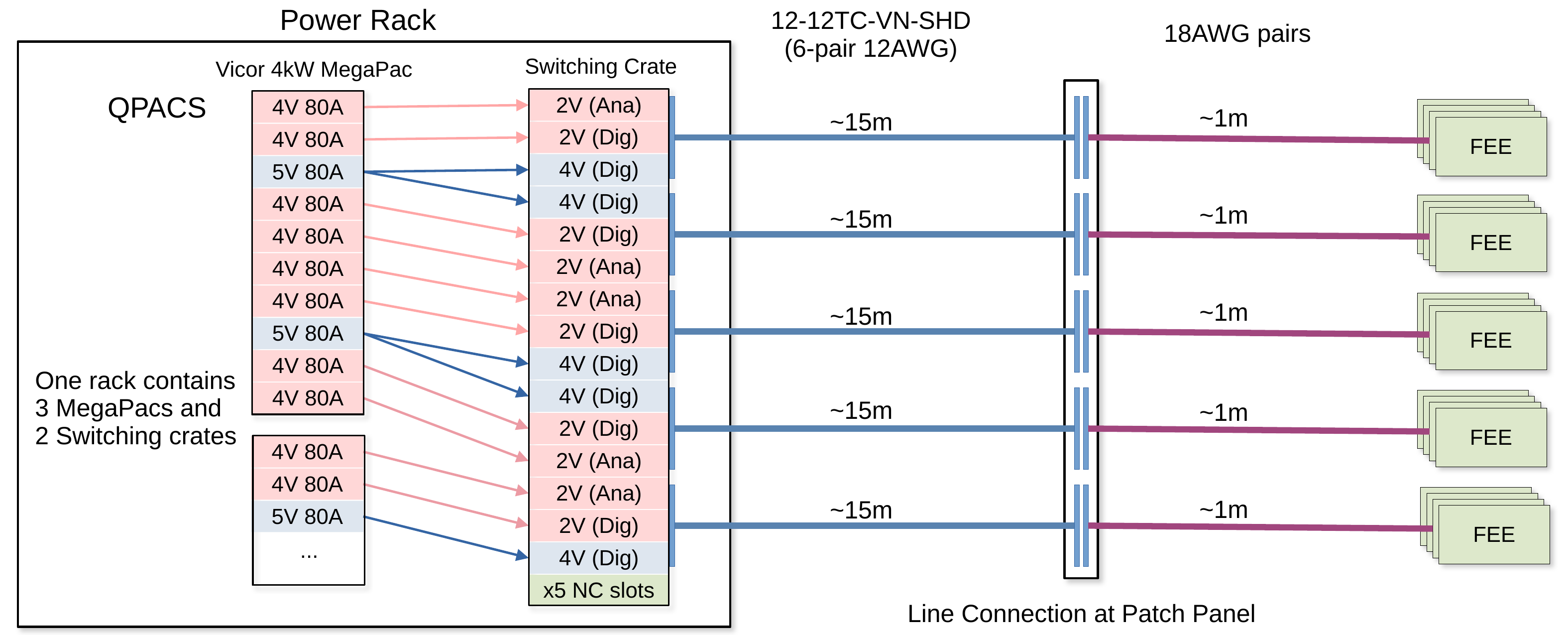}
\caption{\label{fig:TPC_PowerDist} Low voltage power distribution scheme for TPC/TPOT FEE.}
\end{figure}

Figure~\ref{fig:TPC_PowerDist} shows the design of the low voltage power distribution scheme for the TPC/TPOT FEE.
A Vicor MegaPak 4\,kW is used for the bulk power supply, in which ten 400\,W DC-DC converters are installed. 5\,V and 4\,V modules that supply up to 80\,A are used. Considering the significant voltage drop between power distribution boards and FEE, one 5\,V module is used to provide the 4\,V line, and two 4\,V
modules for each 2\,V lines.
The distribution board is designed so that one board powers up to 20 FEEs. 
TPOT uses 12 channels of one distribution board (4 for each TPOT sector).

\subsubsection{Cooling}

\begin{figure}[htb]
 \centering
 \includegraphics[width=0.52\textwidth]{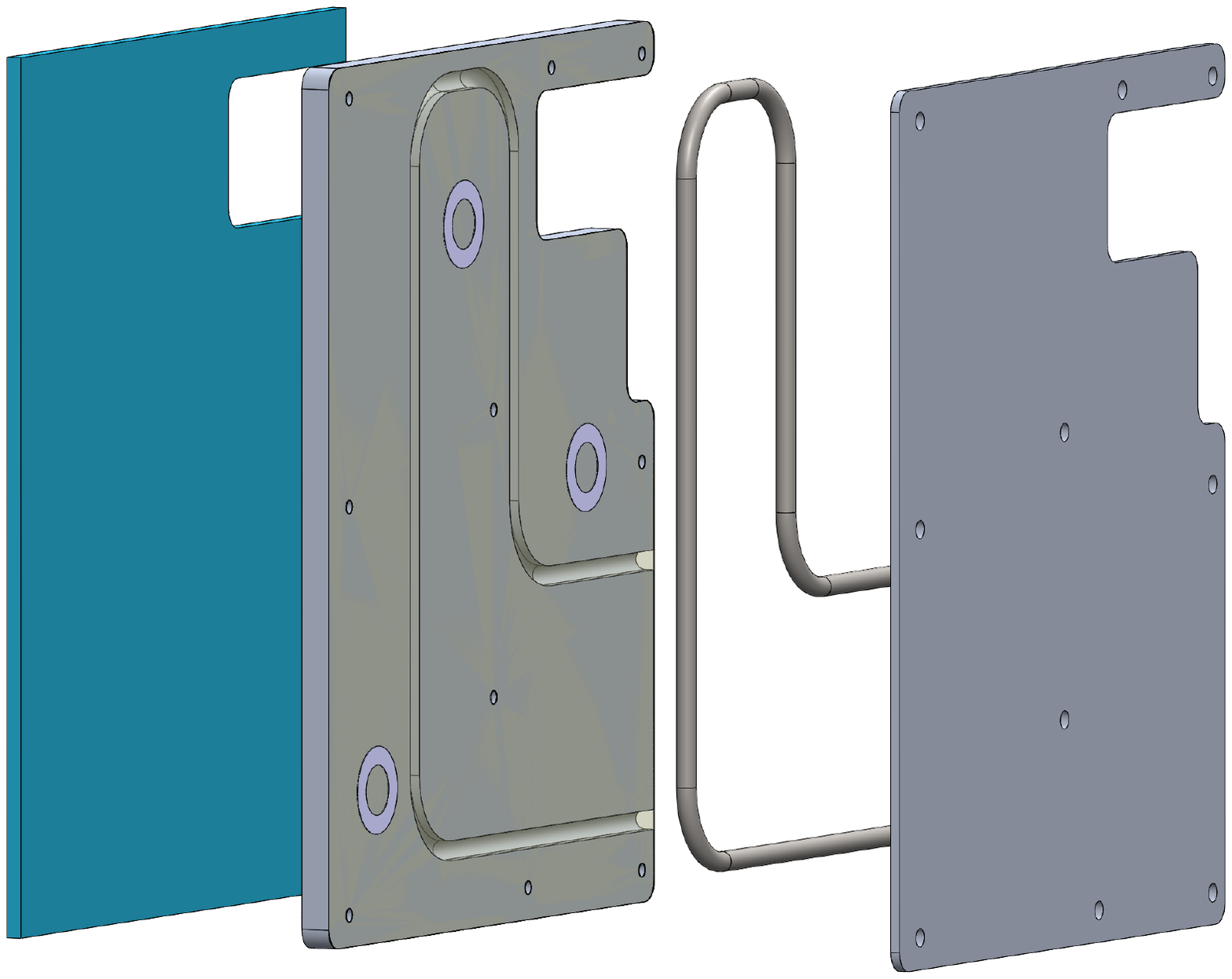}\hspace*{5mm}\includegraphics[width=0.42\textwidth]{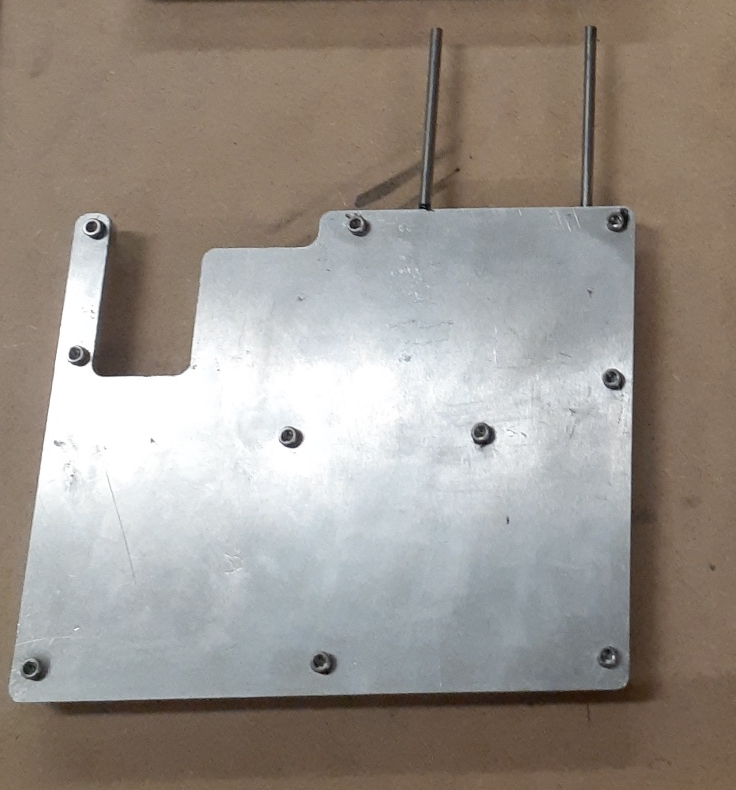}

 \caption{Left: design of the FEE cooling plates. Shown, right-to-left, are: the top aluminum plate, the aluminum tube through which sub-atmospheric cooling water is circulated, a set of washers used to control the thickness of the glue used to assemble the cooling plate (832TC-B thermally conductive epoxy, from MG Chemicals), the bottom aluminum plate, and the thermally conductive 2.5\,mm thick gap-pad material layer (Tflex 300 Series from Laird Technologies); Right: picture of one assembled cooling plate.} 
 \label{fig:cooling_plate}
\end{figure}

The FEE boards are cooled using a sub-atmospheric pressure cold water circuit. The water is circulated through aluminum cooling plates mounted on top of each board and separated by a thermally conductive gap-pad material of thickness 2.5\,mm. A diagram of the cooling plate and a picture of the assembled plate are shown in Figure~\ref{fig:cooling_plate}. The gap-pad material is from Laird Technologies (Tflex 300 Series). It is made of silicone elastomer, has a density of 1.78\,g/cc and a thermal conductivity of 1.2\,W/mK.
A Childyne unit is used to circulate the sub-atmospheric pressure water. Water is provided to each module independently and each line is monitored by its individual flowmeter. The cooling plates of the two FEE boards of a given module are mounted in series. Using sub-atmospheric pressure water ensures that in the event of a leak, air leaks inward but no liquid leaks outward. With this setup, the cooling water input temperature set to 21$\degree$C and the cooling water flow rate to 125\,cc/min, the temperature measured on the FPGA of the FEE boards is around 40$\degree$C.

\subsubsection{Gas system}

\begin{figure}[htb]
 \centering
 \includegraphics[width=0.9\textwidth]{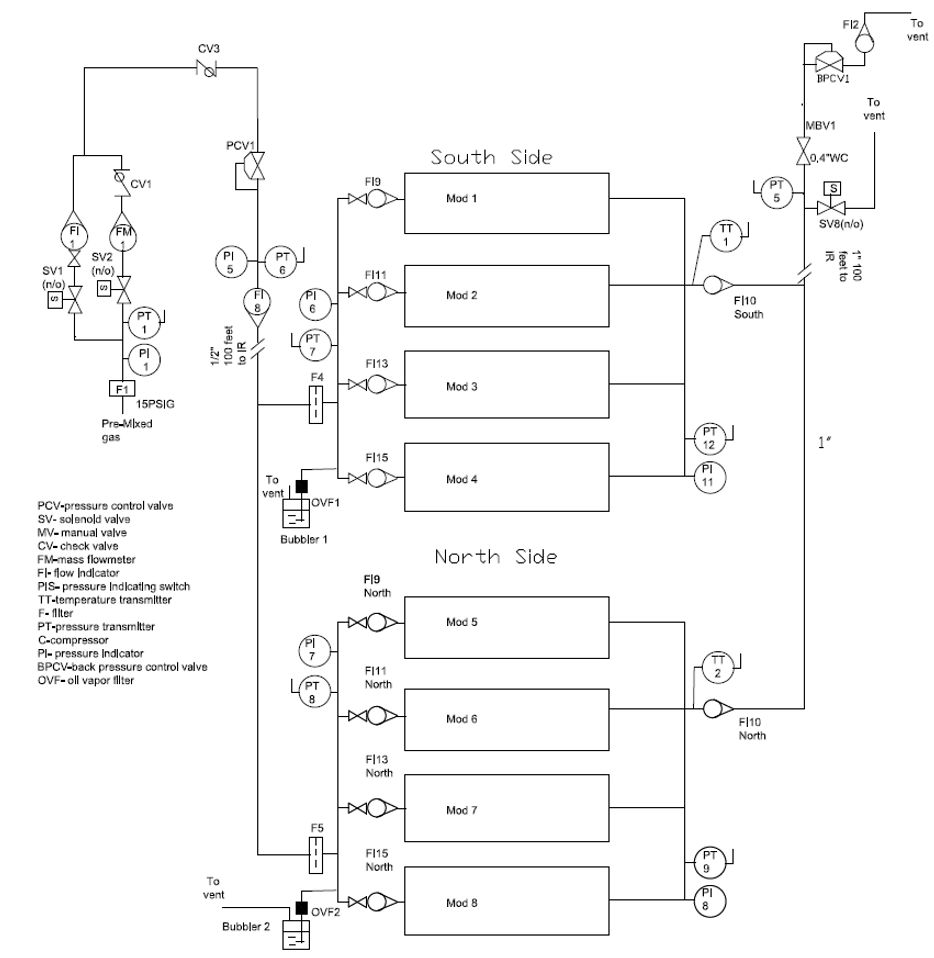} 
 \caption{Schematics of the TPOT gas system.} 
 \label{fig:gas_system}
\end{figure}

Figure~\ref{fig:gas_system} shows the schematics of the TPOT gas system. 
The two chambers in each module are connected in series. Each module has a volume of $1.4$\,l for a total of $11.2$\,l. The gas flow is operated in the 50-100\,cc/min range. Each module is equipped with its own flowmeter. 

TPOT uses premixed gas bottles of 95/5\% Ar/iC\textsubscript{4}H\textsubscript{10}. Each of the eight modules has its own input and output lines that connect to the patch panel, which in turn all connect in parallel to the same gas bottle system. The gas bottle system supplying the gas mixture to the chambers is made up of two 12-bottle packs which switch seamlessly to ensure a continuous and uniform flow.
The gas system operates at a pressure slightly above atmospheric. Six modules maintain proper flow at approximately 1\,inch H\textsubscript{2}O over-pressure. The remaining two modules operate at higher over-pressures of 4 and 6\,inch H\textsubscript{2}O due to higher internal resistance. 

For safety, three more tubes are added to TPOT that connect to the middle of each sector and allow to (i) detect any build-up of gas between TPOT and EMCAL in case of a leak and (ii) inject nitrogen to displace the built-up gas.

\subsection{\label{subsec:grounding}Grounding}

\begin{figure}[htb]
 \centering
 \includegraphics[width=0.8\textwidth]{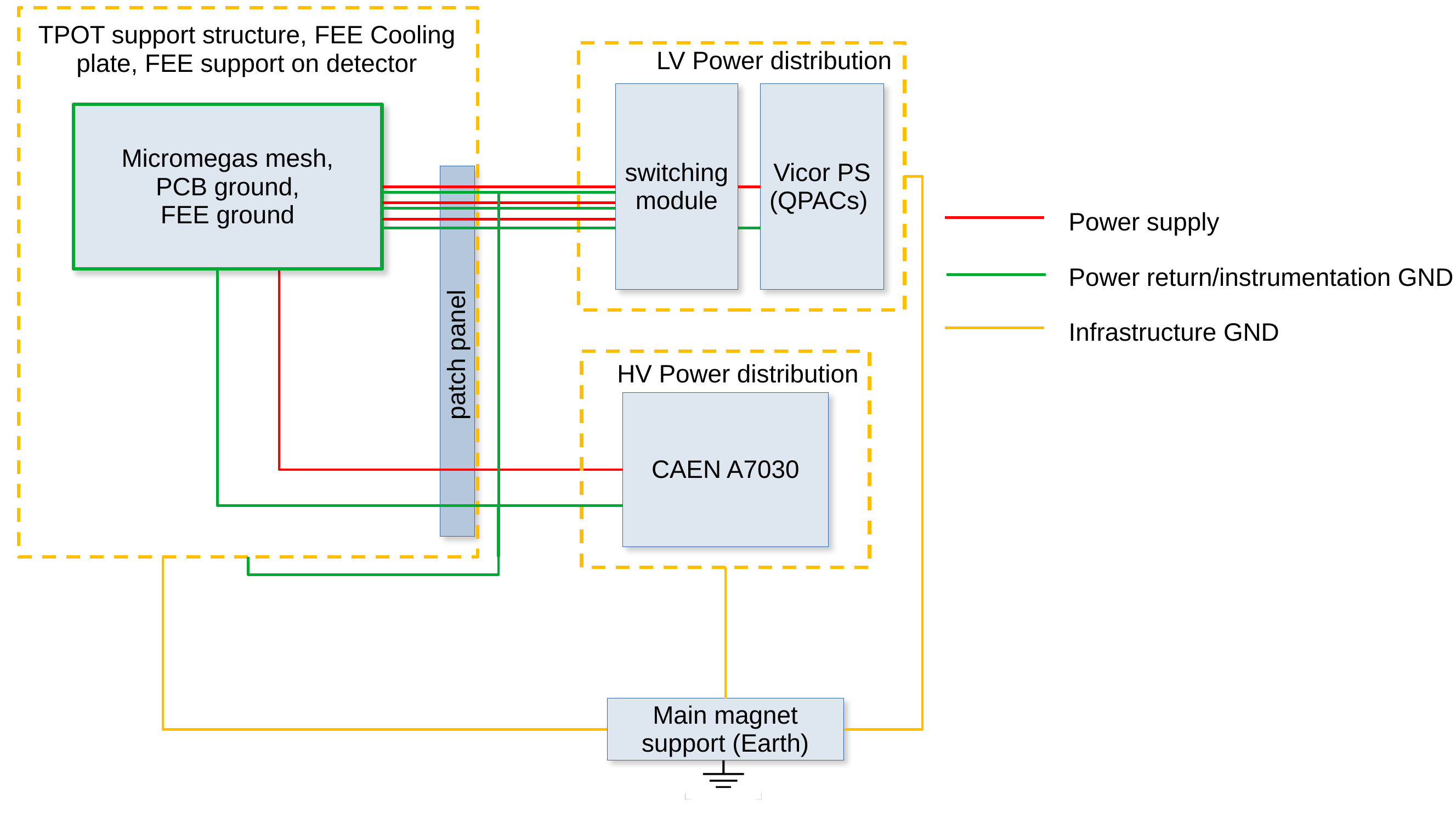} 
 \caption{Schematics of the TPOT grounding scheme.} 
 \label{fig:grounding}
\end{figure}

A simplified description of the TPOT grounding is shown in Figure~\ref{fig:grounding}. Two separate grounds are considered: the infrastructure ground to which all metallic structures of all sPHENIX subsystems connect and the instrumentation ground, which coincides with the return lines of the HV and LV power supply. For TPOT the infrastructure ground connects the support structure, the FEE support plates and the FEE cooling plates. The instrumentation ground connects the LV and HV return lines, the FEE ground, the Micromegas readout PCB ground and the micromesh. By default the infrastructure and instrumentation ground are connected at the sPHENIX earth, located on the main magnet support. In the TPOT case, electromagnetic noise is found to be minimum if another connection between the two grounds is added at the detector patch panel.

\subsection{\label{subsec:slow_control}Slow Control and monitoring}
\begin{figure}[htb]
 \centering
 \includegraphics[width=\textwidth]{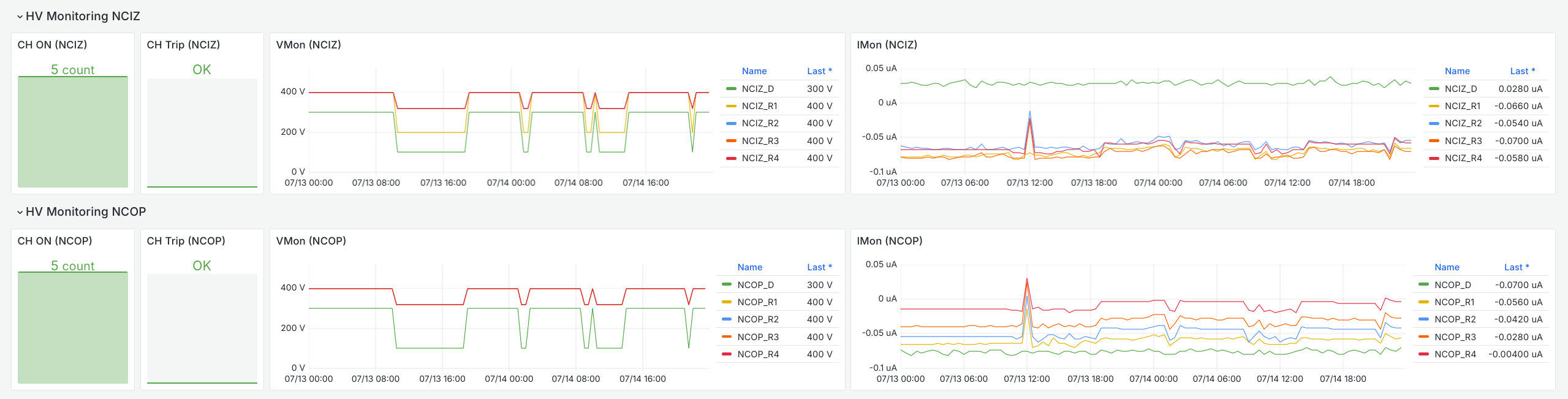} 
 \caption{Example of monitoring of the HV (left) and current (right) of two TPOT chambers over time. The low and high values on the HV panel correspond to TPOT being in either SAFE or ON state, respectively, depending on the presence of beam in sPHENIX.} 
 \label{fig:monitoring}
\end{figure}

The TPOT slow control and monitoring system tracks a number of measurements relevant to detector operation using a Grafana web interface and the time series database Prometheus~\cite{Web:Grafana:Docs}. This includes, but is not limited to: (i) the voltage and current of each HV channel, (ii) the voltage and current of each LV channel, (iii) the log and history of tripped HV channels, (iv) the temperature of the FEE and (v) the recorded data rate. Five temperatures per FEE are measured to monitor the FPGA as well as the FEE PCB. The LV power supply is interlocked to the sub-atmospheric cooling water system, so that it is not possible to turn ON the FEE in absence of cooling. An example of the monitoring of HV and current values over time is shown in Figure~\ref{fig:monitoring}. 

\section{\label{sec:installation}Detector Installation}
\subsection{\label{installation_mechanics}Installation Mechanics}
\begin{figure}[ht]
  \centering
  \includegraphics[width=0.48\textwidth]{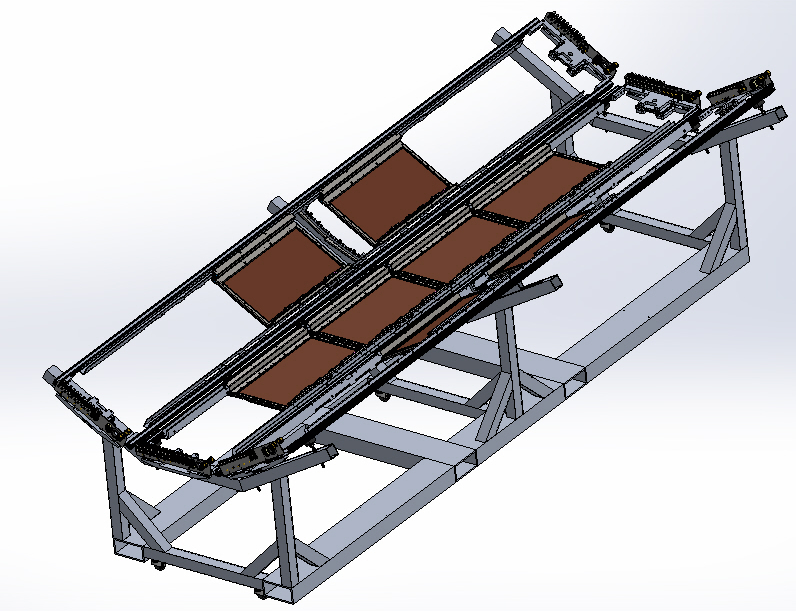}
  \includegraphics[width=0.42\textwidth]{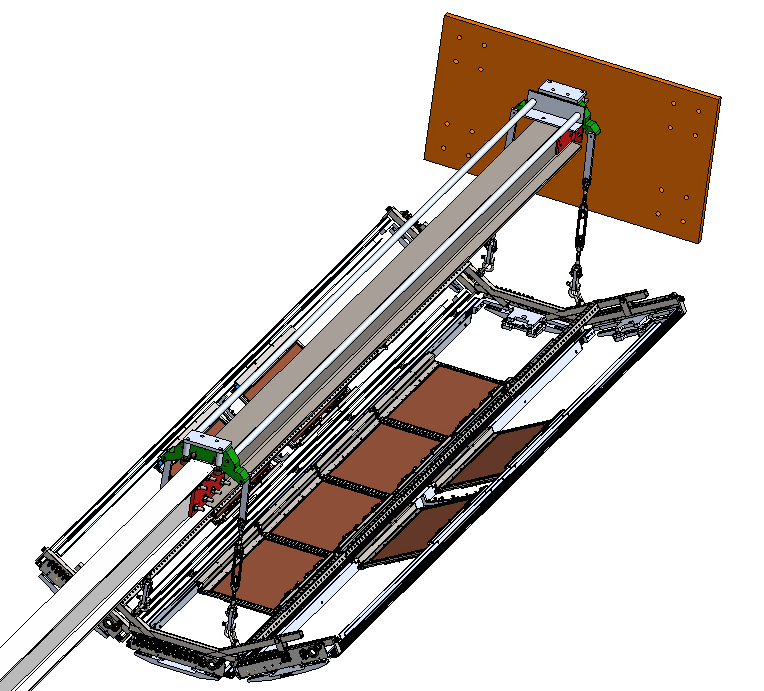}
\caption{Schematics of the TPOT detector, transportation cradle, and installation mechanics.} 
  \label{fig:tpot_transportation_cradle}
\end{figure}

Each of the three TPOT sectors was assembled separately at BNL and later moved on a transportation cradle (Figure~\ref{fig:tpot_transportation_cradle}, left), used to bring the detector to the sPHENIX experimental hall. 

In the experimental hall, the detector was attached to a suspended aluminum I-beam, which became a part of both the lifting and insertion mechanism. This I-beam was then attached to a preexisting fixture in the magnet bore and connected in series with two other beams to allow for the longitudinal translation of the TPOT detector into its final $z$ position (Figure~\ref{fig:tpot_transportation_cradle}, right).

\begin{figure}[ht]
  \centering
  \includegraphics[width=0.58\textwidth]{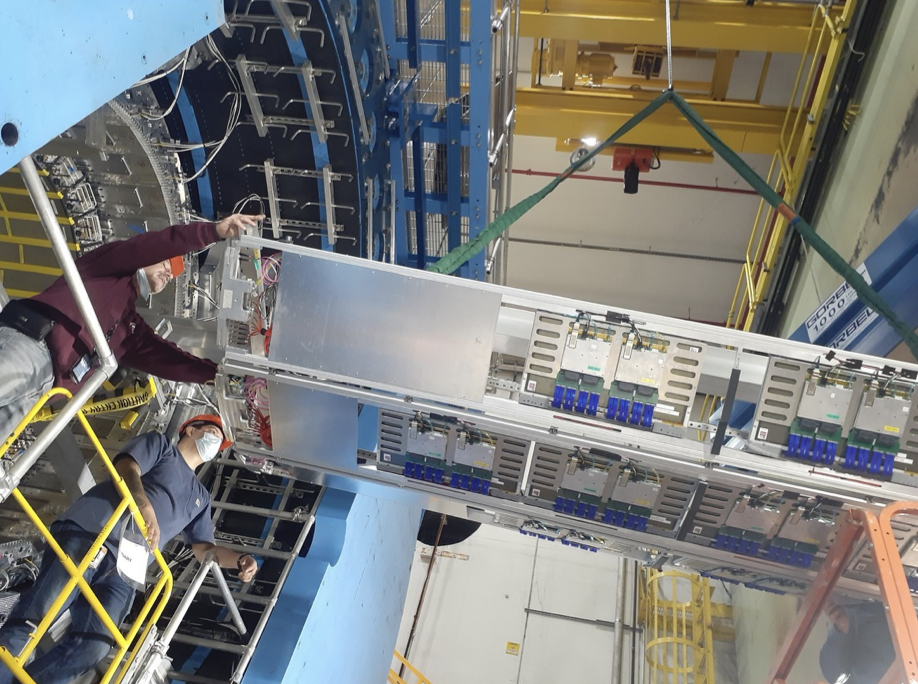}
  \includegraphics[width=0.395\textwidth]{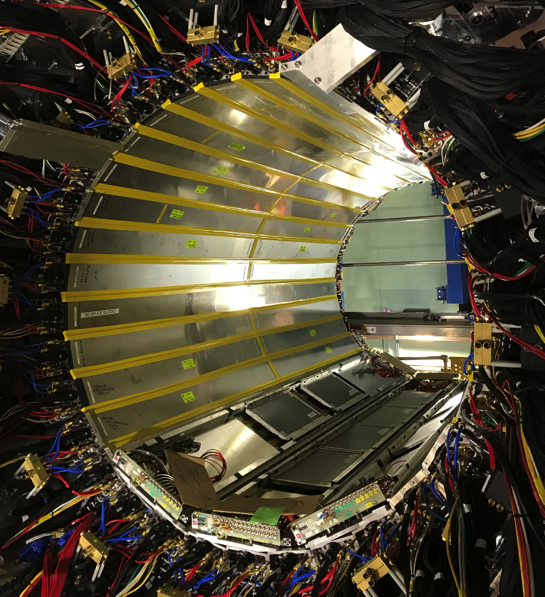}
\caption{Pictures of the TPOT installation inside the sPHENIX EMCAL.} 
  \label{fig:tpot_installation}
\end{figure}

After translation, TPOT was lowered and aligned to the EMCAL using turnbuckles as a
part of the lifting beam (Figure~\ref{fig:tpot_transportation_cradle}, right). It was then secured through its end brackets to hollow EMCAL end frames that rest on the IHCAL. The mid-length support was preset while on the cradle to align with the EMCAL.

Figure~\ref{fig:tpot_installation} shows two pictures of the TPOT installation inside the EMCAL.

During installation, one of the 32 signal transition cables that runs between the detector and its corresponding FEE (Section~\ref{subsubsec:fee_board} and Figure~\ref{fig:tpot_module}) got disconnected from the board. After installation, it was not possible to access and reconnect this cable. This corresponds to an acceptance loss of half a chamber, or 3\%.

\subsection{Metrology and Survey}
Metrology and survey aim to provide the position of the Micromegas strips in the sPHENIX reference frame, after TPOT installation. This is achieved in three steps: (i) after fabrication and assembly of each Micromegas module, one measures the position of the strips with respect to a set of reference targets on the module; (ii) after assembly of the modules on their support structure, one measures their position with respect to a set of targets mounted on the structure and (iii) after installation of the TPOT detector inside the EMCAL, one measures the position of its support structure in the sPHENIX reference frame. 

The first step is referred to as metrology. Each module is surrounded by a tray screwed into the module frame. On the top side there are three space points to insert optical targets into pins for metrology. The pins are either inserted into drill-bushed holes on the detector frame or glued directly on the frame. Measurements of the relative position of the detector strips to the pins were performed at Saclay. First one uses the readout PCB drawings to determine the position of the strips with respect to a set of fixed reference points on the PCB (here, the edges of the MEC8 connectors), then one measures the position of these reference points with respect to the optical targets.

The second and third steps rely on optical targets inserted into drill-bushed holes inside the support structure, at each end of each TPOT sector. The measurements were performed at BNL by the local survey team. 

The combination of these three sets of measurements allows the position of the strips inside sPHENIX to be known at a precision of a few 100\,$\mu$m. This determination is then improved using the reconstructed trajectory of particles passing through sPHENIX tracking detectors, to an accuracy of 100\,$\mu$m or less.
Millepede~\cite{Blobel:2002ax} is used to perform this track-based alignment.

\section{\label{sec:detector_performance}Detector Characterization and Performance}
\subsection{\label{subsec:characterization_saclay}Detector characterization at Saclay}

\begin{figure}[htb]
 \centering
 \includegraphics[height=0.7\textwidth]{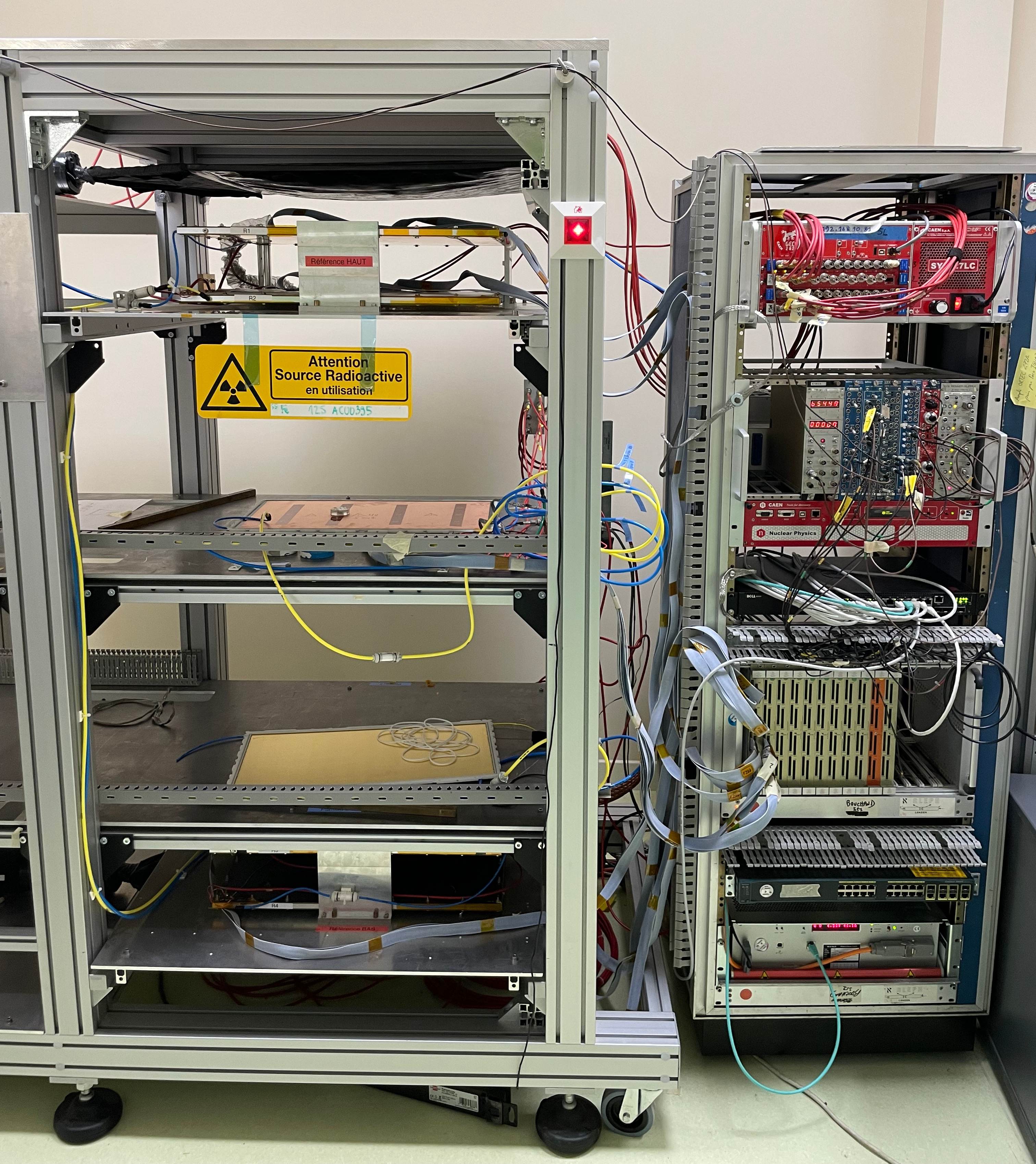}
 \caption{Picture of the cosmic test bench used at Saclay for the characterisation of the TPOT modules during production.} 
 \label{fig:cosmic_testbench}
\end{figure}

Figure~\ref{fig:cosmic_testbench} shows a picture of the cosmic test bench used at Saclay for the characterization of the TPOT modules during production. 
%The cosmic test bench used at Saclay for the characterization of the TPOT modules during production is shown in Figure~\ref{fig:cosmic_testbench}. 
The modules are positioned in the middle trays of the setup between two sets of MultiGen detectors ~\cite{BOUTEILLE2016187} (two detectors above and two detectors below), used for the reconstruction of the reference track. 3D-printed indentation fixtures are attached to the cosmic test bench so all the modules are tested at the same position with respect to the reference detectors. Data acquisition is triggered using a coincidence of scintillating pads mounted on Photo-Multiplier Tubes (PMT). The signal is read with the DREAM electronics~\cite{flouzat2014dream} originally developed for the Micromegas detector of the CLAS12 experiment at the Jefferson Laboratory~\cite{BURKERT2020163419,osti_1607856}. 

\begin{figure}[htb]
 \centering
 \includegraphics[width=0.46\textwidth]{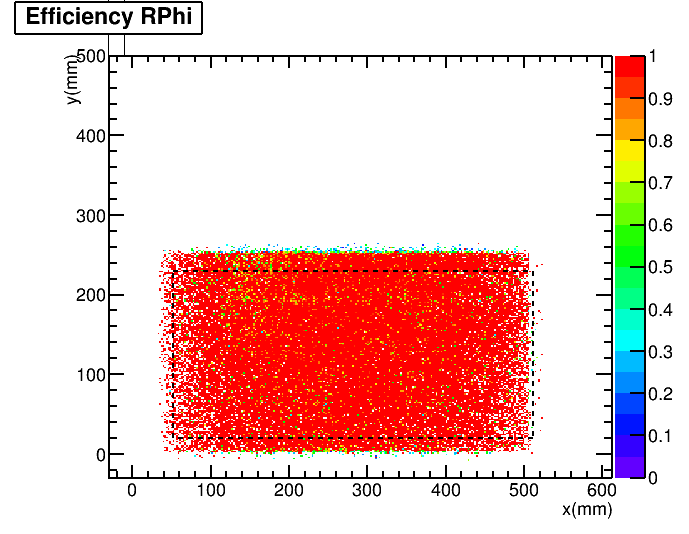}
 \includegraphics[width=0.46\textwidth]{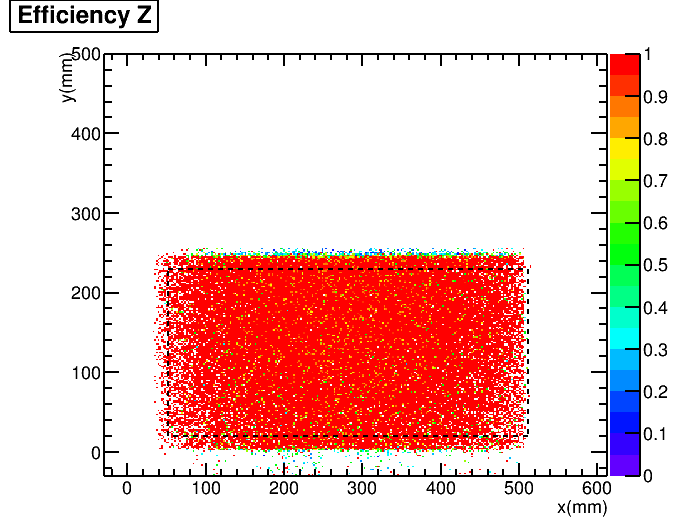}
 \caption{Detection efficiency as a function of position in two chambers of the same TPOT module, for a resistive layer HV of 410\,V. Left: $\phi$ view; right: $z$ view.} 
 \label{fig:efficiency_map}
\end{figure}

The detection efficiency measured as a function of position in the two chambers of a given TPOT module using this setup is shown in Figure~\ref{fig:efficiency_map}. The efficiency is uniform in the entire active area of the two chambers, and above 90\% everywhere.

\begin{figure}[htb]
 \centering
 \includegraphics[width=0.46\textwidth]{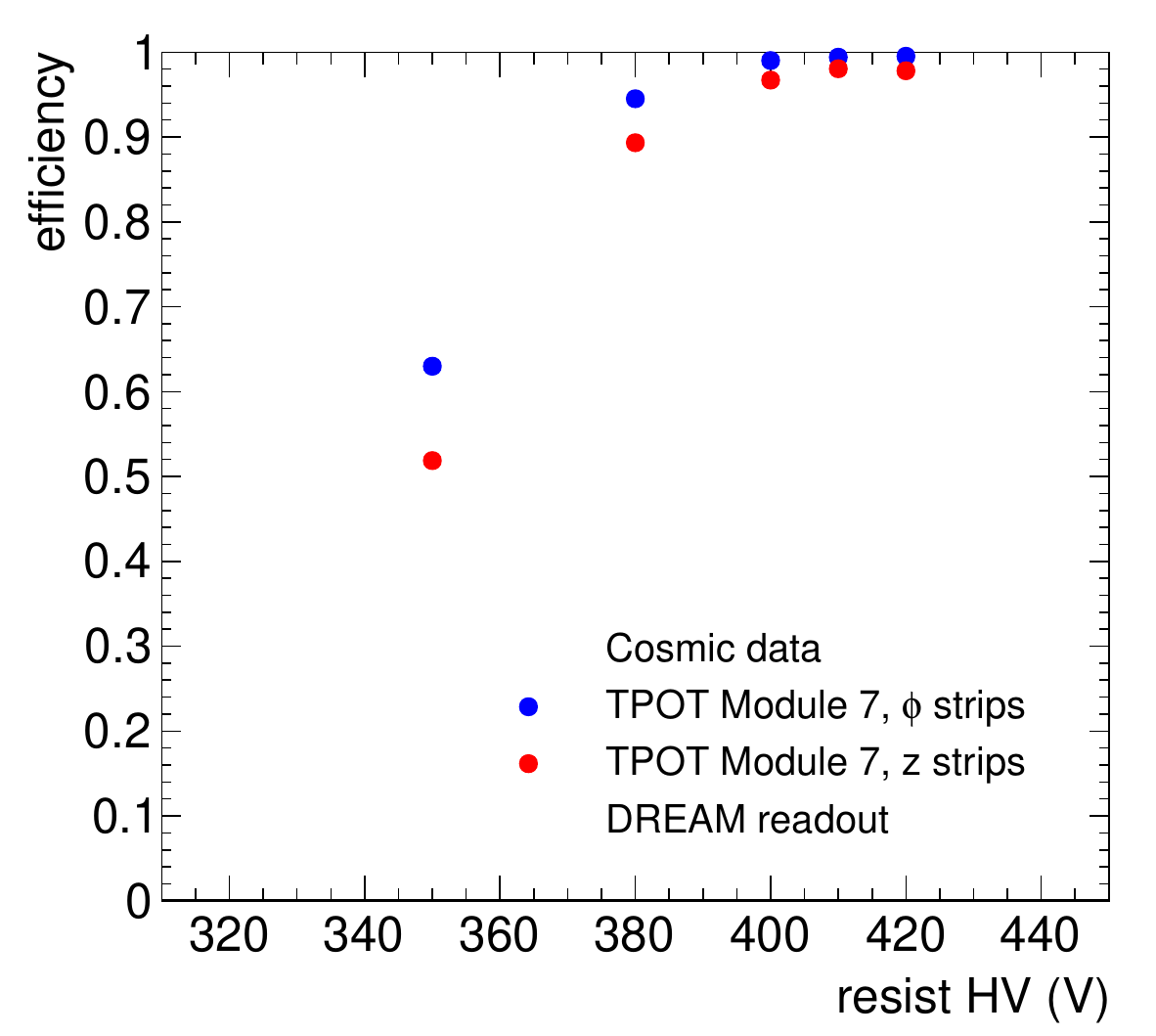}
 \includegraphics[width=0.46\textwidth]{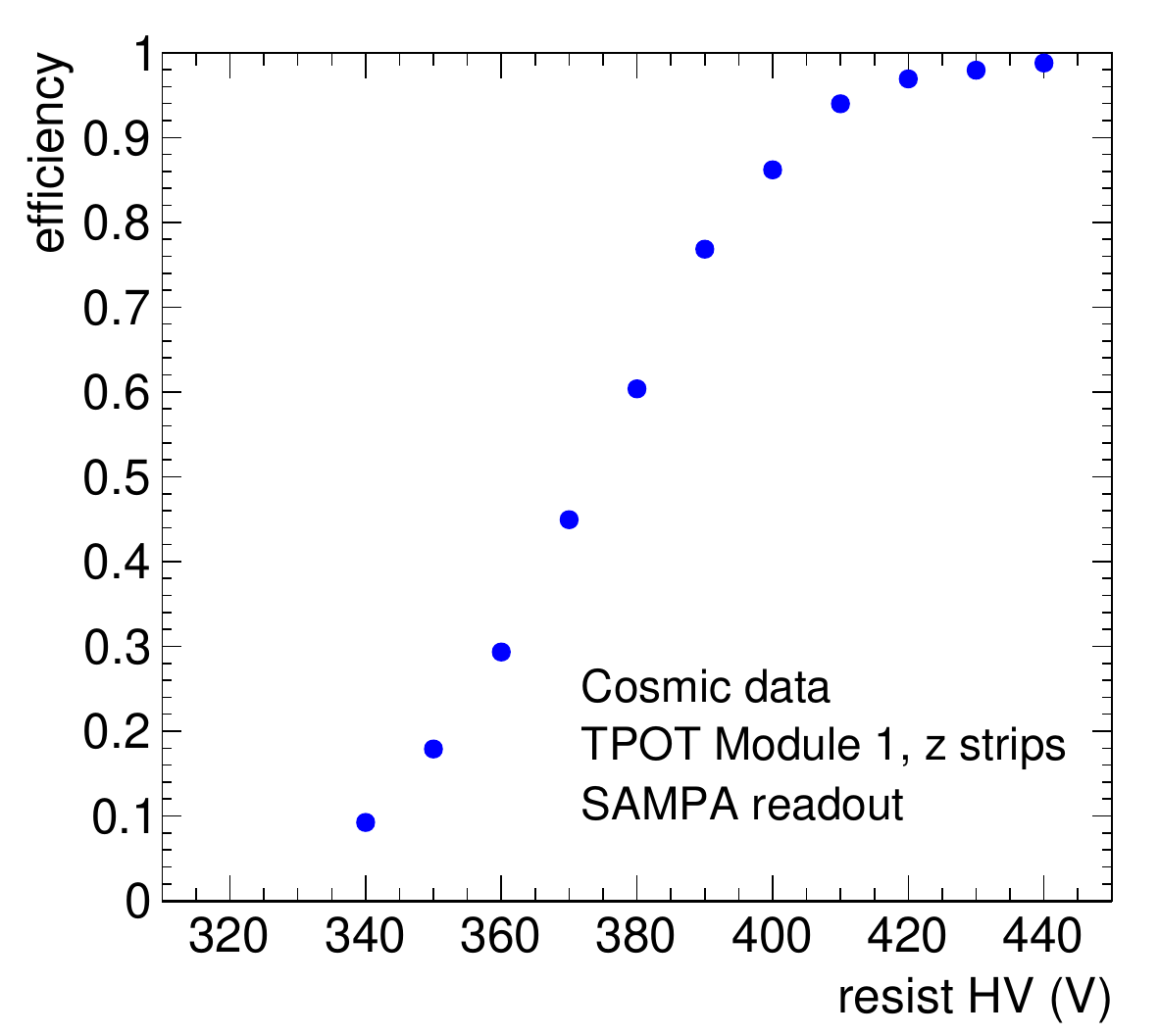}
 \caption{Detection efficiency as a function of resistive layer HV, left: for the two chambers of the same TPOT module, using DREAM electronics at Saclay; right: for the $z$ view of a given TPOT module, measured at BNL using SAMPA.} 
 \label{fig:efficiency_plateau}
\end{figure}

Figure~\ref{fig:efficiency_plateau} (left) shows the detection efficiency measured as a function of the HV applied to the resistive layer in the two chambers of a given TPOT module. Efficiency increases with HV and full efficiency is reached at about 400\,V for both chambers.

\begin{figure}[htb]
 \centering
 \includegraphics[width=0.46\textwidth]{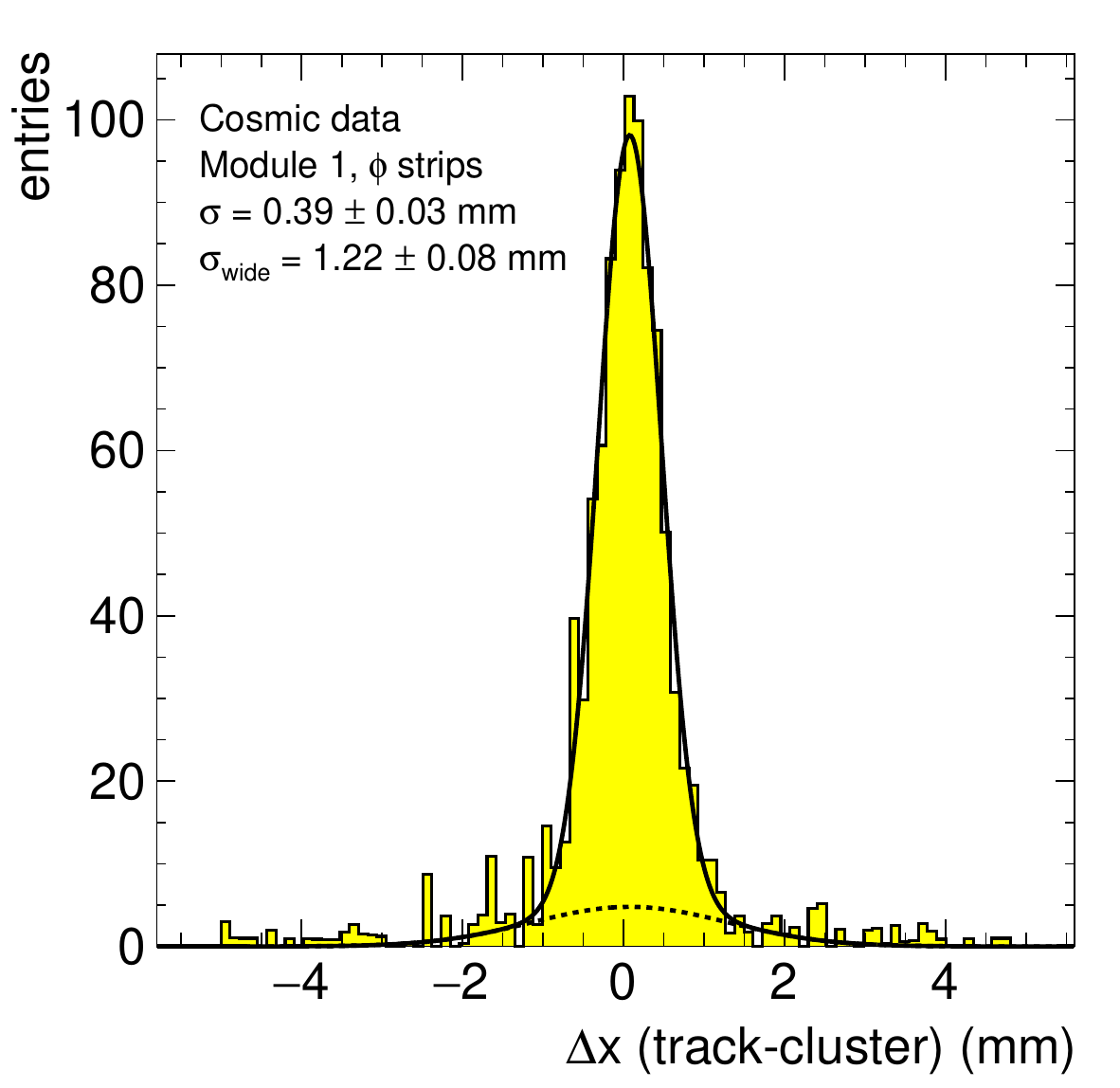}
 \includegraphics[width=0.46\textwidth]{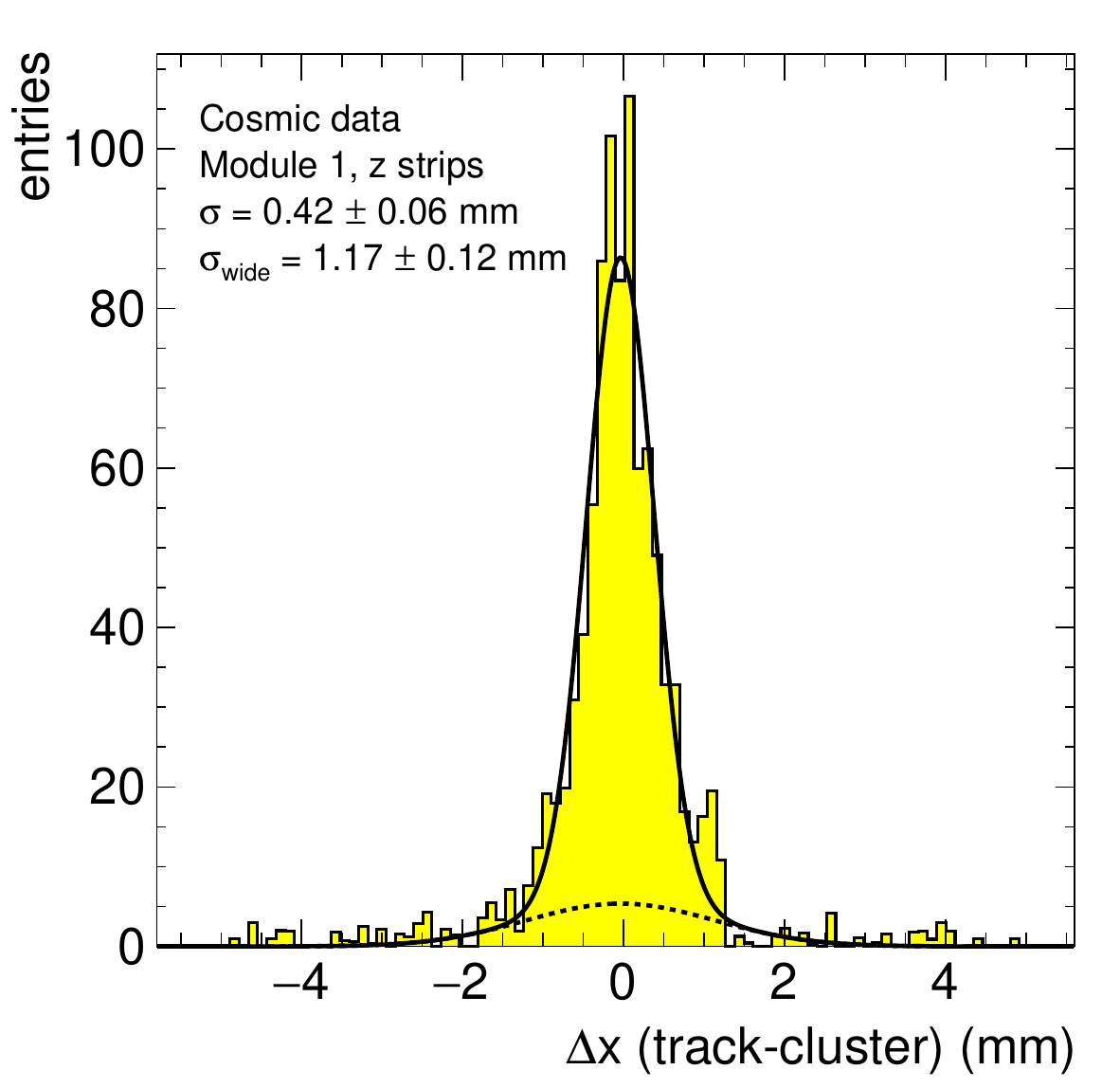}
 \caption{Distribution of the residuals (cluster-reference) in two chambers of the same TPOT module, for a resistive layer HV of 410\,V. Left: $\phi$ view; right: $z$ view.} 
 \label{fig:residuals}
\end{figure}

Figure~\ref{fig:residuals} shows the residuals measured in the two chambers of a given TPOT module, obtained by comparing the position of the measured cluster to that of the track provided by the reference detectors of the cosmic test bench. The width of the residuals distribution is the quadratic sum of the chamber intrinsic spatial resolution, the accuracy of the reference track and the contribution from multiple scattering in the detectors. The combined contribution from multiple scattering and track accuracy is estimated to be 200\,$\mu$m (about 150\,$\mu$m each). The sum of two Gaussian distributions is used to fit the signal. The width of the narrow Gaussian distribution is 390\,$\mu$m (420\,$\mu$m) for the $\phi$ ($z$) view, respectively, corresponding to estimated detector resolutions of 330\,$\mu$m (370\,$\mu$m). The larger value measured for the $z$ view is due to its larger pitch. The wide Gaussian distribution is used to describe the tails of the residuals distribution. It has a width of about 1.2\,mm for both views.

All the Micromegas chambers build at Saclay for the TPOT detector have been characterized on the cosmic test bench shown in Figure~\ref{fig:cosmic_testbench}, using the same, systematic procedure. Performances similar to those shown in Figures~\ref{fig:efficiency_map}, \ref{fig:efficiency_plateau} (left panel) and~\ref{fig:residuals} have been measured for the 16 chambers used in the TPOT detector. Table~\ref{tab:efficiency} shows the detection efficiencies and leak rates measured for nine out of the ten TPOT modules constructed at Saclay. The first module, Module 1, not listed in the table, was sent early on to BNL for testing with the SAMPA electronics (Section~\ref{subsec:characterization_bnl}). It was therefore not characterized in the same manner, nor used for the final TPOT detector. Module 2, which shows a slightly lower efficiency for the $z$ view was not used for the TPOT
detector either.

\begin{table}[htb]
    \centering
    \begin{tabular}{c|c|c|c|c}
    \hline
    Module&Resist HV&Eff. ($\phi$)&Eff. ($z$)& Max. leak rate\\
    &(V)&(\%)&(\%)&(l/h)\\
    \hline
    Module 2&420&99.2&92.9&0.03\\
    Module 3&420&98.7&98.1&0.03\\
    Module 4&430&99.1&98.3&0.02\\
    Module 5&430&99.0&98.1&0.03\\
    Module 6&420&98.2&98.2&0.03\\
    Module 7&420&99.6&97.9&0.01\\
    Module 8&430&99.1&98.6&0.03\\
    Module 9&430&99.1&98.7&0.04\\
    Module 10&430&99.1&98.8&0.04\\
    \hline
    \end{tabular}
    \caption{Detection efficiency and leak rate measured at Saclay for nine out of ten TPOT modules, during detector fabrication.}
    \label{tab:efficiency}
\end{table}

\clearpage
\subsection{\label{subsec:characterization_bnl}Detector characterization at BNL}
%\subsubsection{Detection efficiency with SAMPA electronics}
The detector performance presented in section~\ref{subsec:characterization_saclay} was measured using the data acquisition system available at Saclay, and in particular the DREAM ASIC~\cite{flouzat2014dream}. Compared to SAMPA, primarily designed for low-capacitance detectors (a few 10\,pF), DREAM is better suited to the typical capacitance of Micromegas detectors (up to 300\,pF in the TPOT case) leading to improved signal-to-noise ratio, smaller signal detection threshold and a higher detection efficiency at a given amplification gain.

\begin{figure}[htb]
  \centering
    \includegraphics[width=0.7\textwidth]{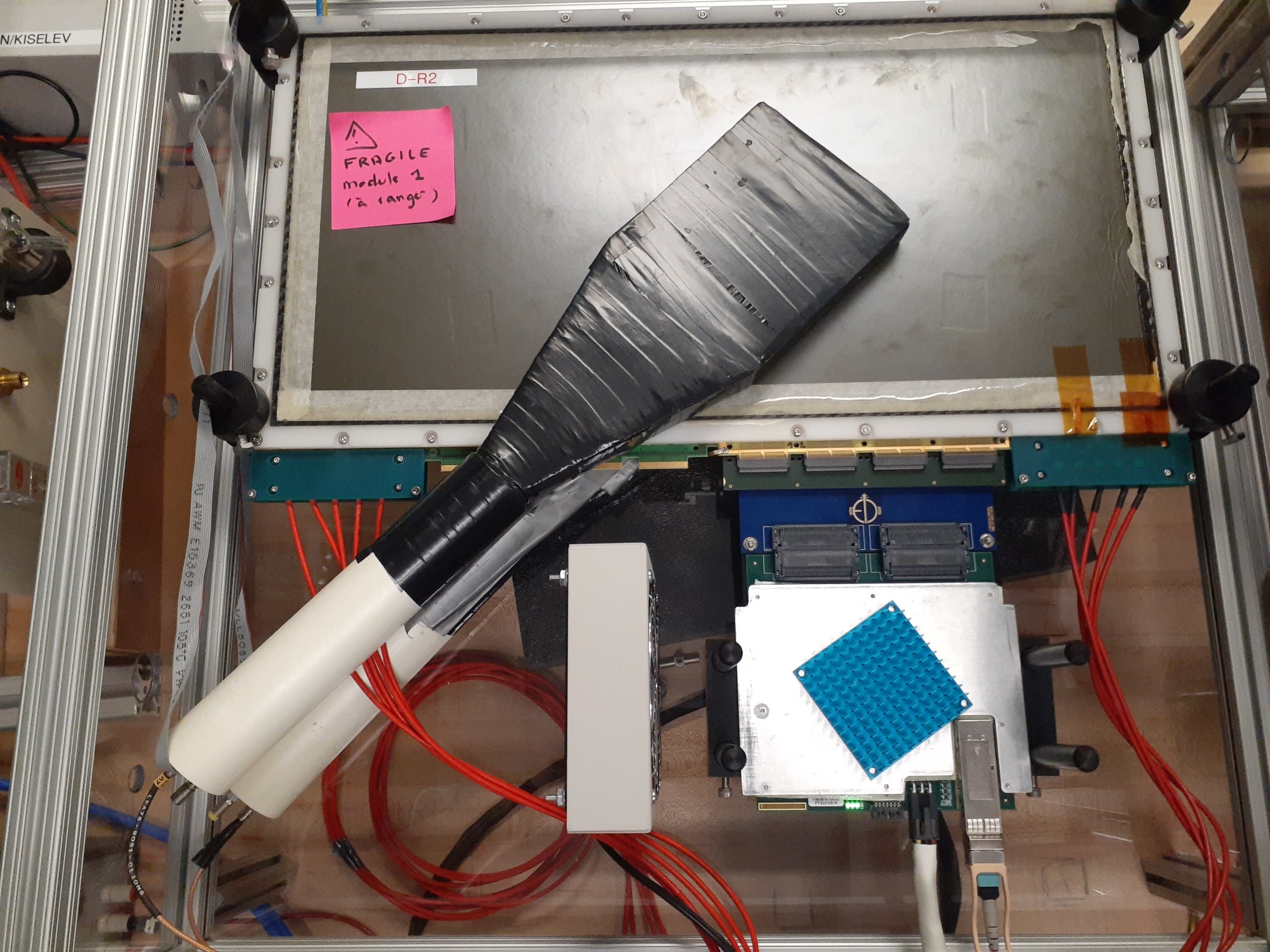}
  \caption{Picture of cosmic test bench used at BNL for the characterisation of the first TPOT module equipped with the SAMPA electronics.} 
  \label{fig:cosmic_testbench_bnl}
\end{figure}

Figure~\ref{fig:efficiency_plateau} (right) shows the detection efficiency measured as a function of the resistive layer HV in the $z$ view of the first TPOT module received at BNL and equipped with the SAMPA electronics. This measurement is performed using a simplified cosmic test bench available at BNL and shown in Figure~\ref{fig:cosmic_testbench_bnl}. Reference cosmic tracks are triggered on using a coincidence of four large scintillating pads mounted on PMTs, to define the region of interest in the detector, one of which is located above the detector and the other three below. Efficiency is defined as the fraction of triggers for which at least one hit above threshold is measured in the detector. The threshold is set to five times the equivalent noise charge (ENC) above pedestal. Compared to figure~\ref{fig:efficiency_plateau} (left), the detection efficiency measured at a given voltage with SAMPA is systematically lower than that measured with DREAM. Full efficiency is reached for voltages larger than 420\,V, as opposed to 400\,V with DREAM.

\clearpage
\subsection{\label{subsec:commissioning}Detector commissioning with beam and cosmic rays}
%\subsection{\label{subsec:commissioning}Detector commissioning with beam and cosmic rays}

\subsubsection{Noise Levels and Timing}
\begin{figure}[htb]
  \centering
    \includegraphics[width=0.48\textwidth]{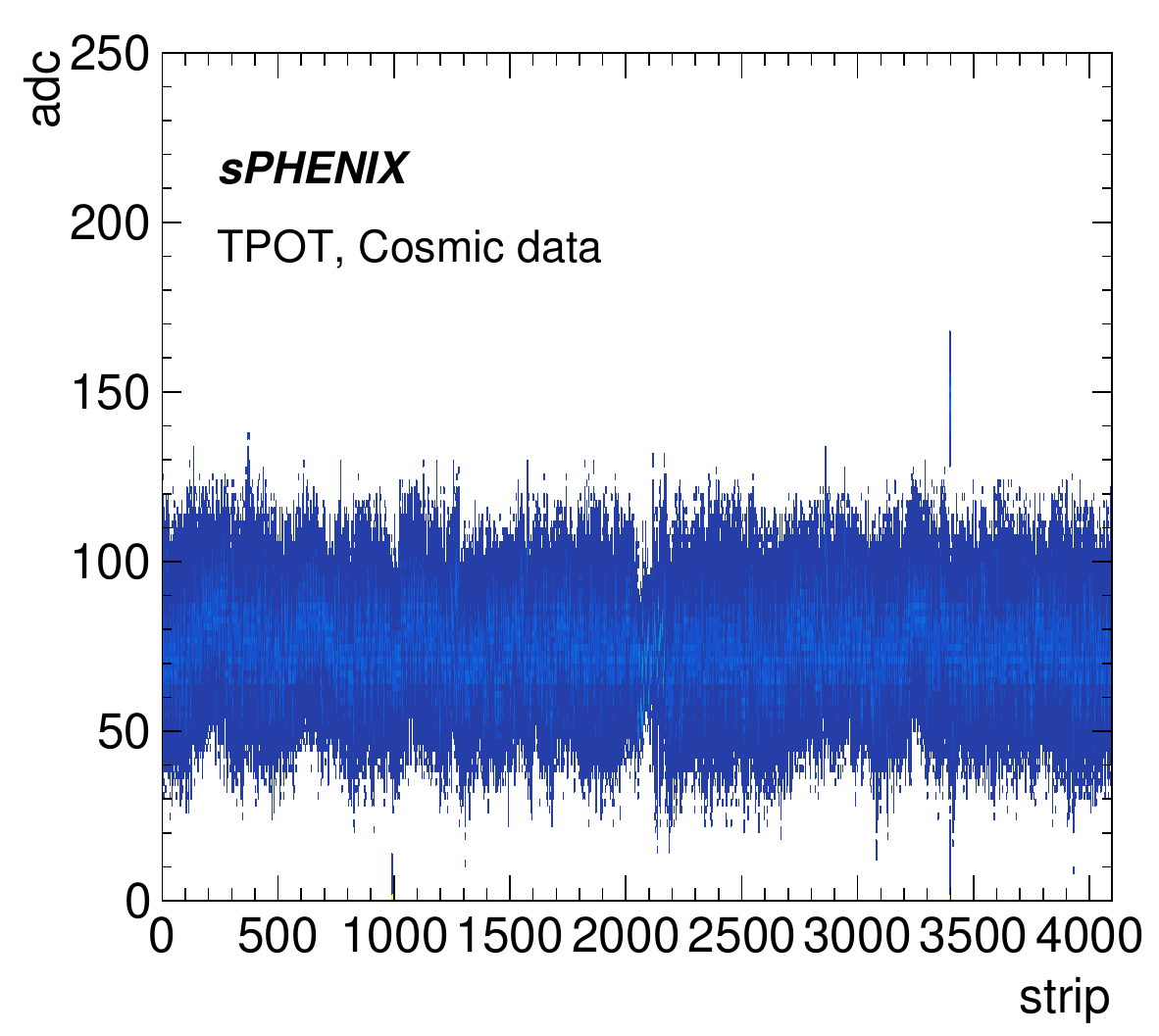}
    \includegraphics[width=0.48\textwidth]{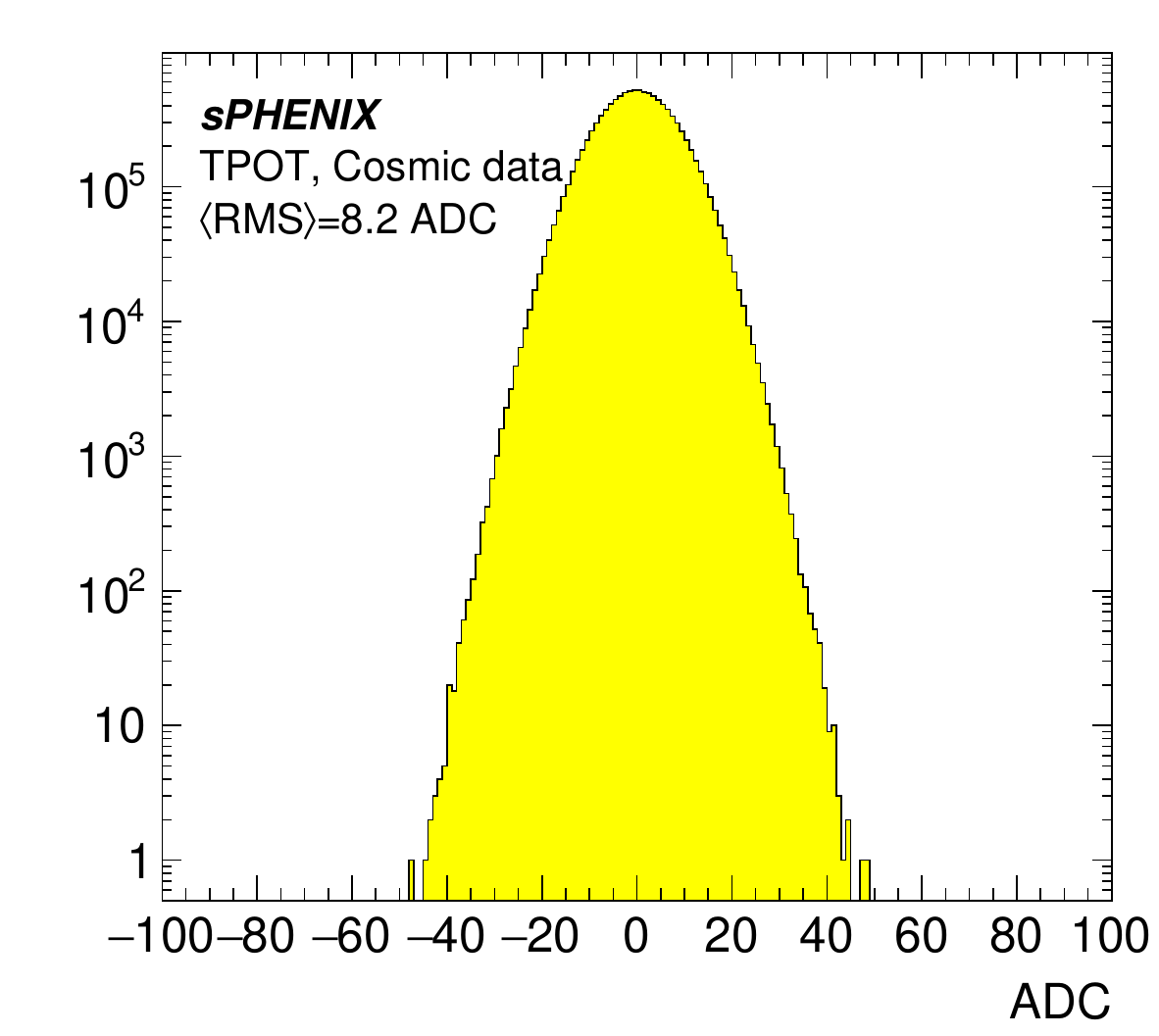}
  \caption{Left: Pedestal and noise levels measured in all TPOT channels during data taking. Right: pedestal-corrected noise count distribution. The RMS of the distribution is approximately 8 ADC counts.}
  \label{fig:noise_levels}
\end{figure}

Noise levels and pedestals measured post-installation inside sPHENIX are presented in Figure~\ref{fig:noise_levels}, on the left as a function of strip number and on the right, integrated over all strips. 
Noise levels are uniform across all 4096 TPOT strips, except for the region around strip 2100 where the electronics is not connected to the detector (section~\ref{installation_mechanics}) and for which the input capacitance is nearly zero. Excluding this region, the Root Mean Square (RMS) of the noise levels
is about 8 ADC counts, corresponding to $7\times10^3$ ($8\times10^3$) electrons ENC for the $\phi$ $(z)$ views, respectively, and using the effective gains quoted in section~\ref{subsubsec:sampa}. For comparison, the typical ENC achieved with DREAM amounts to $6\times10^3$ electrons.

\begin{figure}[htb]
  \centering
    \includegraphics[width=0.7\textwidth]{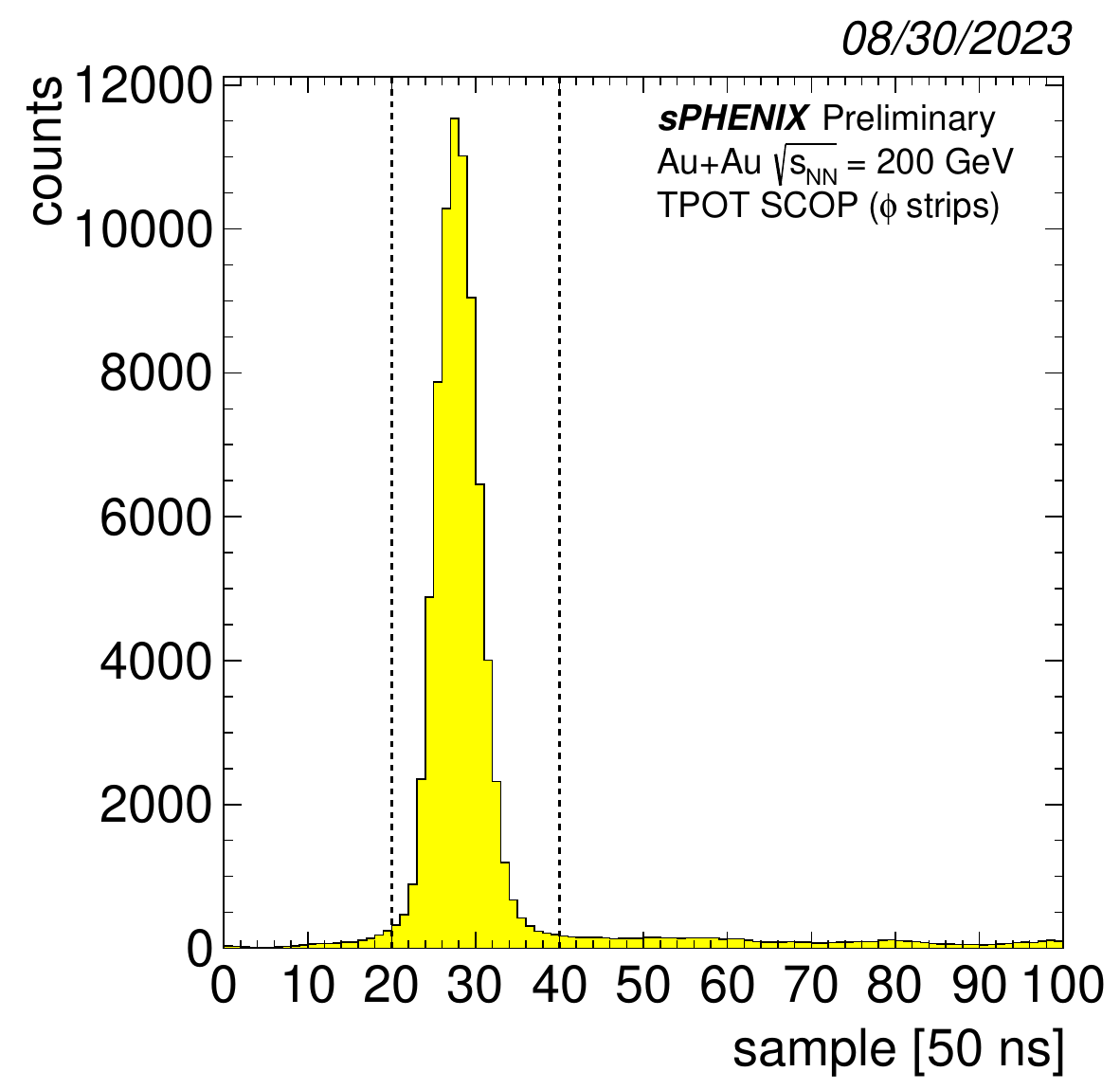}
  \caption{Signal time for particles passing through one TPOT chamber measured with respect to trigger time.}
  \label{fig:timing}
\end{figure}

The timing of the signal with respect to the trigger is verified by counting the number of hits above threshold as a function of time after trigger. The resulting distribution is shown in Figure~\ref{fig:timing}. It is measured using $\snn=200$\,GeV $\auau$ collisions at RHIC, MBD for triggering\footnote{The MBD trigger requires the coincidence of at least one hit in each of the two MBD sections on either side of the interaction point.} and with the TPOT detector operated at nominal voltage (Section~\ref{subsubsec:operation_point}). The majority of signal hits are well contained between samples 20 and 40, corresponding to a time window of 1\,$\mu$s.

\subsubsection{\label{subsubsec:operation_point}Operation Point Determination}
\textit{Without magnetic field:}
a first estimate of the post-installation detection efficiency of the Micromegas chambers is obtained by using the correlation between the number of clusters measured in the two views of the same module. In absence of magnetic field and for a drift HV of -100\,V, full efficiency is reached at a resistive layer HV of approximately 400\,V. This is consistent with the measurements discussed in Sections \ref{subsec:characterization_saclay} and \ref{subsec:characterization_bnl}.

\begin{figure}[htb]
  \centering
    \includegraphics[width=0.95\textwidth]{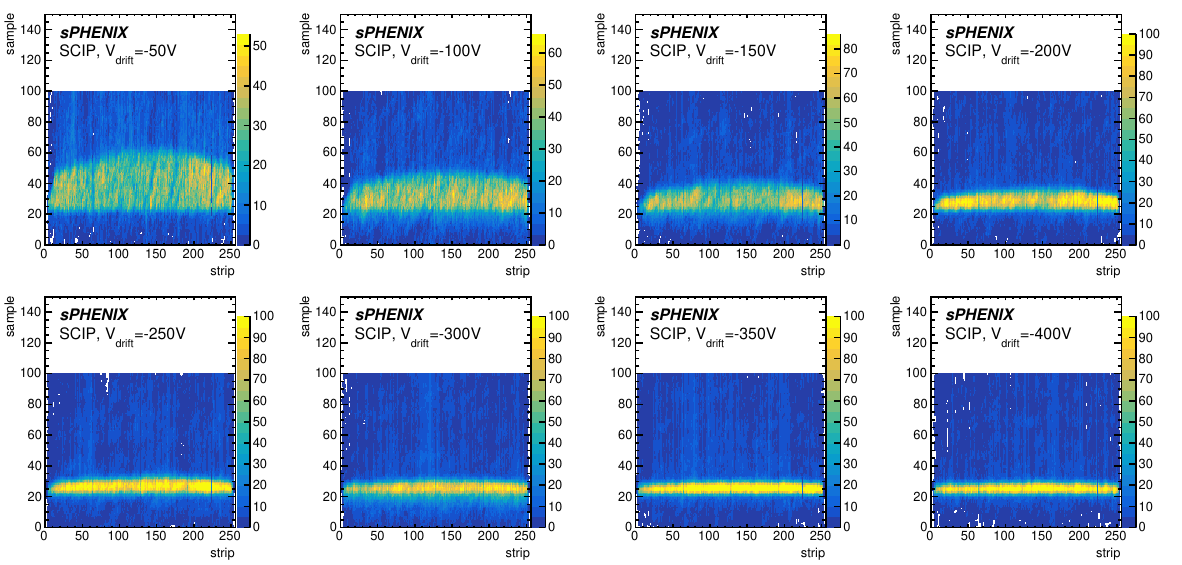}
    \includegraphics[width=0.95\textwidth]{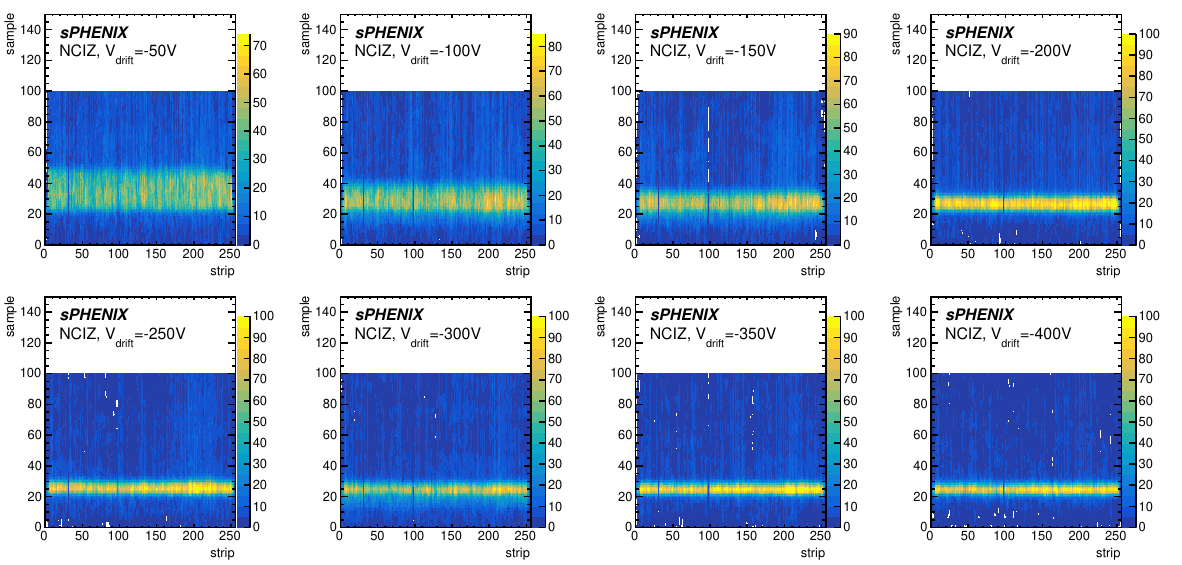}
  \caption{Signal timing distribution as a function of strip number for different values of the drift electrode HV in both chambers of the same TPOT module. Top two rows: $\phi$ view; bottom two rows: $z$ view.}
  \label{fig:drift_scan}
\end{figure}

\textit{With magnetic field:}
the sPHENIX magnetic field is perpendicular to the drift direction of the primary electrons deposited by particles in the Micromegas drift space. Consequently it curves the trajectory of the primary electrons in the transverse plane, resulting in a non-zero Lorentz angle, defined as the angle between the drift direction of the electrons and the direction of the electric field. Too large values of the Lorentz angle result in deteriorated spatial resolution and efficiency due to spread of primary electrons. For a given magnetic field, the Lorentz angle is reduced by increasing the drift electric field. This is achieved by increasing the drift HV. 

Figure~\ref{fig:drift_scan} shows the distribution of the signal time as a function of the position in the detector (labeled as strip number) for scanned values of the drift electrode HV and in two TPOT chambers. The distributions are measured with $\auau$ collisions, MBD trigger and magnet on. For small absolute values of the drift electrode HV (e.g. -50\,V and -100\,V) the timing distributions are significantly wider than that shown in Figure~\ref{fig:timing} and the width reaches 40 sample units (2\,$\mu$s). In addition the distributions exhibit a dependence on the position in the chamber for the $\phi$ view, not observed for the $z$ view. This difference is attributed to the direction along which the electron trajectory is bent with respect to that of the measuring strips: for $\phi$ ($z$) views, the electron trajectory is bent perpendicular to (along) the strip direction.

The width of the distributions decreases with increasing drift electrode HV and the dependence on the position in the chamber vanishes. For values larger (in magnitude) than -300\,V, the distributions become as narrow as that measured without magnetic field, indicating that the Lorentz angle becomes small enough that it does not significantly alter the signal collection in the chamber. With this study, it was decided to operate the drift electrodes at a HV no smaller in magnitude than -300\,V.

\begin{figure}[htb]
  \centering
    \includegraphics[width=0.45\textwidth]{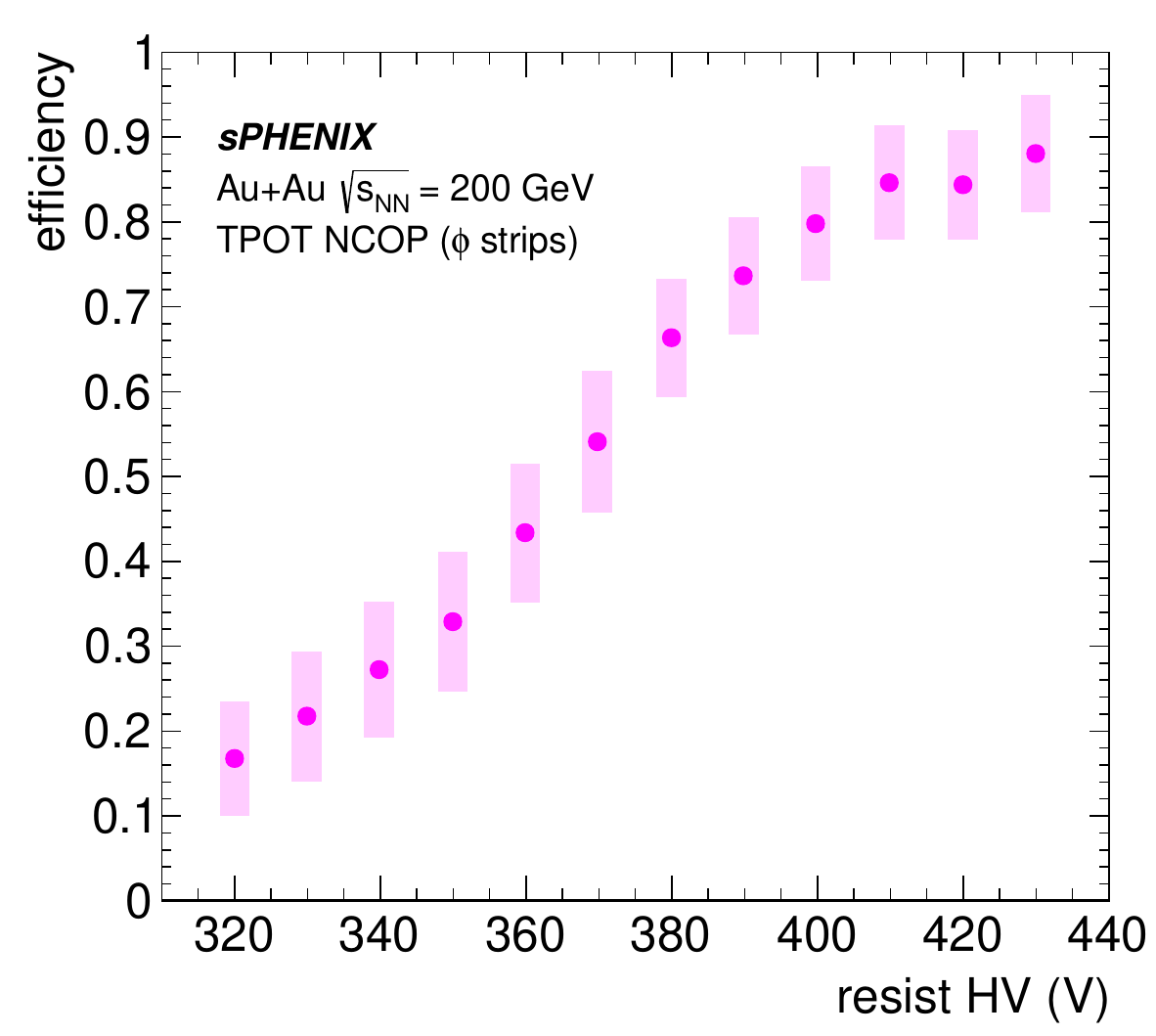}
    \includegraphics[width=0.45\textwidth]{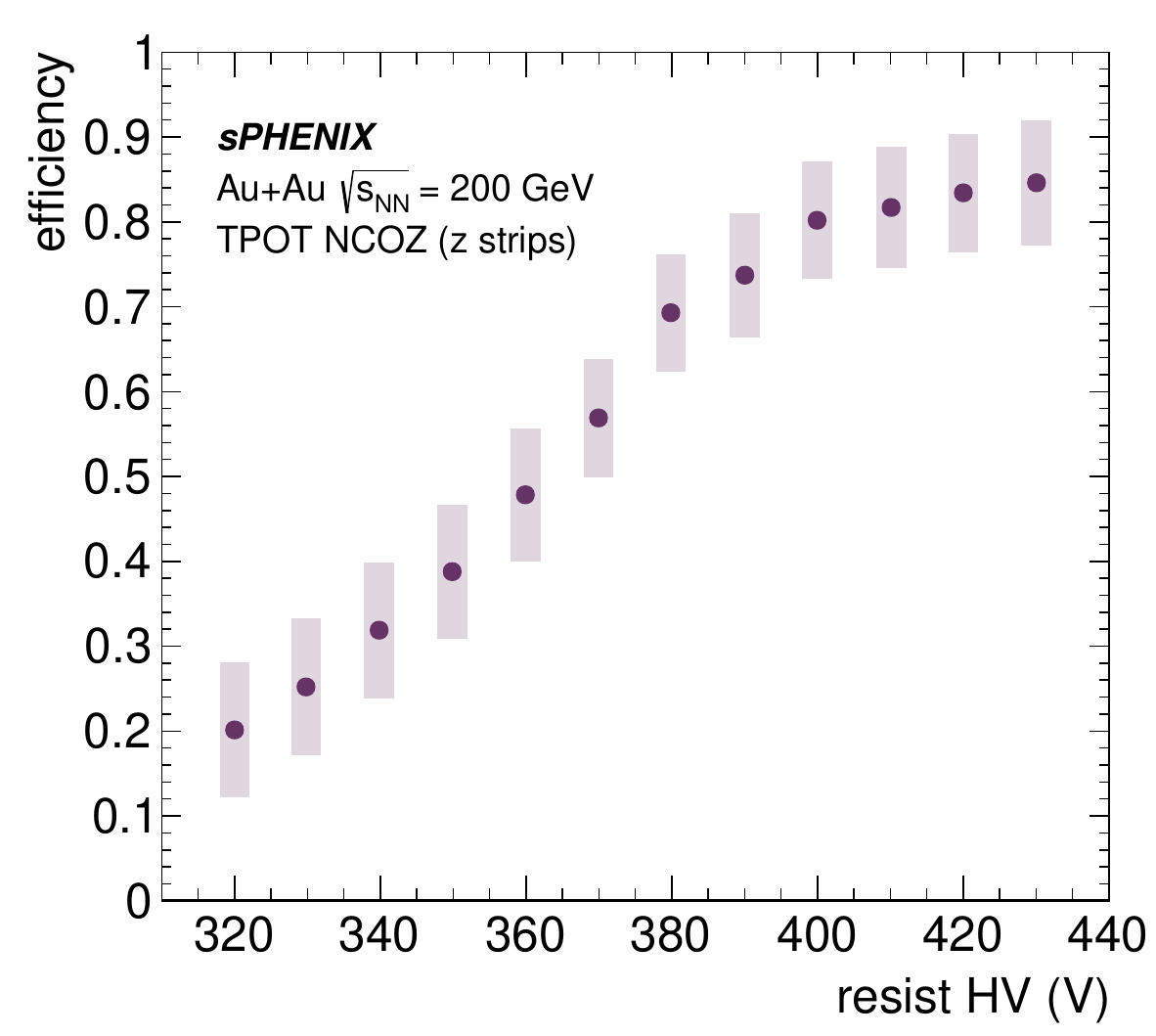}
  \caption{Estimate of the detection efficiency measured in two chambers of the same TPOT module during data taking, as a function of the bias voltage. The boxes around the points represent the systematic uncertainties. See text for details.}
  \label{fig:efficiency_plateau_bnl_2}
\end{figure}

Figure~\ref{fig:efficiency_plateau_bnl_2} shows an estimate of the detection efficiency measured in the two chambers of the same TPOT module as a function of the resistive layer HV with magnetic field on and a drift electrode HV of -400\,V. The boxes around the points represent the systematic uncertainties on the measurement, estimated by varying the selection criteria of either the reference or the measured cluster sample. They correspond to the best accuracy with which the detection efficiency of the TPOT chambers could be measured in $\auau$ collisions and during data taking. They are largely correlated point-to-point. Full efficiency is reached starting from approximately 410\,V. The detection efficiency dependence on the resistive layer voltage is similar to that shown in Figure~\ref{fig:efficiency_plateau} (right) and measured with cosmic tracks. However, i) the efficiency measured on the plateau is lower (about 90$\pm$7\% as opposed to $100$\% in Figure~\ref{fig:efficiency_plateau_bnl_2}) and ii) the efficiency does not drop to zero for small values of the voltage. Both observations are attributed to the poor definition of the reference sample. A more accurate determination of the reference sample and the detection efficiency requires correlating the TPOT measurements to that of other tracking subsystems, in particular the TPC.

\subsubsection{Operation Mode Description}
Based on the results presented in sections~\ref{subsec:characterization_saclay}, \ref{subsec:characterization_bnl} and~\ref{subsubsec:operation_point}, the values chosen as nominal drift and resistive layer HV for detector operation during data taking are: $V_{\rm drift}=-300$\,V, $V_{\rm resist}=400$\,V. This corresponds to a detector amplification gain of about $10^{4}$. Out of the 64 resistive layer HV channels (four for each of the 16 TPOT chambers), 62 are operated at the nominal value while the remaining two are operated at 360\,V and 380\,V because they exhibit high current draw (several $\mu$A) and frequent discharges for larger values. This corresponds to detection efficiencies of about 50 and 70\% respectively (Figure~\ref{fig:efficiency_plateau_bnl_2}). Combining this information with the one signal transition cable disconnected from the chamber during installation, one reaches an overall acceptance for TPOT of about 96\%, assuming 100\% efficiency on the plateau.

\begin{figure}[htb]
  \centering
  \includegraphics[width=0.45\textwidth]{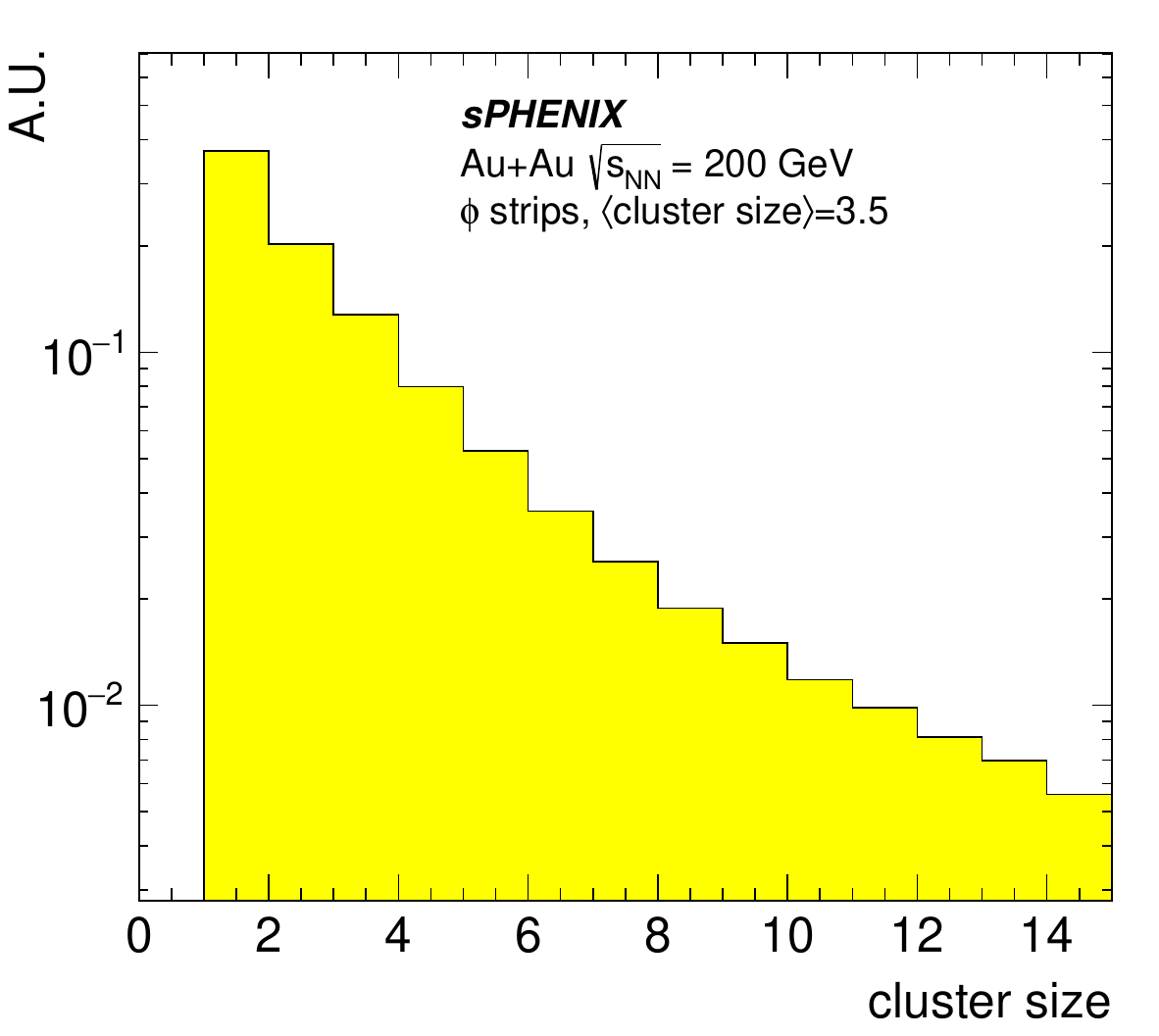}
  \includegraphics[width=0.45\textwidth]{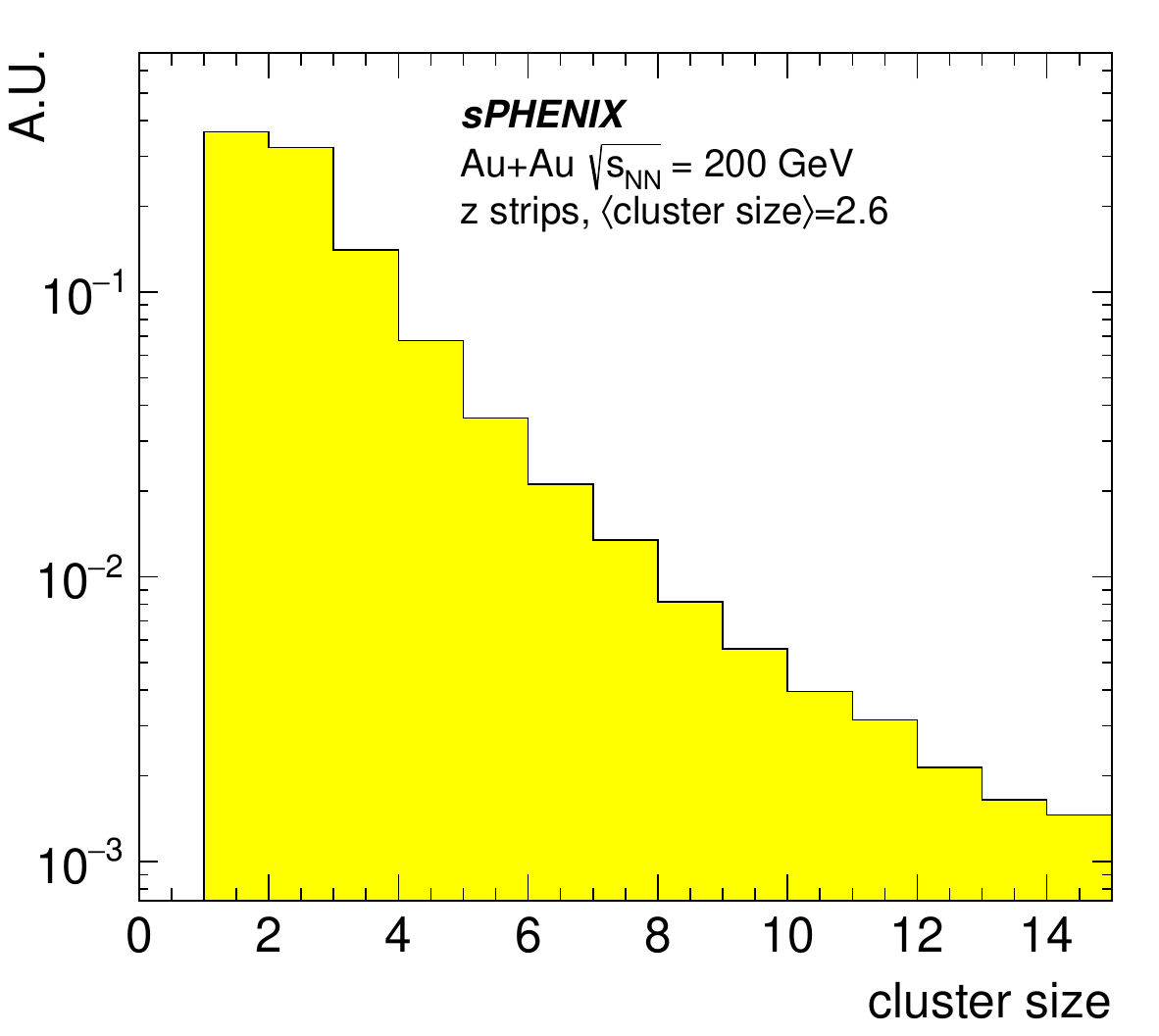}

  \includegraphics[width=0.45\textwidth]{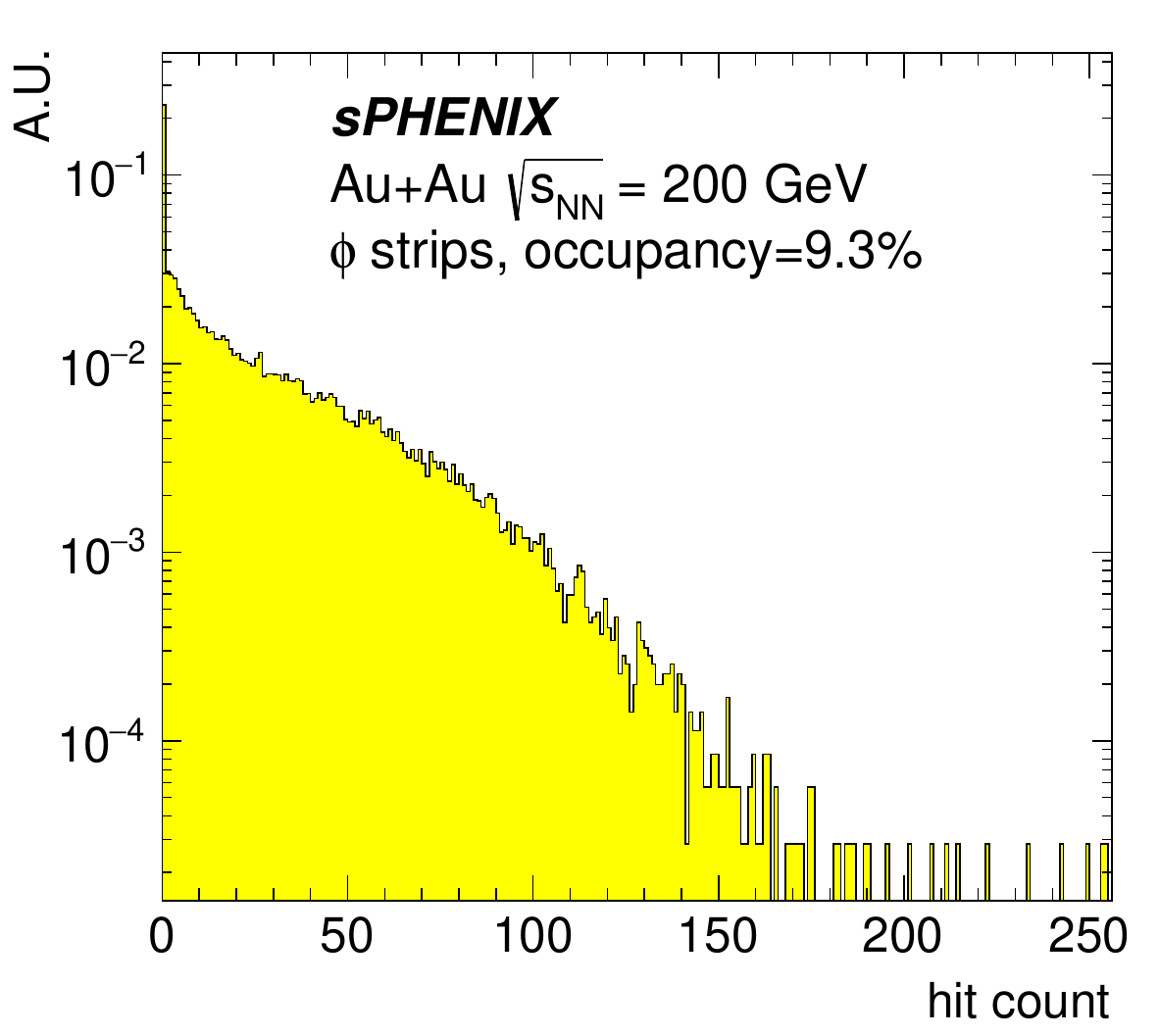}
  \includegraphics[width=0.45\textwidth]{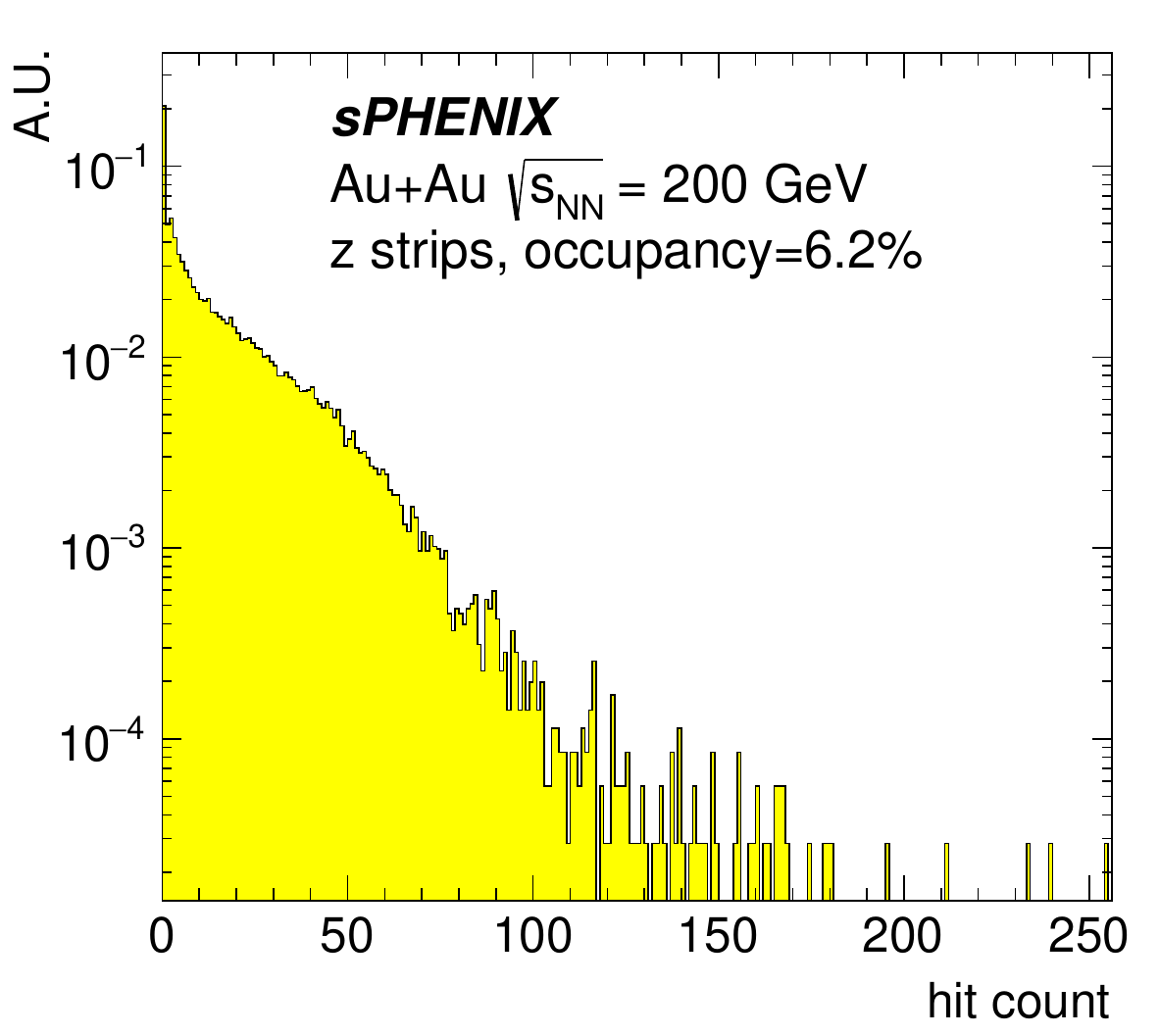}
  
 \caption{Distribution of the cluster size (top) and the number of signal hits per $\snn=200$ GeV $\auau$ collision (bottom) for $\phi$ (left) and $z$ views (right). }
 \label{fig:cluster_size}
\end{figure}

Figure~\ref{fig:cluster_size} shows the cluster size distribution and the number of signal hits per $\auau$ collision in TPOT. The mean cluster size is 3.5 (2.6) for $\phi$ ($z$) views. The larger size measured in the $\phi$ views is attributed to the detector pitch being smaller and the azimuthal angular distribution of the particles passing through the chamber being wider due to the presence of the magnetic field. Similar cluster sizes, within 5\%, are measured for all chambers of the same strip orientation. They are larger than those measured with cosmic data (1.8 and 1.6 for $\phi$ and $z$ views, respectively). This increase is attributed to the angular distribution of particles passing through the detector as well as their mean deposited energy being larger in $\auau$ collisions than with cosmics. The mean occupancy, defined as the mean fraction of strips with a signal above threshold in a given chamber and per $\auau$ collision, is $9.3\%$ ($6.2$) for $\phi$ ($z$) views. The difference between the two is attributed to the larger mean cluster size measured for $\phi$ views. 

\begin{figure}[htb]
  \centering
  \includegraphics[width=0.45\textwidth]{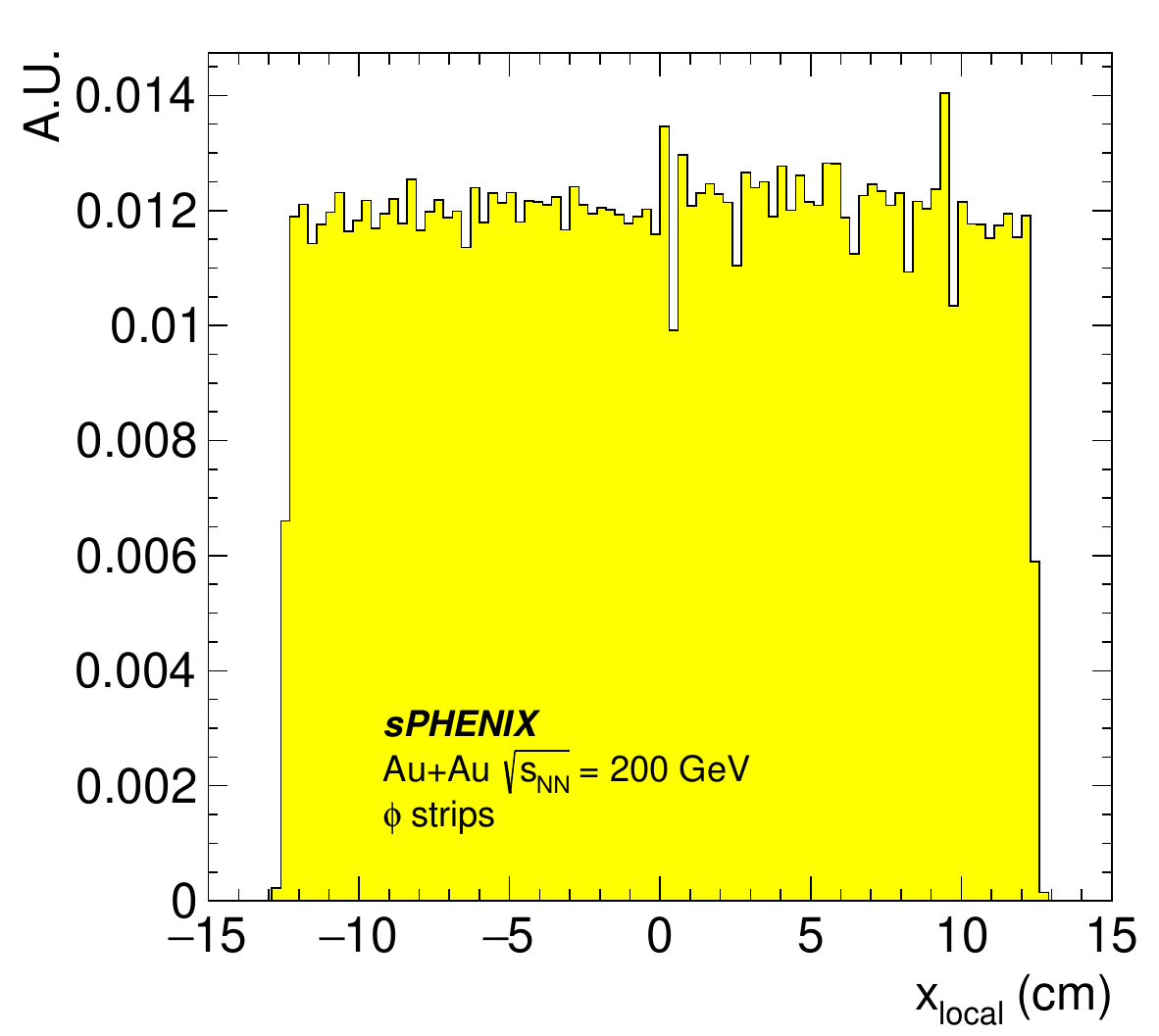}
  \includegraphics[width=0.45\textwidth]{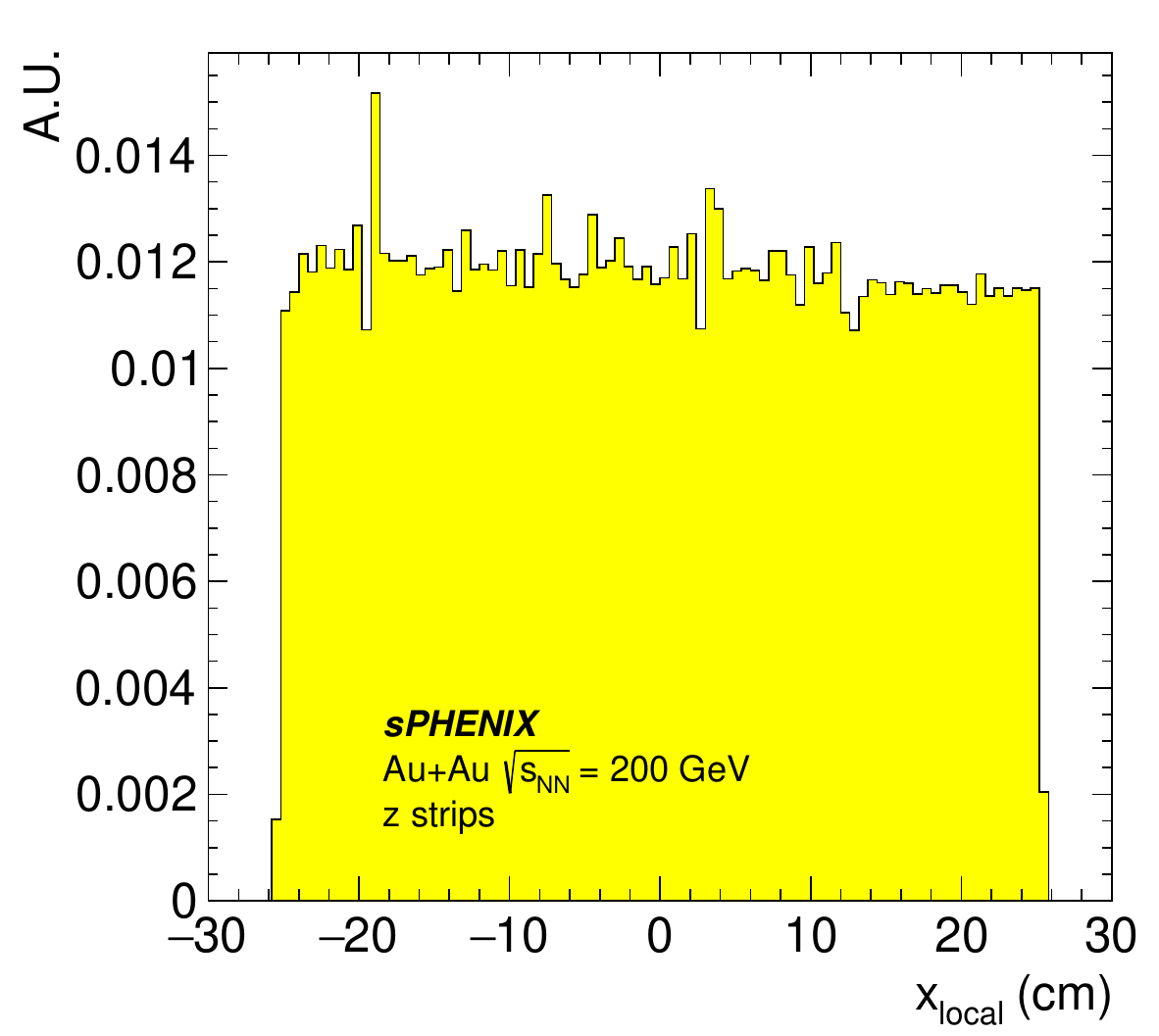}
  \includegraphics[width=0.45\textwidth]{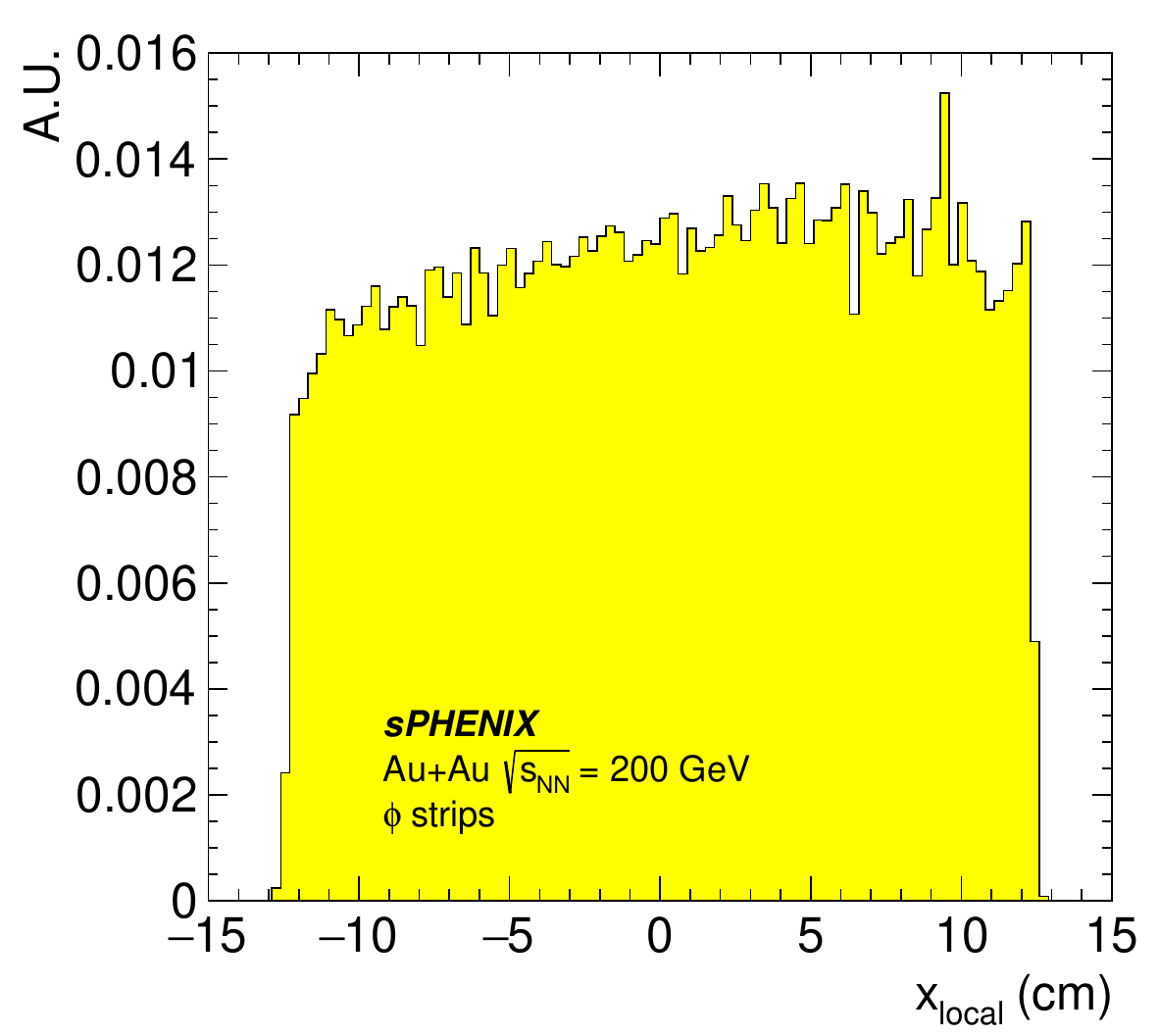}
  \includegraphics[width=0.45\textwidth]{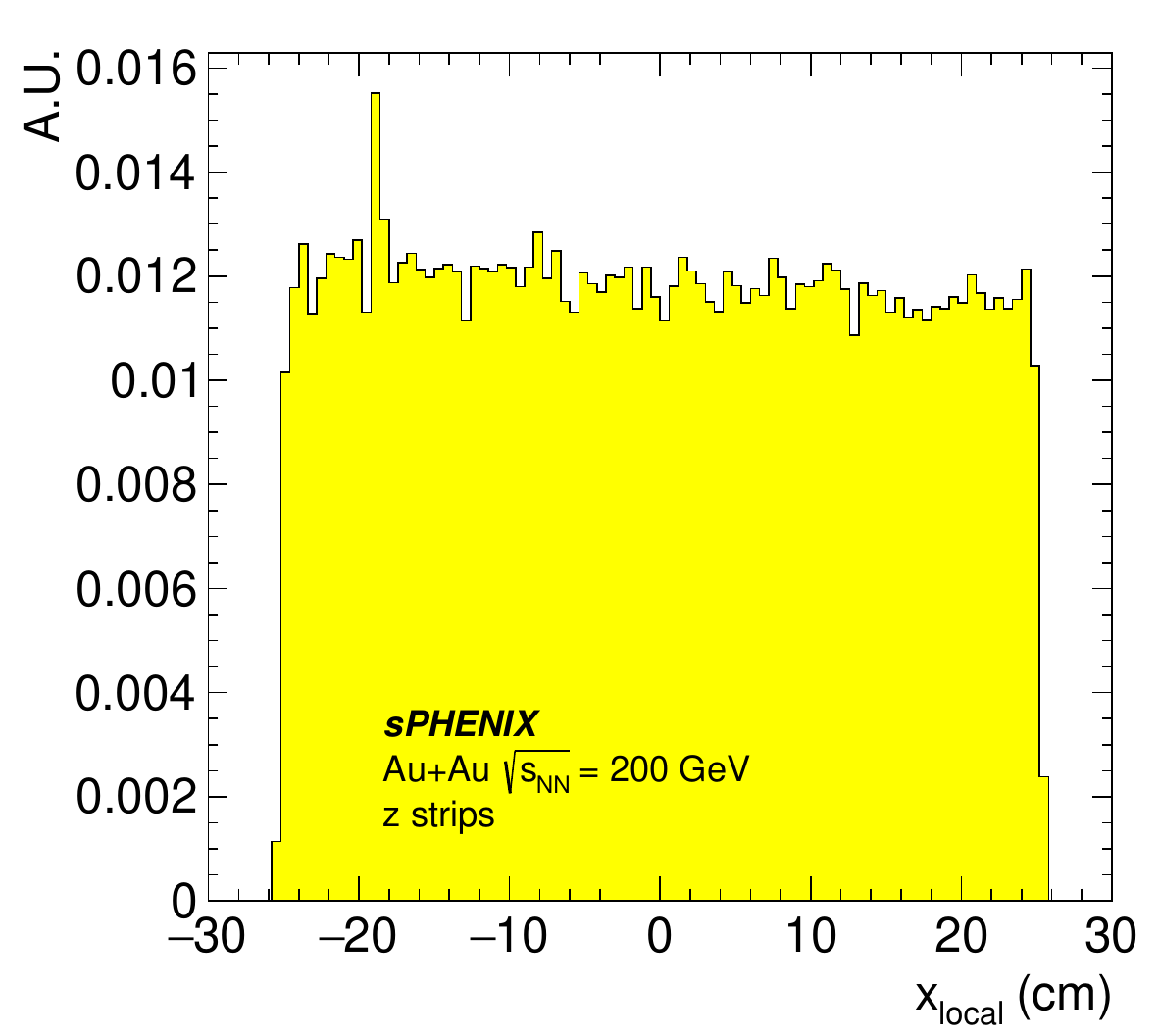}
  \caption{Distribution of the cluster position in the $\Phi$ (left) and $z$ views (right), without magnetic field (top) and with magnetic field (bottom).}
  \label{fig:cluster_profile}
\end{figure}

Figure~\ref{fig:cluster_profile} shows the distributions of the cluster position in the $\phi$ and $z$ views, both with and without magnetic field. For the $z$ views, the distributions show similar $z$ dependence with and without magnetic field. They exhibit a slight slope attributed to the fact that the center of the vertex $z$ distribution is shifted with respect to the center of the TPOT detector.
%the TPOT detector is shifted along $z$ by about 3\,mm with respect to the center of the vertex distribution. 
On the contrary for the $\phi$ views, the distribution is flat without magnetic field, but exhibits an approximately linear dependence on the position with magnetic field. This dependence is attributed to the effect of the magnetic field on the drift of the primary electrons in the chamber, which occurs in the direction parallel to the measurement.

\subsubsection{Correlation with Other Subsystems}

\begin{figure}[htb]
  \centering
    \includegraphics[width=0.9\textwidth]{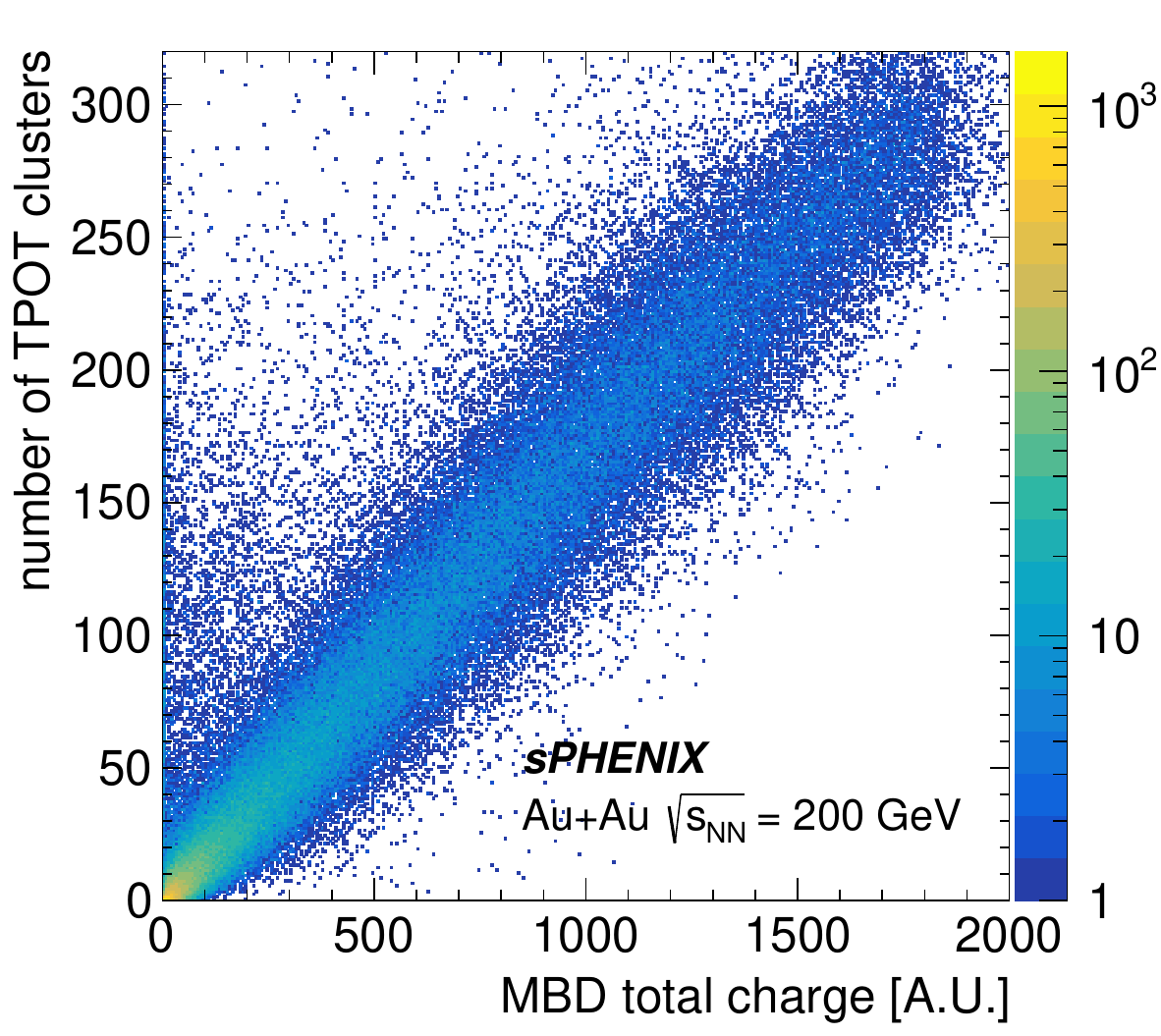}
  \caption{Correlation between the total charge measured in sPHENIX MBD and the total number of clusters measured in TPOT.}
  \label{fig:mbd_correlation}
\end{figure}

Figure~\ref{fig:mbd_correlation} shows the correlation between the total charge measured in the MBD and the total number of clusters measured in TPOT. The two quantities are strongly correlated despite the fact that there is no overlap between the acceptance of the two detectors (Figure~\ref{fig:sphenix_layout}). The correlation proves that the signal measured in TPOT corresponds to particles created in the same $\auau$ collision as that recorded by the MBD. Events with a small number of clusters and small charge in the MBD correspond to peripheral collisions for which the distance between the center of the colliding nuclei is large. Conversely, events with a large number of clusters and a large charge in the MBD correspond to central collisions for which this distance between the two centers is small. 

\begin{figure}[htb]
  \centering
    \includegraphics[width=0.9\textwidth]{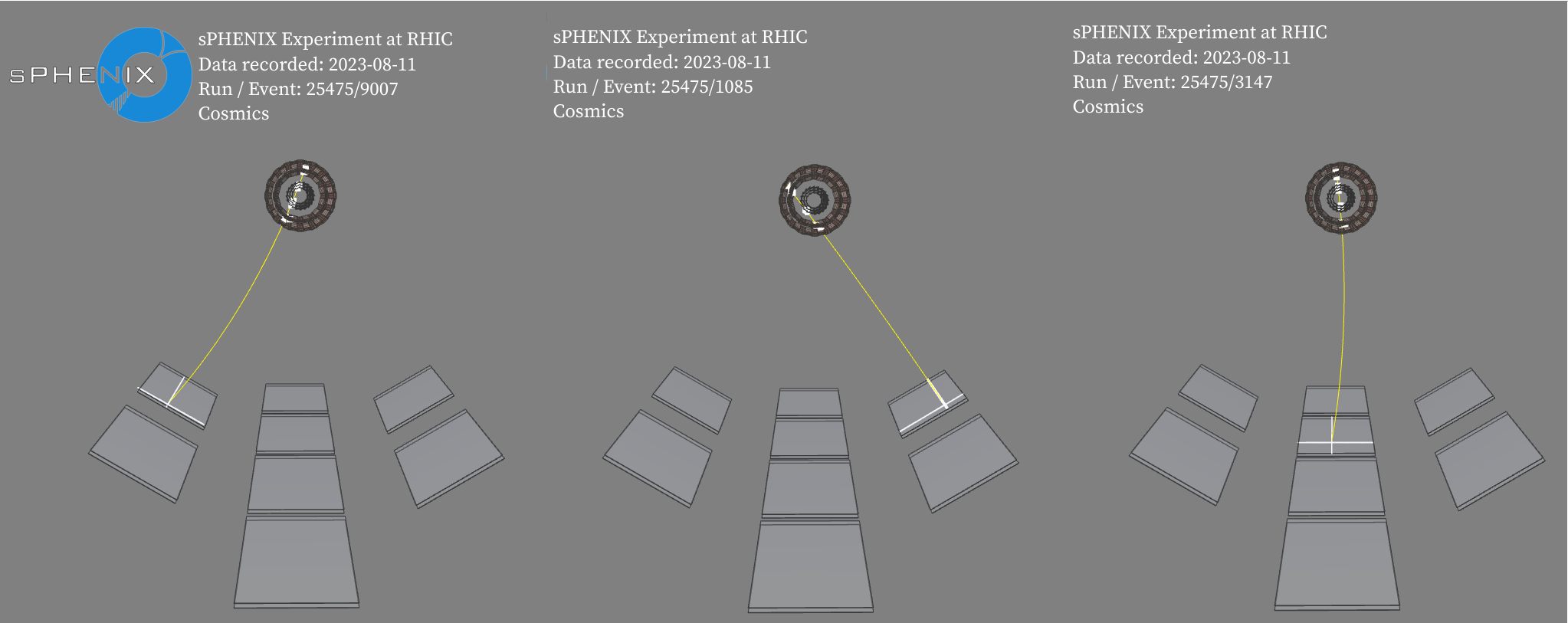}
  \caption{Three examples of cosmic tracks reconstructed in sPHENIX MVX, INTT and TPOT detectors.} 
  \label{fig:cosmic_tracks}
\end{figure}

Three examples of cosmic tracks reconstructed in MVTX, INTT and TPOT are shown in Figure~\ref{fig:cosmic_tracks}. The curvature of the trajectory is due to the presence of the magnetic field. Successfully reconstructing these trajectories with the MVTX, INTT and TPOT and extrapolating it inside the TPC is key to measuring and correcting for the beam-induced distortions of the electron drift inside the TPC.

\section{\label{sec:conclusion}Conclusion}
In this paper, the Time Projection Chamber Outer Tracker (TPOT) installed in the sPHENIX experiment is presented. It sits on the outside of the TPC and provides an additional space point along the particle trajectory to better constrain the tracks inside the TPC and help calibrate out the distortions of the electron drift in the TPC volume.
The detector design, installation, characterization and performance are described. 
TPOT is fully commissioned and ready to collect data for the upcoming RHIC runs. Using cosmic data it has demonstrated the ability to provide reference trajectories to the TPC together with the other tracking detectors.

\section*{\label{sec:acknowledgments}Acknowledgments}
We are grateful to the sPHENIX Collaboration for their support and help during the various phases of the TPOT project. We thank the Brookhaven National Laboratory sPHENIX staff for their help during assembly and installation, and in particular C. Biggs, J. Labounty, M. Lenz, S. Pollizo and F. Toldo. We also thank the MBD, MVTX and INTT teams for providing the data appearing in Figures~\ref{fig:mbd_correlation} and ~\ref{fig:cosmic_tracks}, as well as the staff of the Collider-Accelerator and Physics Departments at Brookhaven National Laboratory. We acknowledge support from the Office of Nuclear Physics in the Office of Science of the Department of Energy, the National Science Foundation, the Commissariat à l’Énergie Atomique (France), and Fundação de Amparo à Pesquisa do Estado de São Paulo (Brasil), processo 2014-12664-3.

This manuscript has been authored by employees of Brookhaven Science Associates, LLC under Contract DE-SC0012704 with the U.S. Department of Energy. The publisher by accepting the manuscript for publication acknowledges that the United States Government retains a non-exclusive, paid-up, irrevocable, world-wide license to publish or reproduce the published form of this manuscript, or allow others to do so, for United States Government purposes.

% bibliography
\bibliographystyle{elsarticle-num} 
\bibliography{references.bib}

\end{document}